\documentclass[letterpaper,twocolumn,10pt]{article}
\usepackage{arxiv}

\usepackage[table]{xcolor}
\usepackage{inconsolata}
\usepackage{graphicx}
\usepackage{booktabs}
\usepackage{multirow}
\usepackage{amsmath}
\usepackage{amssymb}
\usepackage{adjustbox}
\IfFileExists{bbm.sty}{\usepackage{bbm}}{}
\providecommand{\mathbbm}[1]{\mathbb{#1}}
\usepackage{balance}
\usepackage{caption}
\usepackage{url}
\usepackage{hyperref}
\hypersetup{hidelinks}
\usepackage{enumitem}
\usepackage{tikz}
\usetikzlibrary{
    arrows.meta,
    positioning,
    fit,
    calc,
    decorations.pathreplacing
}
\usepackage{pgfplots}
\pgfplotsset{compat=1.18}
\usepgfplotslibrary{groupplots}

\newcommand{\sysname}{MalSets}
\newcommand{\sys}{\textsc{\sysname}}
\newcommand{\dsname}{DeepSets}
\newcommand{\ds}{\textsc{\dsname}}
\newcommand{\dsv}[1]{\textsc{\dsname}\ensuremath{[#1]}}
\newcommand{\mech}{Deep Sets}
\newcommand{\dd}{\textsc{DeepDrebin}}
\newcommand{\ddplus}{\textsc{DeepDrebin+}}
\newcommand{\dr}{\textsc{Drebin}}
\newcommand{\drfe}{Drebin}
\newcommand{\drplus}{\textsc{Drebin+}}
\newcommand{\hcc}{\textsc{HCC}}
\newcommand{\cade}{\textsc{CADE}}
\newcommand{\ramda}{\textsc{RAMDA}}
\newcommand{\drmd}{\textsc{DRMD}}
\newcommand{\tif}{\textsc{TIF}}
\newcommand{\secsvm}{\textsc{Sec-SVM}}
\newcommand{\rf}{\textsc{RandomForest}}
\newcommand{\malcertain}{\textsc{MalCertain}}
\newcommand{\lgbm}{\textsc{LightGBM}}
\newcommand{\monogbt}{\textsc{MonoGBT}}
\newcommand{\lsef}{\textsc{LSE-Forest}}

\newcommand{\MAX}{\operatorname{MAX}}
\newcommand{\MEAN}{\operatorname{MEAN}}
\newcommand{\SUM}{\operatorname{SUM}}
\newcommand{\Pmal}{P(\mathrm{mal})}
\newcommand{\dphi}{d_{\phi}}

\definecolor{segtrain}{HTML}{0072B2}
\definecolor{segcal}{HTML}{E69F00}
\definecolor{segtest}{HTML}{009E73}
\colorlet{segtrainbg}{segtrain!18}
\colorlet{segcalbg}{segcal!24}
\colorlet{segtestbg}{segtest!20}
\DeclareRobustCommand{\segmark}[2]{\tikz[baseline=(hm.base)]\node[fill=#1bg, rounded corners=1.8pt, inner xsep=1.6pt,
   inner ysep=0.6pt, outer sep=0pt, text height=1.6ex, text depth=0.45ex](hm){#2};}
\newcommand{\ultrain}[1]{\segmark{segtrain}{#1}}
\newcommand{\ulcal}[1]{\segmark{segcal}{#1}}
\newcommand{\ultest}[1]{\segmark{segtest}{#1}}

\DeclareRobustCommand{\setting}[1]{\tikz[baseline=(sb.base)]\node[draw=black, line width=0.45pt, rounded corners=1.8pt,
   font=\sffamily\bfseries, inner xsep=2pt, inner ysep=1pt, outer sep=0pt, text height=0.70em, text depth=0.21em](sb){Setting~#1};}
\definecolor{stepgold}{HTML}{8B6508}
\DeclareRobustCommand{\panel}[1]{\tikz[baseline=(pl.base)]\node[rectangle,draw=black!70,line width=0.5pt,inner sep=0pt,minimum size=1.1em,text height=1.4ex,text depth=0pt,font=\bfseries\footnotesize](pl){#1};}
\DeclareRobustCommand{\stepref}[1]{\tikz[baseline=(sr.base)]\node[circle,draw=stepgold,text=stepgold,line width=0.5pt,inner sep=0pt,minimum size=1.1em,font=\footnotesize\bfseries](sr){#1};}
\newcommand{\mkGlyph}[2]{\tikz[baseline=-0.6ex]{\path[draw=black!45,fill=black!12,line width=0.45pt] plot[mark=#1,mark size=#2,mark options={draw=black!45,fill=black!12,line width=0.45pt}] coordinates{(0,0)};}}
\DeclareRobustCommand{\mkLin}{\mkGlyph{square*}{2.3pt}}
\DeclareRobustCommand{\mkTree}{\mkGlyph{triangle*}{3.0pt}}
\DeclareRobustCommand{\mkMLP}{\mkGlyph{*}{2.5pt}}

\title{\bf More Is Never Less: Affordable Robustness to Additive Attacks \\ via Monotone Multiple-Instance Learning}
\author{
    {\rm
        {\bfseries Alexander Herzog}\textsuperscript{\rm$\mathsection$\rm$\parallel$}, 
        {\bfseries Shae McFadden}\textsuperscript{\rm$\dagger$\rm$\ddagger$\rm$\mathsection$}, 
        {\bfseries Lorenzo Cavallaro}\textsuperscript{\rm$\mathsection$\rm$\star$}
    } \\
        \textsuperscript{\rm  $\mathsection$}\textit{University College London},
        \textsuperscript{\rm  $\dagger$}\textit{King's College London}, \\
         \textsuperscript{\rm $\ddagger$}\textit{The Alan Turing Institute},
         \textsuperscript{\rm  $\parallel$}\textit{Core64},
        \textsuperscript{\rm  $\star$}\textit{MBZUAI},
}

\begin{document}
\maketitle

\begin{abstract}

Static malware detectors that represent each sample as a bag of extracted features, such as permissions, API calls, or imported functions, are cheap to deploy and remain state-of-the-art under temporal drift and in selective classification. Yet, attackers can evade them by adding benign-looking content that outweighs malicious evidence. Canonically, detectors encode the presence of a few thousand features, selected during training, as sparse multi-hot vectors over a fixed vocabulary. Features that emerge after training remain invisible. We show that linear models and multi-layer perceptrons (MLPs) over this input are sum-pooled set models in disguise. Sum-pooling lets added benign features cancel malicious evidence.
Instead, we cast malware detection with set-shaped static features as a multiple-instance learning (MIL) problem.
To this end, we propose MalSets, which uses max-pooled Deep Sets models with a monotone head to ensure that any malicious evidence remains undiluted; thus, more is never less.
Under MIL, a general text encoder maps any feature string to an embedding, thus eliminating the long-standing out-of-vocabulary problem at test time. Max-pooling retains the malicious evidence regardless of how many benign features an attacker adds. Since max-pooling can only grow when features are added, constraining the head alone makes the malware score monotone under additions, even against a white-box attacker adapted to this architecture. While monotone detectors are never evaded by state-of-the-art black-box or white-box attacks, they often significantly pay for this robustness in clean performance since benign features cannot count as evidence. However, by combining a non-monotone set encoder with max-pooling and a monotone head, MalSets maintains the robustness of prior monotone classifiers while achieving state-of-the-art performance.

\end{abstract}

\section{Introduction}
\label{sec:intro}

More than twenty years ago, Lowd and Meek showed that adversaries could evade spam filters, using bag-of-words (BoW)-based representations, by embellishing a message with a few ``good words'' that were associated with legitimate email~\cite{lowd2005good}.
However, to this day, the use of static analysis to construct BoW-like feature representations still remains state-of-the-art~\cite{mcfadden2026drmd,chen2023continuous,zheng2026tif,arp2014drebin,li2021ramda} in malware detection as a key first line of defence~\cite{canali2011prophiler,arp2014drebin,sejfia2022amalfi,mcfadden2026replicant} because feature extraction remains lightweight allowing for inexpensive classification at scale.

Like the BoW-based representations considered by Lowd and Meek, these abstractions record the presence or frequency of extracted features.
Commonly considered features for malware detection include:
imported libraries and functions for executables~\cite{schultz2001datamining,anderson2018ember};
keywords and structural paths for PDFs~\cite{biggio2013evasion,srndic2013hidost};
syntax-annotated tokens for JavaScript~\cite{curtsinger2011zozzle};
and permissions, API usage, and metadata for Android applications and software packages~\cite{arp2014drebin,sejfia2022amalfi,ladisa2023crosslang}.
Although individual features may also encode local structure, their aggregation is \emph{permutation invariant}.
We call such feature abstractions, or extractors, \emph{set-shaped}.

Depending on their design, these extractors omit structural or semantic relationships, such as caller-callee dependencies. This creates a \emph{representation gap} between the \textit{sample}, that is, the application, executable, or document under analysis, and the \textit{information} available to the detector.
An attacker can exploit this gap through \emph{additive attacks}: inserting benign-looking code or data, subject to validity and attack-budget constraints~\cite{pierazzi2020intriguing}.
Modifications are commonly restricted to additions in order to avoid removing components that the malware relies on to execute malicious behavior~\cite{biggio2013evasion,hu2017malgan}.

Such attacks have been studied against PDF, executable, and Android detectors, both in feature space and on the samples themselves~\cite{biggio2013evasion,srndic2014practical,hu2017malgan,aldujaili2018adversarial,demetrio2021gamma,grosse2017adversarial,pierazzi2020intriguing}.
Liang et al.\ evaded Chrome's then-deployed phishing-page classifier by inserting benign-associated features~\cite{liang2016cracking}. \v{S}rndi\'c and Laskov reduced PDFRATE's scores from nearly 100\% to a median of approximately 33\% through content insertion~\cite{srndic2014practical}. Pierazzi et al.\ evaded Android malware detectors by transplanting code from benign applications, an attack known as APG~\cite{pierazzi2020intriguing,cortellazzi2025intriguing}.
In each case, features associated with benign samples outweigh evidence of maliciousness that is still present.
To be robust against additive attacks, a detector must retain a malicious classification under every admissible functionality-preserving addition.

Static extractors analyze samples without executing their code.
While the feature categories they produce are fixed, the vocabulary may be open-ended: identifiers, domain names, and structural paths absent from the training data may continue to appear under temporal drift.
Nonetheless, detectors represent each sample as a multi-hot or count vector over a vocabulary selected at training time~\cite{kolter2006learning,srndic2013hidost,aldujaili2018adversarial,ma2009identifying,arp2014drebin}. Alternatively, they hash feature strings into a fixed number of bins at the cost of collisions~\cite{weinberger2009hashing,anderson2018ember}. In both cases, they learn from this feature representation rather than from the raw data.
Therefore, the chosen encoding decides what the detector can actually see. Kolter and Maloof keep only 500 of more than 255 million byte 4-grams~\cite{kolter2006learning}. \v{S}rndi\'c and Laskov retain 6,087 of more than 9 million structural PDF paths~\cite{srndic2013hidost}.

We study this problem in Android malware detection for the following reasons.
While there are several approaches to representing Android samples~\cite{arp2014drebin,wu2019malscan,gao2021gdroid,onwuzurike2019mamadroid,daoudi2021dexray}, the BoW-like \dr{} feature space has the largest set of detectors built on a single extractor~\cite{arp2014drebin,grosse2017adversarial,yang2021cade,chen2023continuous,mcfadden2026drmd,jordaney2017transcend,barbero2022transcendent}
The entire \dr{} representation is set-shaped, whereas in other domains, set-shaped features often form only one part of a broader representation, such as the byte histograms and header values of Windows executables.
\dr{}-based detectors also come with realizable additive attacks~\cite{pierazzi2020intriguing,cortellazzi2025intriguing,bostani2024evadedroid,he2023advdroidzero} and multi-year time-aware benchmarks~\cite{kan2024tesseract,pendlebury2019tesseract,hypercube}.
Together, these let us isolate the effect of the representation itself (Section~\ref{sec:bg-android}).
Over more than a decade, loss functions, training schedules, and feature selection have changed along this line, but the multi-hot encoding over a vocabulary fixed at training time has remained predominate.
We therefore study \textbf{representation and modeling as a coupled design choice}: how should features be represented, and how should their evidence be combined into a decision?

\begin{figure}[t]
    \centering
    \providecommand{\tfDir}{figures/tradeoff_data/}
\pgfplotsset{
  tradeoffcmap/.style={colormap={evade}{rgb255(0)=(26,150,65) rgb255(1)=(166,217,106) rgb255(2)=(254,224,139)
                                       rgb255(3)=(253,174,97) rgb255(4)=(215,25,28)}},
  tfOnly/.style={x filter/.append code={\edef\tfA{\thisrow{paradigm}}\edef\tfB{#1}\ifx\tfA\tfB\else\def\pgfmathresult{nan}\fi}},
  tfDodge/.style n args={3}{x filter/.append code={\edef\tfA{\thisrow{key}}\edef\tfB{#1}\ifx\tfA\tfB\pgfmathparse{\pgfmathresult+(#2)}\fi},
                            y filter/.append code={\edef\tfA{\thisrow{key}}\edef\tfB{#1}\ifx\tfA\tfB\pgfmathparse{\pgfmathresult+(#3)}\fi}},
  tfDodged/.style={tfDodge={drebinplus}{-0.75}{-0.75}, tfDodge={mono_gbt}{0.75}{0.75}, tfDodge={lgbm}{0}{-0.5}},%
  tfPts/.style={scatter, only marks, point meta=explicit, unbounded coords=discard, forget plot,
                scatter/use mapped color={draw=mapped color!62!black, fill=mapped color}, line width=0.45pt},
  tfNum/.style={only marks, mark=none, point meta=explicit, unbounded coords=discard, forget plot,
                visualization depends on={100*\thisrow{txtw} \as \tfW},
                nodes near coords={\pgfmathprintnumber[fixed, precision=0, assume math mode=true]{\pgfplotspointmeta}},
                nodes near coords align={center},
                every node near coord/.append style={font=\sffamily\bfseries\fontsize{4.6}{4.6}\selectfont,
                  text=white!\tfW!black!85, inner sep=0pt}},
  tfLin/.style={mark=square*, mark size=3.5pt},
  tfTree/.style={mark=triangle*, mark size=5.4pt},
  tfMLP/.style={mark=*, mark size=3.9pt},
  tfMIL/.style={mark=diamond*, mark size=4.9pt},
}
\tikzset{
  tfFrame/.style={draw=black, line width=0.8pt, sharp corners, fill=white, inner xsep=2.6pt, inner ysep=2pt,
                  font=\scriptsize\bfseries},
}
\def\tfData{\tfDir tradeoff_data.csv}
\input{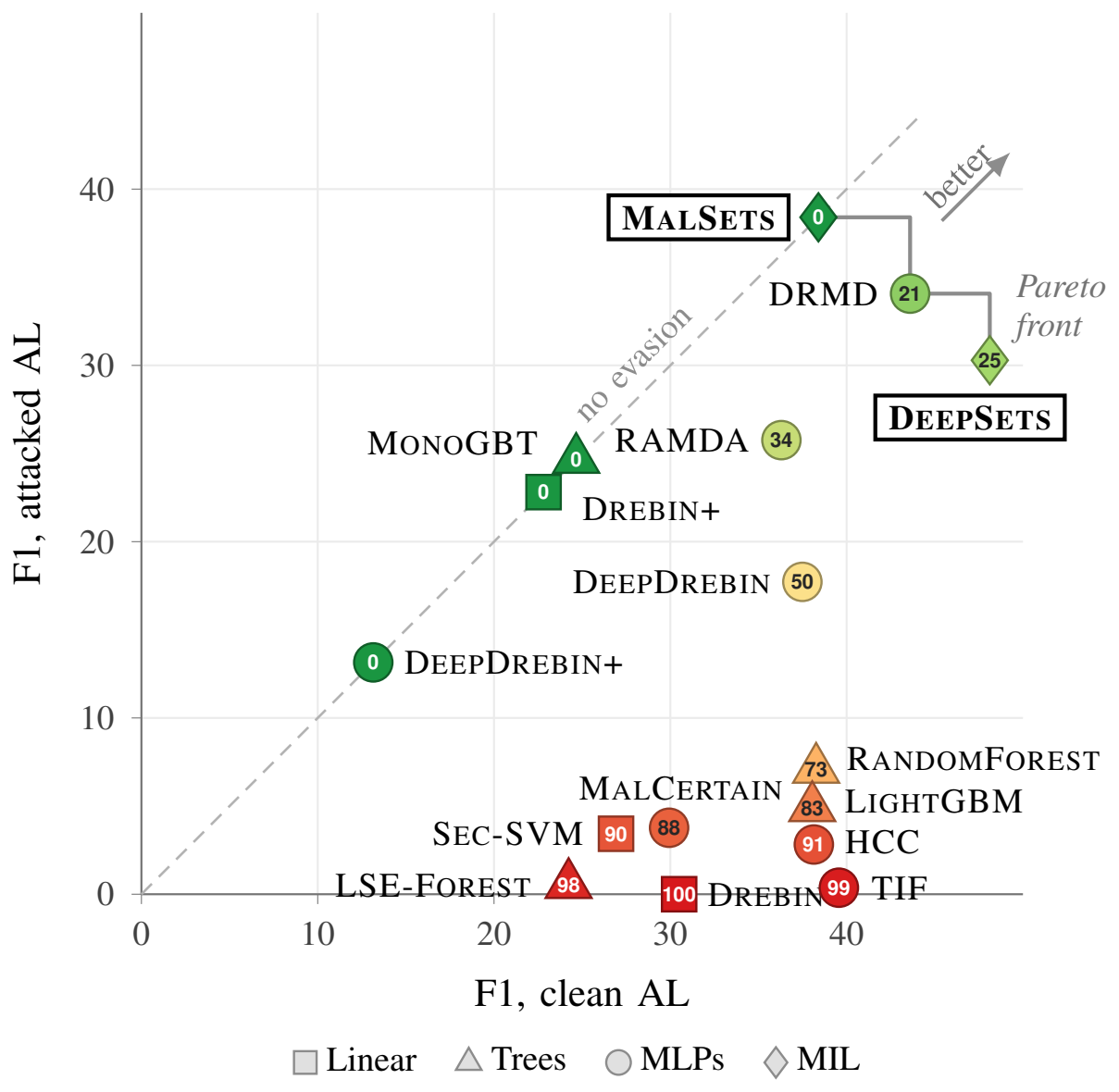}%
\begin{tikzpicture}
\begin{axis}[
  scale only axis, width=6.3cm, height=6.3cm, tradeoffcmap,
  xmin=0, xmax=50, ymin=0, ymax=50,
  axis lines=left, axis line style={black!55, -}, tick style={black!55}, tick align=outside,
  major tick length=2pt,
  xlabel={F1, clean AL}, ylabel={F1, attacked AL},
  xtick={0,10,20,30,40}, ytick={0,10,20,30,40},
  grid=major, major grid style={black!8, line width=0.4pt},
  tick label style={font=\scriptsize, text=black!75}, label style={font=\footnotesize},
  clip=false,
  point meta min=0, point meta max=100,
  legend columns=4,
  legend style={at={(0.5,-0.15)}, anchor=north, draw=none, fill=none, font=\scriptsize,
    /tikz/every even column/.append style={column sep=6pt}},
]
\addplot[black!30, densely dashed, line width=0.5pt, domain=0:44, forget plot] {x};
\node[font=\scriptsize, text=black!45, rotate=45, anchor=south] at (axis cs:29.0,29.0) {no evasion};
\addplot[black!45, line width=0.8pt, line join=round, forget plot] coordinates {\tfPareto};
\node[font=\scriptsize\itshape, text=black!55, anchor=west, inner sep=1pt, align=left] at (axis cs:49.3,33.2) {Pareto\\front};
\draw[-{Latex[length=1.8mm, width=1.5mm]}, black!45, line width=0.7pt] (axis cs:45.4,38.2) -- (axis cs:49.3,42.1)
  node[pos=0.45, sloped, above, font=\scriptsize, text=black!55, inner sep=1.5pt] {better};
\addplot[tfPts, tfLin]  table[x=clean, y=attacked, meta=ev, col sep=comma, tfDodged, tfOnly=lin]  {\tfData};
\addplot[tfPts, tfTree] table[x=clean, y=attacked, meta=ev, col sep=comma, tfDodged, tfOnly=tree] {\tfData};
\addplot[tfPts, tfMLP]  table[x=clean, y=attacked, meta=ev, col sep=comma, tfDodged, tfOnly=mlp]  {\tfData};
\addplot[tfPts, tfMIL]  table[x=clean, y=attacked, meta=ev, col sep=comma, tfDodged, tfOnly=mil]  {\tfData};
\addplot[tfNum] table[x=clean, y=ynum, meta=ev, col sep=comma, tfDodged] {\tfData};
\addlegendimage{only marks, mark=square*,   mark size=2.3pt, draw=black!45, fill=black!12, line width=0.45pt}\addlegendentry{Linear}
\addlegendimage{only marks, mark=triangle*, mark size=3.0pt, draw=black!45, fill=black!12, line width=0.45pt}\addlegendentry{Trees}
\addlegendimage{only marks, mark=*,         mark size=2.5pt, draw=black!45, fill=black!12, line width=0.45pt}\addlegendentry{MLPs}
\addlegendimage{only marks, mark=diamond*,  mark size=3.1pt, draw=black!45, fill=black!12, line width=0.45pt}\addlegendentry{MIL}
\begin{scope}[every node/.style={font=\scriptsize, inner sep=1.5pt}]
\node[anchor=west, inner xsep=0.6pt] at (axis cs:31.85,0.0) {\dr{}};
\node[anchor=east]       at (axis cs:22.6,0.53)  {\textsc{LSE-Forest}};
\node[anchor=east]       at (axis cs:25.6,3.43)  {\secsvm{}};
\node[anchor=south]      at (axis cs:30.6,4.9)   {\malcertain{}};
\node[anchor=west]       at (axis cs:40.9,0.37)  {\tif{}};
\node[anchor=west]       at (axis cs:39.5,7.7)   {\rf{}};
\node[anchor=west]       at (axis cs:39.4,3.0)   {\hcc{}};
\node[anchor=east]       at (axis cs:35.0,25.76) {\ramda{}};
\node[anchor=east]       at (axis cs:42.3,34.07) {\drmd{}};
\node[anchor=east]       at (axis cs:36.2,17.72) {\dd{}};
\node[anchor=east]       at (axis cs:23.0,25.6)  {\textsc{MonoGBT}};
\node[anchor=west]       at (axis cs:24.5,21.9)  {\textsc{Drebin+}};
\node[anchor=west]       at (axis cs:39.4,5.3)   {\textsc{LightGBM}};
\node[tfFrame, anchor=north, align=center] at (axis cs:47.0,28.4) {\ds{}};
\node[anchor=west]       at (axis cs:14.4,13.15) {\ddplus{}};
\node[tfFrame, anchor=east] at (axis cs:36.82,38.40) {\sys{}};
\end{scope}
\end{axis}
\end{tikzpicture}

    \caption[Clean versus attacked active learning]{\textbf{Clean versus Attacked Active Learning.}
    Each detector is evaluated over a $30$ month test period, on the Hypercube dataset, under both the clean  ($x$ axis) and attacked ($y$ axis) active learning settings (For more information see Sections~\ref{sec:setup}-\ref{sec:results}).
    With \sys{} being the only detector on the Pareto front that is never evaded.
    The numbers assigned to each detector reflect the percentage of malware evaded, whereas the shape corresponds to the model paradigm.
    }
    \label{fig:intro-tradeoff}
\end{figure}

We first address the representation. Since the multi-hot vector encodes an application's feature set, every detector computes a function of that set. Linear models~\mkLin{} and Multi-Layer Perceptrons (MLPs)~\mkMLP{}, two of the four detector classes in Figure~\ref{fig:intro-tradeoff}, share a specific structure. For an application with feature set $S$, the multi-hot vector is the indicator vector $\mathbbm 1_{V(S)}$ over the fixed vocabulary $V$. The first layer of these models compute $W_1\mathbbm 1_{V(S)}=\sum_{s\in S\cap V}\mathbf w_s$, where $\mathbf w_s$ is the column of $W_1$ assigned to feature $s$. This layer is therefore a lookup table of learned feature embeddings followed by implicit $\SUM$ pooling; in linear models, the embeddings reduce to scalar weights. Tree ensembles~\mkTree{} also compute a function of the set, but they split on individual coordinates instead of summing embeddings. Hence, linear and MLP-based detectors on binary representations are set models in disguise, namely $\SUM$-pooled \mech{}~\cite{deep_sets} models over a vocabulary fixed at training time. This explains both weaknesses discussed above. First, strings outside the training vocabulary are never encoded, so these models are blind to out-of-vocabulary (OOV) strings. Second, $\SUM$ pooling allows added benign features to outweigh malicious evidence, which is exactly how additive benign attacks succeed. We break with this tradition and embed each feature, $s\in S$, with a general text encoder instead.
Since \dr{} features are set-shaped and the cardinality varies between samples, we cast detection as multiple-instance learning (MIL)~\cite{dietterich1997mil}. MIL has modeled network traffic~\cite{pevny2016discriminative} and spam emails~\cite{jorgensen2008mil} as bags before. We instantiate MIL via \mech{}~\cite{deep_sets} with $\MAX$ pooling, which retains malicious evidence of the sample, regardless of how much benign evidence is added. While Android malware is our evaluation domain, our findings are principled and hold wherever static extractors yield set-shaped features (see Section~\ref{sec:bg-sets}).

$\MAX$-pooled \mech{} models preserve malicious evidence in the latent representation. However, a downstream classifier head may still map a larger latent representation to a lower score, which an attacker can exploit. To retain a malware classification under any addition that maintains malicious evidence, the score must never decrease when features are added. Gimenez et al.\ show that this monotonicity is both sufficient and necessary for robustness to additive attacks~\cite{gimenez2025erdalt}. Monotone malware detectors exist~\cite{incer2018monotonic,fleshman2019nonnegative}, but so far they have come at a twofold cost. First, they operate on a fixed vocabulary, so features seen after the training period add no evidence; thus, they are blind to OOV features. Second, they lose detection performance. Since no feature may lower the score, benign features cannot count as evidence. Incer et al.\ report a drop of approximately 13\% in detection rate, which they consider the cost of security~\cite{incer2018monotonic}. We summarize our contributions as follows:

\begin{enumerate}

    \item \textbf{Problem Aligned Design through MIL.}
    We cast static malware detection as MIL and operationalize it via \ds{}. A general text encoder embeds every feature string, which eliminates the out-of-vocabulary (OOV) problem by design. $\MAX$ pooling retains the malicious evidence of the sample, regardless of how many benign features an attacker adds.

    \item \textbf{Functional Monotonicity.}
    We devise \sys{} by equipping $\MAX$-pooled \ds{} models with a monotone head, over an open vocabulary representation. Since $\MAX$ pooling never decreases under benign addition, neither does the malware score.
    \sys{} thereby addresses the cost of security paid by prior monotone detectors, as it jointly maximizes both clean and attacked AL performance.

    \item \textbf{Adversarial Robustness.}
    For \sys{}, \textit{more is never less} as it is never evaded by any of the five attack families in our benchmark, ranging from black-box to white-box attacks. Under attack, \sys{} also retains the highest F1 of all detectors (Figure~\ref{fig:intro-tradeoff}).
\end{enumerate}

\section{Background}
\label{sec:background}

\noindent In this section, we introduce all core concepts for our work. First, we describe set-shaped static features (Section~\ref{sec:bg-sets}) and their multi-hot encoding over a fixed vocabulary (Section~\ref{sec:bg-multihot}). We then provide an overview of multiple-instance learning (Section~\ref{sec:bg-mil}), our threat model (Section~\ref{sec:meth-threat}), and the motivation for Android malware detection as our choice of domain (Section~\ref{sec:bg-android}).

\subsection{Set-Shaped Static Features}
\label{sec:bg-sets}

\noindent A static feature extractor $E$ maps a sample $z$, such as an application, to a finite set of strings $S=E(z)\subset\Sigma^*$, where $\Sigma^*$ denotes all strings~\cite{pierazzi2020intriguing}. Each string, $s\in S$, represents one feature in $z$, such as a requested permission or an API call. The extractor, $E$, may yield features of different categories. For example, the \drfe{} extractor~\cite{arp2014drebin} considers multiple categories, where a prefix, \texttt{api\_permissions}, denotes the category of the feature: \texttt{api\_permissions::android\_permission\_SEND\_SMS}.
These features are set-shaped, and their order carries no information (permutation-invariant). Furthermore, the number of features, $|S|$, varies among samples (varying set-cardinality). Feature categories are known in advance, but the vocabulary is \emph{open}: $\Sigma^*$ is infinite, whereas the set of strings observed during the training-phase of detectors is finite. We call a feature category \emph{closed} if its strings come from a finite list known in advance, such as the permissions an operating system defines, and \emph{open} otherwise. During the test-phase, a sample contains out-of-vocabulary (OOV) strings whenever $S$ includes strings never observed during training; Wressnegger et al.\ measure their expected fraction as the \emph{perturbation} of a feature space~\cite{wressnegger2013ngrams}.

\textbf{Set-Based Modalities in Security.} This \textit{set-based} modality recurs across security domains, often as one block of a larger feature vector.
The import block of a Windows executable is the set of its libraries and \texttt{library:function} pairs~\cite{anderson2018ember}; Al-Dujaili et al.\ represent executables solely by such indicators over $22{,}761$ distinct API calls~\cite{aldujaili2018adversarial}.
A PDF maps to the structural paths of its objects, whose number is unbounded~\cite{xu2016evademl}.
The lexical block of a URL types each token by its position (hostname, path, top-level domain), so that ``com'' in the top-level domain is a different token from ``com'' elsewhere. After $100$ days Ma et al.\ counted $1.79$ million such tokens~\cite{ma2009identifying}.
A browser extension declares permissions from a closed list of $70$~\cite{jagpal2015webeval} and host permissions as arbitrary string patterns, which Rosenzweig et al.\ cannot enumerate and truncate to the $400$ most popular~\cite{rosenzweig2025noteasy}. Yet, in these domains, the set-shaped feature block is only part of the representation. Evasion there also packs binaries, renames sections \cite{anderson2018rl} or deletes content~\cite{xu2016evademl}. Our work applies to the set-shaped part. Hence, an evaluation in these domains would also need to account for the interplay with remaining features and for non-additive modifications.

This work builds on Android malware detection as our domain of choice: the \drfe{} extractor~\cite{arp2014drebin} emits strings in nine typed categories, and for the Hypercube~\cite{hypercube} dataset, $85\%$ of distinct strings in the test period never occur in the training year (Section~\ref{sec:setup}).

\subsection{Multi-Hot Encoding over a Fixed Vocabulary}
\label{sec:bg-multihot}

\noindent Detectors on set-shaped features rarely consume $S$ directly. Instead, the canonical approach during preprocessing follows two steps. (1) During Feature selection, each detector selects a vocabulary $V$ from the strings observed in its training data. Feature selection can be supervised or unsupervised. To preserve temporal training consistency~\cite{pendlebury2019tesseract,arp2022dos}, selection must precede the test-period. (2) During \textit{multi-hot encoding}, each sample is encoded as a binary vector $\mathbf x\in\{0,1\}^{|V|}$ with one coordinate per string in $V$, set to $1$ if the string occurs in $S$.

\textbf{Limitations of Fixed Vocabularies.} The encoding inherits the permutation invariance of $S$ and maps features $s\in S$ to the selected $|V|$ coordinates. This discards features outside $V$, therefore, two samples that differ only by strings outside $V$ are encoded with the same feature-vector, $\mathbf x$. Since $V$, a string first observed after training is unseen by a detector trained on $V$. Yang et al.\ mention this constraint: a temporal split ``means we cannot use any new features that only appear in the testing set''~\cite{yang2021cade}. On the Hypercube~\cite{hypercube} dataset considered in this work, the shared top-$10{,}000$ vocabulary (see Section~\ref{sec:setup}) retains a median of $28\%$ of an application's strings in the training year and $27\%$ over the test period.

\textbf{Alternatives to a Fixed Vocabulary.} \textit{Feature hashing} maps every string, including strings unseen during training, to one of a fixed number of bins. If hashing maps two strings to the same coordinate, feature collisions can occur~\cite{weinberger2009hashing,anderson2018ember}. Feature \textit{abstraction} maps features to a closed alphabet, such as API calls to their packages~\cite{onwuzurike2019mamadroid} or to clusters of related APIs~\cite{zhang2020apigraph}. This closes a category by design but requires strings of a known structure, which URLs and component names lack.

\textbf{Classifiers.} Besides notable examples, including the linear SVMs of \dr{}, \secsvm{}~\cite{arp2014drebin,demontis2019secureml} and \rf{}~\cite{breiman2001random}, the underpinning classification methods most frequently used are based on multi-layer perceptrons (MLPs):
\dd{}~\cite{grosse2017adversarial} is a two-layer MLP (with two hidden layers of $200$ ReLU units); \hcc{}~\cite{chen2023continuous} trains an MLP encoder with a hierarchical contrastive objective and classifies with an MLP; \cade{}~\cite{yang2021cade}, a drift detector, trains a contrastive MLP autoencoder on the same features as the classifier it monitors; \ramda{}~\cite{li2021ramda} combines a variational autoencoder with an MLP that shares an encoder; \drmd{}~\cite{mcfadden2026drmd}, a deep reinforcement learning (DRL)-based detector, uses MLPs for both its actor and critic networks; and \tif{}~\cite{zheng2026tif} trains existing detectors (on \drfe{} features, \dd{}) with temporal-invariance objectives.

\textbf{Choosing the Vocabulary.} Most classifiers differ at the input only by how $V$ is chosen. \dd{} was introduced on the full \drfe{} feature set of $545{,}333$ strings~\cite{grosse2017adversarial}. Later work restricts this space to $10{,}000$ strings: the budget originates with \secsvm{}, which ranks strings by the difference in their frequencies in malware and benign training applications~\cite{demontis2019secureml}; Pierazzi et al.\ select with an $\ell_2$-regularized linear model~\cite{pierazzi2020intriguing}; and Kan et al.\ keep the strings with the largest absolute weights of a linear SVM~\cite{kan2024tesseract}, adopted by \drmd{}~\cite{mcfadden2026drmd}. \hcc{}~\cite{chen2023continuous} filters by variance exceeding a threshold, and \cade{}~\cite{yang2021cade} applies a variance filter. \ramda{} encodes manually chosen permissions, intent actions, and sensitive API calls~\cite{li2021ramda}. We follow the procedure of Kan et al.~\cite{kan2024tesseract} (see Section~\ref{sec:setup}) to select a shared top $10{,}000$ vocabulary previously applied by~\cite{mcfadden2026replicant} to the majority of the baselines in our experiments.

\textbf{Key Takeaway:} While feature-selection may differ among methods, the representation, $\mathbf x$, and its dependence on a fixed vocabulary, $V$, at training time do not.

\subsection{Multiple-Instance Learning}
\label{sec:bg-mil}

\noindent Multiple-instance learning (MIL) labels bags of instances rather than single instances, and under its standard assumption, a bag is positive if at least one of its instances is positive~\cite{dietterich1997mil}.
Since set-shaped features are bags by construction, with the sample as the bag $S$ and each string as an instance, neural MIL embeds every instance and aggregates the embeddings by pooling into one bag representation~\cite{deep_sets}.
Under $\MEAN$ and $\SUM$ pooling, when instance embeddings can be negative, added instances can lower every coordinate of the set representation and thereby dilute the evidence of a positive instance.
$\MAX$ pooling~\cite{qi2017pointnet} preserves the standard assumption under which a bag's label is the maximum of its instance labels in the embeddings.
Hence, each coordinate of the bag representation is the largest value among its instances. Lowering the value is not possible; values either increase or leave the representation unchanged.

\begin{figure*}[t]
    \centering

    \resizebox{\textwidth}{!}{%
        \input{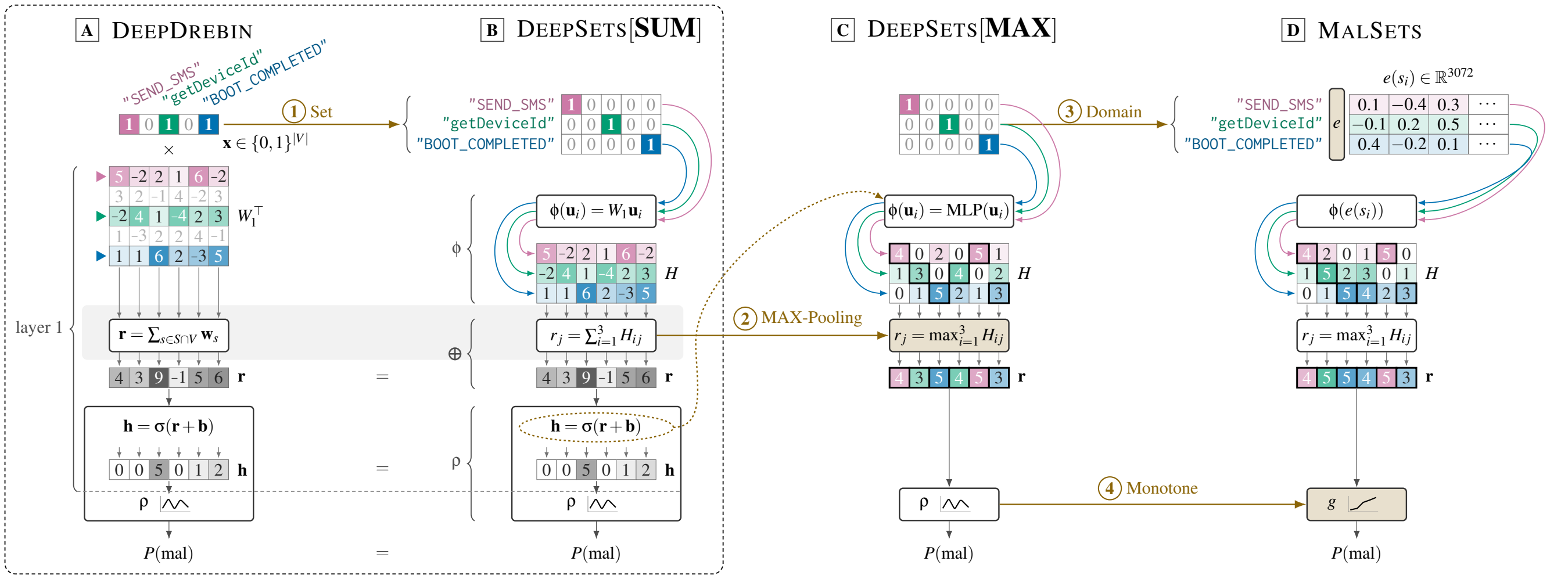}%
    }

    \caption[From \dd{} to \sys{}]{%
        {\textbf{From \dd{} to \sys{}:}
        Step~\stepref{1} splits $\mathbf x$ into the set $U$ of one-hot elements;
        Step~\stepref{2} replaces $\SUM$ by $\MAX$, and $\phi$ is an MLP from panel~\panel{C} onward;
        Step~\stepref{3} maps every feature string with a frozen general text encoder $e$ to an embedding;
        Step~\stepref{4} classifies with a monotone head $g$. Panels~\panel{A} to~\panel{D} are rows~1, 1a, 3 and~11 of Table~\ref{tab:ablation}.}
    }
    \label{fig:ffn_deepsets_equivalence}
\end{figure*}

\subsection{Threat Model}
\label{sec:meth-threat}

\noindent We assume an adversary who controls the sample (Android APKs in our evaluations) but cannot change the deployed feature extractor or the trained model. The goal is \emph{evasion}: to classify a malware sample as benign. A \emph{feature-space} attack that perturbs the feature vector directly provides only a theoretical upper-bound, since an arbitrary vector does not necessarily correspond to a working APK. Hence, we evaluate in the \emph{problem space} following Pierazzi~et~al.~\cite{pierazzi2020intriguing}, where an attacker applies a transformation $T$ to the APK $z$, subject to four constraints: (i) available transformations, (ii) preserved semantics, (iii) plausibility, and (iv) robustness to preprocessing.

\textbf{Problem-Space Attacks.} We take the \emph{available-transformations} constraint to be \emph{addition only}, $E(z)\subseteq E(T(z))$, preserving the host's malicious functionality. To this end, we adopt \emph{organ transplantation}~\cite{barr2015transplantation}, where bytecode slices harvested from benign \emph{donor} apps are grafted into a malicious \emph{host}. An \emph{organ} $o$ is the self-contained bytecode slice of a harvested gadget, and an \emph{ice-box} $O$ is the bank of organs assembled ahead of the attack time. A \emph{graft} is the set union $S\mapsto S\cup o$, where $o$ also denotes the strings that the organ adds, including \emph{side-effect features} that satisfy the problem-space constraints~\cite{pierazzi2020intriguing}. Removal and obfuscation are feasible~\cite{demontis2019secureml,maiorca2015stealth,scano2026droidbreaker} but violate $E(z)\subseteq E(T(z))$; we treat them as out of scope. The attackers of Section~\ref{sec:meth-attacks} share the same capability set, differing only in strategy and knowledge.

\subsection{Why Android?}
\label{sec:bg-android}

\noindent Our findings apply to set-shaped static features in general (Section~\ref{sec:bg-sets}), however, we chose Android malware detection (AMD) as the domain for our evaluations because it fulfills the following four requirements:

\begin{enumerate}[leftmargin=*]
\item \textbf{Set-Shaped Features with an Open Vocabulary.} The \drfe{} extractor emits strings in nine fixed categories. Their vocabulary is open (Section~\ref{sec:bg-sets}).
\item \textbf{One Representation with Many Detectors.} Linear, tree-based, and neural detectors, including drift-aware ones, share the multi-hot vector $\mathbf x$ over \drfe{} features (Section~\ref{sec:bg-multihot}). Hence, the representation is the only variable we change.
\item \textbf{Realizable Additive Attacks.} Organ transplantation (Section~\ref{sec:meth-threat}) yields an evasive variant for each malware sample in all four configurations studied by Pierazzi et al.~\cite{pierazzi2020intriguing}. Its ice-box and implementation are available via contacting the original authors~\cite{cortellazzi2025intriguing}. Such are query-based attacks, including \textsc{EvadeDroid} and \textsc{AdvDroidZero}~\cite{bostani2024evadedroid,he2023advdroidzero}.
\item \textbf{Time-Aware Evaluation at Scale.} Time-aware evaluation protocols are established for Android~\cite{pendlebury2019tesseract,kan2024tesseract}, and Hypercube~\cite{hypercube} provides $224{,}965$ applications from a single market (Google Play) over four years, each dated by its upload and labeled with its number of VirusTotal detections (VTT).
\end{enumerate}
Therefore, in the \dr{} feature space, the entire representation is set-shaped and fulfills all four requirements, so our experiments isolate the effect of the representation itself.

\section{\texorpdfstring{From \dd{} to \sys{}}{From DeepDrebin to MalSets}}
\label{sec:pillars}

\begin{figure*}[t]
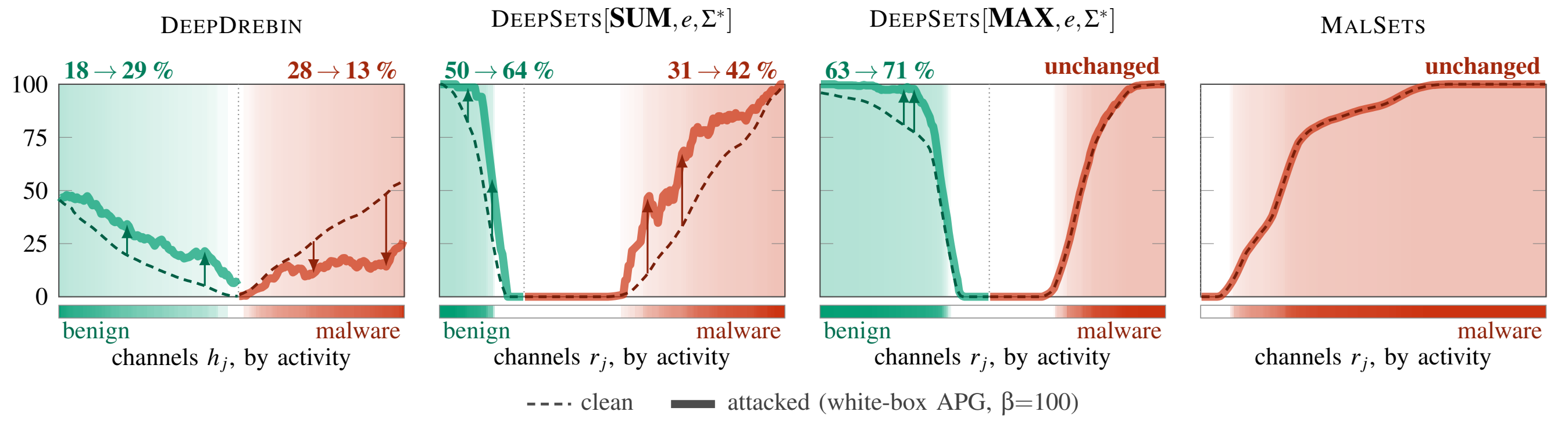

    \centering
    \providecommand{\polData}{figures/polarity_data}
\input{\polData/summary.tex}
\input{figures/polarity_style}
\begin{tikzpicture}
\begin{groupplot}[group style={group size=4 by 1, horizontal sep=0.40cm, y descriptions at=edge left},
                  polax, width=0.31\textwidth, height=4.0cm]
\nextgroupplot[title={\dd{}}, xlabel={channels $h_j$, by activity}]                \polPanel{dd}{DD}
\nextgroupplot[title={\dsv{\mbox{\small\bfseries SUM}, e, \Sigma^{*}}}, xlabel={channels $r_j$, by activity}] \polPanel{dssum}{DSS}
\nextgroupplot[title={\dsv{\mbox{\small\bfseries MAX}, e, \Sigma^{*}}}, xlabel={channels $r_j$, by activity}] \polPanel{dsfree}{DSF}
\nextgroupplot[title={\sys{}}, xlabel={channels $r_j$, by activity}]                \polPanel{dscert}{DSC}
\end{groupplot}
\polLegend{($(group c2r1.south)!0.5!(group c3r1.south)$)}
\end{tikzpicture}
    \caption[The Effect of Attacks on Channels]{\textbf{The Effect of Attacks on Channels.} Share of applications (\%) in which a channel of $\mathbf h$ (\dd{}) or $\mathbf r$ (\mech{} models) is active, split by polarity (benign left, malware right) and ordered by their average polarity over the training set. The shading reflects average polarity magnitude. Lines are smoothed as a running mean over $15$ channels. Models are rows~1, 6, 5 and 11 of Table~\ref{tab:ablation}.
    }
    \label{fig:polarity}
\end{figure*}

\noindent This section focus on the design of \sys{}, which is show through panels \panel{A}--\panel{D} and steps \stepref{1}--\stepref{4} in Figure~\ref{fig:ffn_deepsets_equivalence}. We derive \sys{} from \dd{} for two primary reasons. First, it makes no structural assumptions about its input. Grosse et al. choose a plain feed-forward network because the binary indicator vector $\mathbf x$ does not require translation invariance or ordering, as in convolutional and recurrent neural networks, respectively~\cite{grosse2017adversarial}. However, $\mathbf x$ is not fully unstructured, as it encodes a set. Detectors on $\mathbf x$ are implicitly permutation invariant: \dd{}'s first layer represents each element by one column of $W_1$ and combines all elements by summation (\panel{A}).
Second, \dd{} is trained via standard stochastic gradient descent without relying on bespoke assumptions. Whereas all other MLP-based detectors, mentioned in Section~\ref{sec:bg-multihot}, add objectives or specialized architectures such as contrastive terms \cite{chen2023continuous}, auxiliary networks \cite{li2021ramda}, or reward structures \cite{mcfadden2026drmd}.
Therefore, \dd{} acts as a natural foundation to ablate the decisions of \sys{} shown in panels~\panel{A}--\panel{D}.

\textbf{\dd{}} (\panel{A}) is a feed-forward network on input $\mathbf x$. At its first layer, it computes the vector $\mathbf r=W_1\mathbf x$ with the weights $W_1\in\mathbb R^{200\times|V|}$. The classification head $\rho$ comprises all subsequent operations. It starts with the first-layer bias $\mathbf b$ and activation $\sigma$, which give the first hidden layer $\mathbf h=\sigma(\mathbf r+\mathbf b)$ and maps $\mathbf r$ to a malware logit, which a sigmoid converts to $\Pmal$. We focus on these layers as $\mathbf x$ treats the first layer as a lookup table (\stepref{1}) that is implicitly sum-aggregated (\stepref{2}) over a closed vocabulary (\stepref{3}) while $\rho$ is easily overwhelmed by contrary evidence (\stepref{4}).

\textbf{\ds{}}~\cite{deep_sets} (panels~\panel{B}--\panel{D}) is multiple-instance learning (MIL) applied to learning functions over sets, where a shared element-wise MLP $\phi:\mathbb{R}^{d}\to\mathbb{R}^{d_\phi}$, the element network, maps each embedded feature to a latent representation. These representations form the rows of a matrix $H$.
The set of these representations is then aggregated into a single vector, the pooled vector $\mathbf r$, by a pooling operator $\oplus$,
\begin{equation}
  \mathbf r(S) := \bigoplus_{s\in S}\phi\bigl(e(s)\bigr)\in\mathbb{R}^{d_\phi}
  \quad\text{for any finite } S\subset\Sigma^*,
  \label{eq:pooled-map}
\end{equation}
with $\mathbf r=\mathbf r(S)$ for the application, $e:\Sigma^*\to\mathbb R^{d}$ encoding each string,
$\bigoplus_{s\in S}$ denoting the pooling operator $\oplus$, applied to all feature vectors $\phi(e(s))$. The classification head $\rho$ then maps $\mathbf{r}(S)$ to a malware logit.
This formulation makes the individual feature representations explicit (\stepref{1}), allowing us to change the pooling operator (\stepref{2}), replace the fixed vocabulary with an encoder for unseen strings (\stepref{3}), and replace the unconstrained classification head $\rho$ with a monotone head $g$ (\stepref{4}).

\textbf{Constraints for Invariant Set-Learning.}
According to~\cite{deep_sets}, there are two requirements for the pooling operator $\oplus$.
First, because $S$ is unordered, its output must not change when the arguments are permuted.
Secondly, $\oplus$ must accept any number of arguments, as set cardinality may vary.
Coordinate-wise $\SUM$, $\MEAN$ and $\MAX$ pooling all satisfy these conditions and are common choices.

\textbf{Encoding Strings.}
In \dd{} (panel~\panel{A}) the multi-hot vector $\mathbf x=\mathbbm 1_{V(S)}\in\{0,1\}^{|V|}$ records which vocabulary features are present in the sample.
In \ds{} (panels~\panel{B} and \panel{C}), the string encoder maps each retained string $s\in S\cap V$ to its one-hot vector $\mathbbm 1_{V(s)}$.
In \sys{} (panel~\panel{D}), $e$ is a frozen general text encoder that is applied to every string in $S$ (see Section~\ref{sec:p-domain}).

\textbf{Design Choices towards \sys{}.}
Outside of $e$, Equation~\ref{eq:pooled-map} also leaves three other key choices open.
(i) the pooling operator for $\oplus$, as they determine how $\mathbf r$ responds to the composition of $S$ (\stepref{2});
(ii) Which elements of $S$ reach the encoder, i.e. the vocabulary $V$ (\stepref{3});
and, (iii) the parameterization of the networks $\phi$ and $\rho$ (\stepref{4}).
We write \dsv{\oplus, \dots} for a \ds{} variant with pooling $\oplus$ followed by the components it adds, in step order:
the frozen open-vocabulary encoder $e$ applied to all strings;
$\Sigma^{*}$ (Section~\ref{sec:p-domain});
and, a monotone head $g$ (Section~\ref{sec:p-cert}).
Therefore, \sys{} is \dsv{\MAX, e, \Sigma^{*}, g} and \ddplus{}, used in later comparisons, is \dd{} with non-negative weights in every layer (Section~\ref{sec:p-cert}).

\subsection{\texorpdfstring{\stepref{1}~Features as Sets}{(1) Features as Sets}}
\label{sec:p-set}

\noindent We show that MLPs including \dd{} on multi-hot encoded features can compute the same function as \ds{} models on one-hot features, following our construction.

\textbf{Encoding.} First, we convert the multi-hot encoding $\mathbf x$, which encodes the feature strings of a sample $S\cap V=\{s_1,\dots,s_k\}$, to a set of one-hot vectors $U=\{\mathbf u_1,\dots,\mathbf u_k\}$ (\stepref{1}), encoding each feature $s_i$ as $\mathbf u_i$.

\textbf{SUM Pooling Construction}. Second, we construct a \ds{} model, where we parameterize $\phi$ with a linear layer, $W_1$ (see panels \panel{A} and \panel{B}). The product $W_{1}\mathbf x$ in \dd{} then equals $W_1$ applied to each $\mathbf u_i\in U$ and summed:

\begin{equation}
\begin{aligned}
  \mathbf r &= \sum_{i}\phi(\mathbf u_i)
             = \sum_{s\in S\cap V}\mathbf w_s
             = W_1\sum_{i}\mathbf u_i
             = W_1\mathbf x,\\
\end{aligned}
\end{equation}

s.t. $\mathbf u_i=\mathbbm 1_{V(s_i)}$, $\lVert\mathbf u_i\rVert_1=1$, and $\sum_{i}\mathbf u_i=\mathbf x\in\{0,1\}^{|V|}$.
While MLPs implicitly compute a sum over the selected row-vectors to obtain $r$, we explicitly select $\SUM$ pooling for our \ds{} model in this context.

\textbf{Classifier Construction:} MLPs then compute a latent representation by applying an activation function and adding a bias-term: $h=\sigma(r + b)$. This marks a key-difference with respect to \ds{} models, which usually parameterize $\phi$ with an MLP~\cite{deep_sets}, applying $\sigma$ and $b$ to all vectors comprising $H$. Hence, to achieve parity with \dd{}, we construct a classifier, $\phi$, which applies $\sigma$ and $b$ at the input (see panel \panel{B}).

\textbf{Parametrization of $\mathbf{\phi}$.} For all other \ds{}-based models in our paper (panels \panel{C}--\panel{D}), we do not maintain this parity. We instead parameterize $\phi$ via an MLP as it aids the detector in learning a rich $\mathbf u_i$-level representation while freeing the choice for the pooling operator $\bigoplus$ (Section~\ref{sec:p-join}).

\subsection{\texorpdfstring{\stepref{2}~Max Pooling}{(2) Max Pooling}}
\label{sec:p-join}
\label{sec:p2}

\noindent After having established \dd{} as a $\SUM$-pooled set model, we replace the $\SUM$ pooling with a coordinate-wise $\MAX$ pooling, $r_j=\max_i H_{ij}$ (\stepref{2}).

\textbf{Polarity.} We call the coordinates of \dd{}'s first hidden layer $\mathbf h$ and of the pooled vector $\mathbf r$ of \ds{} models channels. For each channel, we compute the gradient attribution, $\bar w_j=\mathbb E[\partial f/\partial r_j]$, as the average gradient of the score over the training set~\cite{olah2020zoom}. In the context of binary malware classification, we refer to its sign as the channel's \emph{polarity}: \emph{malware} if $\bar w_j>0$ and \emph{benign} if $\bar w_j<0$. As polarity is an average, it need not hold for every application. Figure~\ref{fig:polarity} shows the polarity of the channels before and after an attack (Section~\ref{sec:ev-join}).

\textbf{Losing Evidence under SUM Pooling.} Additive benign attacks dilute malicious evidence under $\SUM$ pooling (panels \panel{A} and \panel{B}), by: (i) lowering a malware-polarity channel or (ii) raising a benign-polarity channel.

\textbf{Cancellation.} First, evidence is lost due \emph{to cancellation}. In \dd{}, sample features map to columns $\mathbf w_s$, aggregated by $\SUM$ pooling prior to activation. Since entries in $\mathbf w_s$ can be either positive or negative (Section~\ref{sec:p-set}), they may cancel out evidence. Specifically, when a benign organ $o$ is grafted into a host $S$, malicious evidence in $\mathbf h$ may be diminished.
For \dd{}, the ReLU activation function switches channels ``off'' once their values fall to or below zero, we visualize this cancellation effect in Figure~\ref{fig:polarity}.
For \dsv{\MAX, e, \Sigma^{*}}, $\MAX$ pooling prevents cancellation, as channels preserve the values of the strongest features, which cannot be lowered by benign addition. Attacking $\MAX$-pooled models can, therefore, only raise channels; thus, the model can only be evaded if the classifier $\rho$ lowers the score in response.

\noindent\textbf{Accumulation.} Second, the evidence is outweighed by feature \emph{accumulation}.
Suppose every contribution is non-negative and no \emph{cancellation} occurs. A channel's value no longer distinguishes one informative feature from many features with small contributions.
Attacks can therefore raise benign-polarity channels by adding more benign features to outweigh the host's malicious features.
Figure~\ref{fig:polarity} shows that in all non-monotone models, the benign attacks raise benign-polarity channels. Under $\SUM$ pooling, the values of added features accumulate. While under $\MAX$ pooling, each added feature increases the probability that a channel's value is surpassed.

\textbf{Absorption and Witnesses in MAX Pooling.} In our \ds{} models, $\phi$ embeds features individually and independently. Hence, grafting benign organs $S\mapsto S\cup o$ leaves the host's rows of $H$ unchanged. Values of the $\MAX$ pooled vector, $r$, therefore, can only increase~\cite{davey2002lattices},
\begin{equation}
  \mathbf r(S\cup o)=\mathbf r(S)\vee\mathbf r(o)\ge\mathbf r(S).
\label{eq:wb-decomp}
\end{equation}
Which requires that perturbations are additive and that $\phi$ maps every string of the host to a row of $H$.
For each channel $j$, the feature that sets its value is referred to as the channel's \emph{witness}, $s^\star_j\in\arg\max_{s\in S}\phi_j(s)$, as per ~\cite{qi2017pointnet}. At most $\dphi$ witnesses produce the pooled vector $\mathbf r$~\cite{qi2017pointnet}.
An added feature changes a channel only if it exceeds that channel's witness. Therefore, the largest impact that an attack can have on a channel is bound by the organ $o\in O$, which produces the greatest value in that channel.
Eq.~\ref{eq:wb-decomp} also makes white-box attacks on $\MAX$-pooled models more efficient, as every organ is scored from $\mathbf r(S)$ and $\mathbf r(o)$ alone, without re-encoding the application (Section~\ref{sec:meth-apg}).

\textbf{The Classifier Bottleneck.} Eq.~\ref{eq:wb-decomp} only holds for the pooled vector $\mathbf r$ and not for the score, which the classifier $\rho$ computes from $\mathbf r$ (\panel{C}). Therefore, if the weights of the classifier $\rho$ are not strictly positive, then raising a channel can lower the score.
The classifier is thus the only part of a $\MAX$-pooled \ds{} model that an additive attack can exploit, as can be seen in Figure~\ref{fig:polarity}. A monotone head $g$, whose score never falls when a channel rises, leaves the attacker with nothing to exploit (panel \panel{D}, Section \ref{sec:p-cert}).

\subsection{\texorpdfstring{\stepref{3}~Open-Vocabulary Encoder}{(3) Open-Vocabulary Encoder}}
\label{sec:p-domain}
\label{sec:p1}

\noindent MLP-based detectors trained on multi-hot representations, $\mathbf x=\mathbbm 1_{V(S)}$ over a fixed vocabulary $V$ (Section~\ref{sec:bg-multihot}), map input-features to columns of the weight-matrix $W_1$; features outside of $V$ have no input coordinates and therefore no row in $H$.
We illustrate this in \panel{A} through \panel{C}.

\textbf{Eliminating the OOV Problem.} We remove this fixed-vocabulary constraint, see \stepref{3} and \panel{D}, and replace $V$ with a general text encoder $e:\Sigma^*\to\mathbb R^{d}$ that embeds each feature string:
\begin{equation}
  \mathbf r(S)=\max_{s\in S}\phi\bigl(e(s)\bigr).
  \label{eq:pooled-map-e}
\end{equation}
Which requires an open-vocabulary encoder, $e$, that is defined over all features of $\Sigma^*$. Hence, every feature that the extractor, $E$, emits receives an embedding.

Therefore, the shared trainable instance network $\phi$ and the classifier $\rho$ of our \ds{} models remain independent of the vocabulary $V$ fixed at training time.

\subsection{\texorpdfstring{\stepref{4}~Monotone Head}{(4) Monotone Head}}
\label{sec:p-cert}
\label{sec:p3}
\label{sec:p4}

\noindent Robustness requires that additions preserve a malicious decision~\cite{gimenez2025erdalt}. \sys{} guaranties the stronger property that additions cannot decrease the malware score by replacing $\rho$ with a non-decreasing, monotone, head $g$ such that $f(S)=g(\mathbf r(S))$.
Since eq.~\ref{eq:wb-decomp} holds coordinate-wise in $\mathbf r$:
\begin{equation}
  f(S\cup o)\;\geq\;f(S)\qquad\text{for every finite } o\subset\Sigma^*.
  \label{eq:certificate}
\end{equation}
Therefore, under monotonicity, additive benign attacks can never lower a classifier's score.
This requires that the classifier's input never decreases under additions (see $\mathbf r$ in Figure~\ref{fig:ffn_deepsets_equivalence}).

\textbf{Criteria for Pooling Operators.} $\MAX$ pooling provides this for any $\phi$ (Eq.~\ref{eq:wb-decomp}), whereas $\SUM$ pooling only suffices if $H$ is non-negative; for example, when $\phi$ is ReLU activated (Section~\ref{sec:ev-cert}).
Since \dd{} implicitly $\SUM$ pools and the columns of $W_1$ can have negative entries, it does not suffice.
Our proposed \ddplus{} therefore constrains every layer, including $W_1$, to be strictly positive.

\textbf{Monotone Head.} Since any choice of a monotone head suffices, we choose a trivial linear layer with non-negative weights for \sys{} and $g_{\text{lin}}(\mathbf r)=\mathbf w^{\top}\mathbf r+b$, while $\mathbf w=\operatorname{softplus}(\boldsymbol\theta)\geq 0$ and $\boldsymbol\theta\in\mathbb R^{d_\phi}$ are unconstrained trainable vectors.

\textbf{Cost of Security.} However, monotonicity comes at the cost of all benign evidence being ignored, whether added by an attacker or not. Therefore, if malware were only differentiable from benign samples by its absence of benign evidence, then per Eq.~\ref{eq:wb-decomp}, the monotone head would score benign samples equally to those of malware. Thus, this would make the resulting detector unable to flag malware without also misclassifying benign samples.

\section{Experimental Settings}
\label{sec:setup}
\begin{figure}[t]
\centering
\begin{tikzpicture}[x=\columnwidth/51, y=1cm, font=\footnotesize]
  \def\hs{0.00014}%
  \draw[black!25, dash pattern=on 1pt off 1pt] (0,{5000*\hs}) -- (49.47,{5000*\hs});
  \node[anchor=south east, text=black!60, inner sep=1pt] at (49.47,{5000*\hs}) {5,000 apps/month};
  \foreach \n [count=\c from 0] in {5626,5707,5681,5722,5661,5666,5663,5548,5613,5625,5775,5666,
      5600,5531,5587,5557,5522,5597,5593,5490,5471,5569,5438,5420,5428,5373,5421,5317,5268,5253,
      5144,5308,5247,5266,5015,4079,2550,2254,3650,2790,1847,1955,2453,3296,2146,1895,1997,1685}{%
    \pgfmathtruncatemacro{\yr}{\c/12}%
    \pgfmathsetmacro{\xl}{\c+0.55*\yr}%
    \ifnum\c<12 \def\cc{segtrain}\else\ifnum\c<18 \def\cc{segcal}\else\def\cc{segtest}\fi\fi
    \fill[\cc!70] (\xl,0) rectangle ++(0.82,{\n*\hs});}
  \draw[black!60, line width=0.4pt] (-0.2,0) -- (49.67,0);
  \foreach \k in {0,1,2,3,4}{\draw[black!60] ({12.55*\k-0.365},0) -- ({12.55*\k-0.365},-0.09);}
  \foreach \yr [count=\k from 0] in {2021,2022,2023,2024}
    \node[anchor=north, text=black!70, inner sep=1.5pt] at ({12.55*\k+5.91},-0.02) {\yr};
  \node[anchor=south, align=center, inner sep=1pt] at (5.91,0.86)
    {\textbf{\color{segtrain!75!black}Training}\\{\color{black!60}67,953 apps}};
  \node[anchor=south, align=center, inner sep=1pt] at (15.46,0.86)
    {\textbf{\color{segcal!75!black}Validation}\\{\color{black!60}33,394 apps}};
  \node[anchor=south, align=center, inner sep=1pt] at (30.5,0.86)
    {\textbf{\color{segtest!75!black}Test ($T{=}30$)}\\{\color{black!60}123,618 apps}};
  \draw[decorate, decoration={brace, amplitude=3pt}, black!70] (12.55,1.54) -- (49.47,1.54)
    node[midway, above=2pt, text=black] {36 months of active learning, 400 labels per month};
\end{tikzpicture}
\caption{\textbf{Hypercube (VTT2) timeline.} Every detector is trained on \ultrain{2021} and then updated monthly by active learning from January 2022 to December 2024. The first six months (\ulcal{January to June 2022}) are the validation period with all results being reported on the remaining 30 test months (\ultest{July 2022 to December 2024}), following Chen et al.~\cite{chen2023continuous}. Bar height: applications per month. Malware: 10.7\%, 10.8\% and 9.5\% over the three periods.}
\label{fig:timeline}
\end{figure}
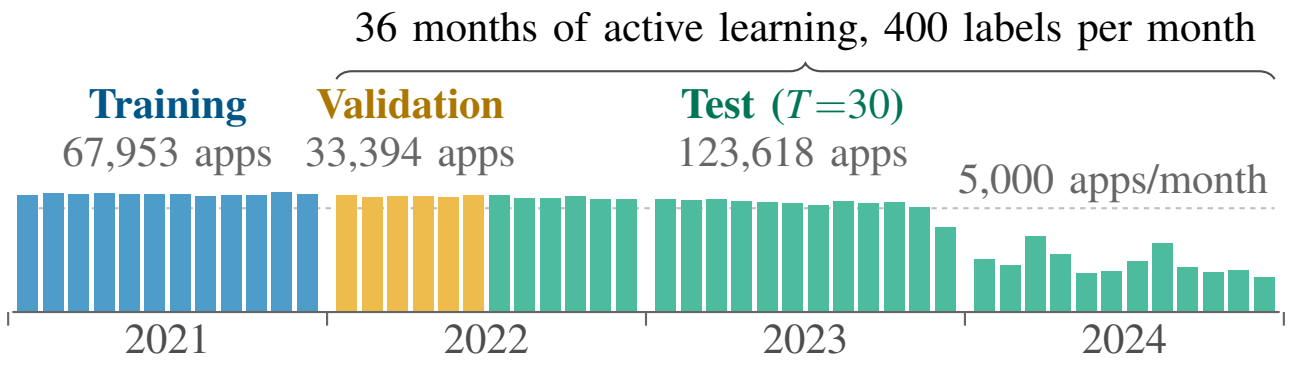

\noindent This section provides a breakdown of the core experimental settings used in the remainder of this paper.

\subsection{Evaluation Procedure}

\textbf{Dataset.}
We evaluate on Hypercube~\cite{hypercube} as it is, to the best of our knowledge, the least biased public dataset for Android malware detection.
It is compiled from a single market (Google Play) and time-stamped at upload time. Applications with at least two VirusTotal detections (VTT2) are labeled as malware, at a benign-to-malware ratio of 9:1.
For a longer temporal evaluation window, we use the extended release of Hypercube spanning 2021--2024 with 224,965 applications.
Regarding cardinality, all apps contain between $0$ and $36{,}265$ strings (median $199$; $24$ to $636$ between the 1st and 99th percentiles).

\textbf{Dataset Splits.}
We work with three temporal splits (see Figure~\ref{fig:timeline}).
First, we train on the first year (\ultrain{2021}), for which we assume clean data and all labels.
Second, following~\cite{chen2023continuous}, we use the following six months (\ulcal{Jan to Jun 2022}) as a validation period, where active learning runs during it, but we report no results on it.
Baselines keep the settings of their public implementations (Appendix~\ref{app:impl-detectors}).
Third, we report all results on the remaining 30 months (\ultest{Jul 2022 to Dec 2024}), our test period.

\textbf{Active Learning.}
For active learning we follow the protocol used in~\cite{chen2023continuous,mcfadden2026replicant,pendlebury2019tesseract,mcfadden2023poster,mcfadden2024impact,mcfadden2026drmd}.
Every month, each detector scores the new applications, and the $400$ with the lowest Maximum Softmax Probability (MSP)~\cite{hendrycks2017baseline,aurora} are labeled, simulating a human analyst~\cite{settles2009active,yang2021cade}. \hcc{} and \ramda{} use their own scores (Appendix~\ref{app:impl-detectors}).

\textbf{Metrics.}
For headline performance, we report F1, FPR, and FNR. Additionally, we report \textbf{reliability} via E-AURC, Cap$_{400}$ and F1$^{*}$, and \textbf{stability} via Sen[F1] following~\cite{aurora}.
For robustness, we report the attack success rate (ASR) as the share of detected malware that an attacker turns benign. We limit the attack-budget to at most $\beta{=}100$ organs per application. We average all metrics over five seeds.

\subsection{Malware Detectors}
\label{sec:setup-detectors}

\textbf{Proposed Detectors.} All \ds{}-based models share a common base design, where:
(i) $e$ is OpenAI's text-embedding-3-large~\cite{openai2024embeddings} with all $3{,}072$ dimensions, with empty feature groups skipped, and Appendix~\ref{app:ablation-encoder} providing a comparison with three other encoders that pre-date the dataset.
(ii) The element network $\phi$ is a two-layer MLP with ReLU and dropout $0.3$.
(iii) The classifier head is a linear layer for $\rho$ and a linear layer with softplus for the monotone head $g_{\text{lin}}$.
(iv) Training uses Adam (learning rate $0.002$, batch size $64$, $10$ epochs, gradients clipped at norm $1$) on binary cross-entropy, weighting malware by the 2021 class ratio, and is finetuned for an additional $5$ more epochs on 2021 before the first AL month.
(v) In each AL month, the model is fine-tuned for $10$ epochs on the $400$ new labels only, with a learning rate of $0.001$.

\textbf{Baseline Detectors.} We compare our proposed detectors against twelve state-of-the-art malware detectors: \dr{}~\cite{arp2014drebin}, \secsvm{}~\cite{demontis2019secureml}, \rf{}~\cite{onwuzurike2019mamadroid}, \dd{}~\cite{grosse2017adversarial}, \ramda{}~\cite{li2021ramda}, \hcc{}~\cite{chen2023continuous}, \drmd{}~\cite{mcfadden2026drmd}, \tif{}~\cite{zheng2026tif}, \malcertain{}~\cite{li2024malcertain}, \lsef{}~\cite{calzavara2023verifiable}, \lgbm{}~\cite{ke2017lightgbm}, and \monogbt{}~\cite{incer2018monotonic}. With all baselines using the same top-10,000 Drebin features, $V$, and Appendix~\ref{app:impl-detectors} giving the implementation and hyperparameters of each detector. Appendix~\ref{app:ablation-vocab} repeats experiments for monotone detectors on a larger vocabulary, $V_{100\text{k}}$.

\subsection{Attackers}
\label{sec:meth-attacks}
\label{sec:meth-wb}

\noindent Following our threat model (Section~\ref{sec:meth-threat}), we consider additive attacks that graft organs, benign code fragments, onto malware applications. The attacks in this paper are conducted using an organ ice-box, $O$, harvested by prior problem-space work~\cite{pierazzi2020intriguing, cortellazzi2025intriguing, bostani2024effectiveness, jigsaw} comprising $33{,}280$ organs alongside their pre-computed feature-space effects. We call the number of grafted organs the dose $\beta$ and report the attack success rate (ASR) at $\beta{=}100$, using the pre-computed feature-space effects, to provide a high-dose stress test. Using this $\beta$, we consider five different attackers: \textsc{B-Rnd}, \textsc{I-Rnd}, \textsc{EvD}, \textsc{ADZ}, and \textsc{APG}.

\textbf{Attacker Knowledge.} Organ selection depends on an attacker's knowledge (Table~\ref{tab:attackers}). For \textit{Blind Random} we assume no knowledge of a detector's feature space. Here we draw from all $33{,}280$ organs, although many make the same change to detectors for our baselines as they are trained on the top-$10{,}000$ features, $V$. All other attackers assume knowledge of the feature space and draw only from a subset of $O$ that \emph{adds} at least one feature the target detector sees, keeping one organ per distinct change. The effective pool, therefore, depends on the target's feature space. After deduplication, $7{,}694$ distinct organs remain against our models and only $3{,}143$ against models trained on $V$. Except for our \textit{Blind Random}, we follow the realizability constraints provided in problem-space attack repository of \cite{cortellazzi2025intriguing}.

\begin{table*}[t]\centering
\renewcommand{\arraystretch}{1.1}
\newcommand{\kY}{\tikz[baseline=-0.7ex]\draw[line width=0.5pt,fill=black] (0,0) circle (3pt);}%
\newcommand{\kN}{\tikz[baseline=-0.7ex]\draw[line width=0.5pt,fill=white] (0,0) circle (3pt);}%
\newcommand{\kH}[1]{\multicolumn{1}{c}{\rotatebox{90}{#1}}}%
\caption{
    \textbf{The Five Attackers.} Knowledge: \protect\kY{} known to the attacker, \protect\kN{} unknown. $V$: the top-$10{,}000$ features of the baselines. $\Sigma^{*}$: all strings. All attackers except for \textit{Blind Random} drawn from the \emph{effective} pool requiring knowledge that the \dr{} feature extractor was used. Deduplication leaves $7{,}694$ \emph{effective} organs for $\Sigma^{*}$ and $3{,}143$ for  $V$.
}
\begin{tabular*}{\textwidth}{@{}l@{\extracolsep{\fill}}cccc l cc c l@{}}
\toprule
 & \multicolumn{4}{c}{Knowledge} & & \multicolumn{2}{c}{Organ pool} & & \\
\cmidrule(lr){2-5}\cmidrule(lr){7-8}
Attacker & \kH{Labels} & \kH{Features} & \kH{Score} & \kH{Model} & Organs selected by & $V$ & $\Sigma^{*}$ & Budgets & Origin \\
\midrule
\textsc{B-Rnd} & \kY & \kN & \kN & \kN & uniform draw across all organs & $33{,}280$ & $33{,}280$ & -- & \cite{cortellazzi2025intriguing} \\
\textsc{I-Rnd} & \kY & \kY & \kN & \kN & uniform draw across the effective pool & $3{,}143$ & $7{,}694$ & \checkmark & \cite{mcfadden2026replicant} \\
\textsc{EvD} & \kY & \kY & \kY & \kN & random draw, kept only if the score drops & $3{,}143$ & $7{,}694$ & \checkmark & \cite{bostani2024evadedroid} \\
\textsc{ADZ} & \kY & \kY & \kY & \kN & query-guided search, plus own permission and intent edits & $3{,}143$ & $7{,}694$ & \checkmark & \cite{he2023advdroidzero} \\
\textsc{APG} & \kY & \kY & \kY & \kY & best-ranked organ, exact form for $\MAX$ pooled \ds{} models & $3{,}143$ & $7{,}694$ & \checkmark & \cite{pierazzi2020intriguing,cortellazzi2025intriguing} \\
\bottomrule
\end{tabular*}
\label{tab:attackers}
\end{table*}

\textbf{Random Attackers.} Our \textit{Blind Random} (\textsc{B-Rnd}) and \textit{Informed Random} (\textsc{I-Rnd}) attacks differ in how organs are drawn from $O$. Both add one organ at a time until the malware sample is classified as benign \underline{or} we reach $\beta{=}100$. \textsc{B-Rnd} draws uniformly at random from all of $O$ and may draw the same organ twice. As de-duplicating organs requires knowledge of which feature-space the detectors operate on. \textsc{I-Rnd} only draws organs from the effective pool that \emph{add} at least one new feature to the malware sample without replacement, while respecting realizability budgets and assuming knowledge about the feature-space and effective model-mask ($V$ vs. $\Sigma^{*}$).

\textbf{Query-Based Attackers.} \textit{AdvDroidZero} (\textsc{ADZ}), proposed by He et al.~\cite{he2023advdroidzero}, is a query-based attack, which performs additions using a selection tree based on their own set of manifest modifications and code insertions.
To maintain parity across attackers, we build these sets of manifest and code perturbations through our effective pool, based on the target's feature space.
\textit{EvadeDroid} (\textsc{EvD}), proposed by Bostani and Moonsamy~\cite{bostani2024evadedroid}, is the second query-based attack used in this work. At each step, EvD draws one organ at random without replacement, and only retains that organ if its grafting reduces the detector's score, otherwise it is discarded.

\textbf{White-Box Attacker.}
\label{sec:meth-apg}
\textit{Adversarial Program Generation} (\textsc{APG})~\cite{pierazzi2020intriguing} is a white-box attack proposed by Pierazzi et al. that, at each step, selects the organ that lowers a detector's score the most. In linear models, organs are selected deterministically and in descending order of their score-reducing impact via the weight-space.

\textit{APG for \sys{}}: The cost of computing APG depends on the model. (i) The greedy procedure for MLPs \cite{mcfadden2026replicant} such as \dd{} costs about $10^{9}$ multiply-add operations per step of every organ-addition. (ii)~Applied naively to \ds{} models, all candidates and their features need to be re-encoded by $\phi$ at about $10^{12}$ operations per step. (iii) $\MAX$ pooling reduces this cost since $\phi$ encodes features independently, where we compute $\mathbf r(S)$ once for the host and $\mathbf r(o)$ for organs $o \in O$. By Eq.~\ref{eq:wb-decomp}, every organ-addition costs $\mathbf r(S)\vee\mathbf r(o)$. Organs that replace no witnesses are removed, and only $\rho$ re-computes for candidates, reducing the cost to about $10^{8}$ operations per step. In our experiments, a median of $5{,}138$ organs ($67$\%) replace no witnesses and leave the score unchanged, reducing the cost by an additional two thirds. (iv) For \sys{}, parameterizing $\rho$ with a linear layer eliminates the need for re-scoring candidates by $\rho$, since the score changes by $\Delta(o\mid S)=\sum_j w_j\max(0,\,r_j(o)-r_j(S))$. Ignoring the cost of computing $r$ for the host and all organs once at the start of the procedure, the remaining cost reduces to that of APG, as introduced for linear models~\cite{pierazzi2020intriguing}, at about $10^{6}$ operations per step. Hence, ranking organs in weight space reduces the cost by three orders of magnitude for \sys{} compared to the greedy procedure for MLPs, while remaining exact.

\section{Results}
\label{sec:results}

\noindent In this section we evaluate \sys{} both in terms of its step-wise derivation (Section~\ref{sec:pillars}) and in comparison with the prior state-of-the-art detectors, across the following three settings.

\textbf{\setting{1} Clean AL:} no attacker. We report clean performance over the 30 test months (\ultest{Jul 2022 to Dec 2024}). \underline{Why?} A robust detector is only useful if it continues to detect malware without an attacker. This experiment measures each detector's clean performance under drift and serves as a baseline to measure the cost of security under attack.

\textbf{\setting{2} Attacking over Time:} no feedback into learning. We perform each of the five attacks monthly. Attackers only target malware, which was detected in that month, and the attacked samples never enter re-training. Here, we report the area-under-time of the ASR (AUT-ASR)~\cite{pendlebury2019tesseract,mcfadden2026replicant} from \ultest{Jul 2022 to Dec 2024}. \underline{Why?} Active learning retrains the detector every month; hence, its robustness may change with every update. This experiment isolates the robustness of each deployed model from the training loop.

\textbf{\setting{3} Attacked AL:} adversarial samples feedback into AL. From \ulcal{Jan 2022} onward, the attacker modifies every malware sample with APG. The adversarial samples then replace the their original sample in active learning.
\begin{itemize}
    \item In \setting{3\,\textbar\,A}, the attacker stops once the sample is classified as benign.
    \item In \setting{3\,\textbar\,B}, the attacker continues until the sample is both classified as benign and beyond the threshold for AL sampling based on a rolling average of the prior six months~\cite{aurora}.
\end{itemize}
\underline{Why?} In deployment, attacked samples are selected for labeling and influence AL model updates. These experiment tests whether robustness holds when adversarial sample enter AL.

\textbf{Metrics for Setting 3.} We report F1, E-AURC, and ASR over the 30 test months from \ultest{Jul 2022 to Dec 2024}. Since some detectors detect only a handful of malware samples per month, AUT-ASR is too unstable. Hence, we pool ASR over these months, $\mathrm{ASR} = \sum_m E_m / \sum_m D_m$. Here, $D_m$ is the malware of month $m$ that the deployed detector flags before modification, and $E_m$ is the fraction that the attacker turns benign. Malware that the detector already misses without an attack is not a success for the attacker and is thus reported in F1 instead.

\subsection{Ablations}
\label{sec:evidence}
\label{sec:ev-setup}

\noindent Table~\ref{tab:ablation} presents the results for the transformation of \dd{} (row~1) into \sys{} (row~11).
Metrics and settings follow Section~\ref{sec:setup}. Full results are in Appendix~\ref{app:ablation-vocab}--\ref{app:ablation-domain}.

\providecommand{\tDD}{\dd{}}
\providecommand{\tDS}{\ds{}}
\providecommand{\tSYS}{\sys{}}
\begin{table*}[t]
\centering
\setlength{\tabcolsep}{3.0pt}%
\color{black}
\definecolor{ablationChg}{RGB}{255,236,140}
\colorlet{ablationNew}{stepgold!22}
\definecolor{ablationReuse}{RGB}{212,212,212}
\providecolor{tabbaseband}{RGB}{240,240,240}
\providecolor{tabreliband}{RGB}{230,245,220}
\providecolor{tabstabband}{RGB}{235,240,250}
\providecolor{basetext}{RGB}{90,90,90}
\providecolor{relitext}{RGB}{46,125,50}
\providecolor{stabtext}{RGB}{40,53,147}
\providecolor{tabevband}{RGB}{238,229,248}\providecolor{evtext}{RGB}{106,27,154}
\providecolor{tabbasebest}{RGB}{216,216,216}\providecolor{tabrelibest}{RGB}{196,230,178}
\providecolor{tabstabbest}{RGB}{208,218,243}\providecolor{tabevbest}{RGB}{216,198,242}
\providecommand{\tH}[2]{\shortstack{\rule[-0.6ex]{0pt}{2.6ex}#1\\$#2$}}
\newcommand{\ablationBand}[2]{\multicolumn{1}{>{\columncolor{#1}[\tabcolsep][\tabcolsep]}c}{#2}}
\let\ablationPlainCheck\checkmark\newcommand{\ablationBoldCheck}{\pmb{\ablationPlainCheck}}%
\newcommand{\ablationCellFont}{\bfseries\boldmath\let\checkmark\ablationBoldCheck}
\newcommand{\ablationNewCell}[1]{\cellcolor{ablationNew}{\ablationCellFont#1}}
\newcommand{\ablationReuseCell}[1]{\cellcolor{ablationReuse}{\ablationCellFont#1}}
\newcommand{\ablationSD}[1]{\ensuremath{_{\pm#1}}}%
\def\ablationBestCol{white}\newcommand{\ablationBest}[1]{\edef\ablationBestTmp{\noexpand\cellcolor{\ablationBestCol}}\ablationBestTmp\textbf{#1}}
\newcommand{\ablationRun}{\textit{run}}
\newcommand{\ablationPillarRule}{\specialrule{0.5pt}{2pt}{1pt}}
\newcommand{\ablationEq}{\raisebox{0.15ex}{\footnotesize$=$}}
\renewcommand{\arraystretch}{1.22}

\caption[Step ablations]{\textbf{Ablations} (\setting{1}, \setting{2}) according to \stepref{1}--\stepref{4} and panels \panel{A}--\panel{D} of Figure~\ref{fig:ffn_deepsets_equivalence}.
    Gold: our modifications. Grey: a contribution reused from an earlier row. Shaded cells: best value per results column.
    Perf.: performance. Rel.: reliability. Stab.: stability. Ev.: evasion. F1 and ASR in \%. $\pm$: SD over five seeds. Columns are defined in Section~\ref{sec:ev-setup}.
}
\adjustbox{width=\textwidth}{
\begin{tabular}{@{}l >{\let\ablationCellFont\relax}l cccccc
  >{\def\ablationBestCol{tabbasebest}}c
  >{\def\ablationBestCol{tabrelibest}}c
  >{\def\ablationBestCol{tabstabbest}}c
  *{4}{>{\def\ablationBestCol{tabevbest}}c}
  >{\def\ablationBestCol{tabevbest}\columncolor{white}[\tabcolsep][0pt]}c@{}}%
\toprule
\multicolumn{8}{@{}c}{\textbf{Method and configuration}} & \multicolumn{8}{c@{}}{\textbf{Results}} \\
\cmidrule(r){1-8}\cmidrule(l){9-16}
 & & \multicolumn{6}{c}{Configuration} & \multicolumn{3}{c}{\setting{1}} & \multicolumn{5}{c@{}}{\setting{2}} \\
\cmidrule(lr){3-8}\cmidrule(lr){9-11}\cmidrule(l){12-16}
 & & & & & & & & \ablationBand{tabbaseband}{\color{basetext}\textbf{Perf.}} & \ablationBand{tabreliband}{\color{relitext}\textbf{Rel.}} & \ablationBand{tabstabband}{\color{stabtext}\textbf{Stab.}} & \multicolumn{5}{>{\columncolor{tabevband}[\tabcolsep][0pt]}c@{}}{\color{evtext}\textbf{Ev.} (AUT-ASR, \%)} \\
\# & Detector & \rotatebox{90}{Element} & $\phi$ & $\oplus$ & \rotatebox{90}{Domain} & \rotatebox{90}{OOV (\%)} & \rotatebox{90}{Head} & \ablationBand{tabbaseband}{\tH{F1}{\uparrow}} & \ablationBand{tabreliband}{\tH{E-AURC}{\downarrow}} & \ablationBand{tabstabband}{\tH{Sen[F1]}{\uparrow}} & \ablationBand{tabevband}{\tH{B-Rnd}{\downarrow}} & \ablationBand{tabevband}{\tH{I-Rnd}{\downarrow}} & \ablationBand{tabevband}{\tH{EvD}{\downarrow}} & \ablationBand{tabevband}{\tH{ADZ}{\downarrow}} & \ablationBand{tabevband}{\tH{APG}{\downarrow}} \\
\midrule
\addlinespace[2pt]
\multicolumn{16}{@{}l}{\stepref{1}~\textbf{Features as Sets.} } \\
1\,\panel{A} & \tDD & $\mathbf x$ &  & $\SUM$ & $V$ & 75 & $\rho$ & 37.5\ablationSD{1.0} & 4.4\ablationSD{0.0} & $-$1.19\ablationSD{0.12} & 49.5\ablationSD{2.6} & 76.8\ablationSD{1.0} & 99.4\ablationSD{0.1} & 30.2\ablationSD{3.4} & 100.0\ablationSD{0.0} \\
1a\,\panel{B} & \ablationNewCell{\tDS} & \ablationNewCell{$U$} & \ablationNewCell{$W_1$} & $\SUM$ & $V$ & 75 & $\rho$ & 37.5\ablationSD{1.0} & 4.4\ablationSD{0.0} & $-$1.19\ablationSD{0.12} & \ablationEq & \ablationEq & \ablationEq & \ablationEq & \ablationEq \\
1b\,\panel{B} & \ablationReuseCell{\tDS} & \ablationReuseCell{$U$} & \ablationReuseCell{$W_1$} & $\SUM$ & $V$ & 75 & $\rho$ & 37.3\ablationSD{0.5} & 4.3\ablationSD{0.0} & $-$1.22\ablationSD{0.18} & 49.3\ablationSD{0.8} & 76.9\ablationSD{0.7} & 99.4\ablationSD{0.2} & 29.1\ablationSD{1.3} & 100.0\ablationSD{0.0} \\
\multicolumn{16}{@{}l}{\quad\itshape $\phi$, $\rho$ and training recipe of \ds{} (panel~\panel{C}), with $\SUM$ pooling.} \\
2 & \ablationReuseCell{\tDS} & \ablationReuseCell{$U$} & \ablationNewCell{MLP} & $\SUM$ & $V$ & 75 & $\rho$ & 40.9\ablationSD{1.4} & 5.5\ablationSD{0.2} & $-$0.96\ablationSD{0.09} & 66.1\ablationSD{5.8} & 97.1\ablationSD{1.2} & 95.9\ablationSD{1.3} & 61.1\ablationSD{6.8} & 99.9\ablationSD{0.1} \\
\ablationPillarRule
\multicolumn{16}{@{}l}{\stepref{2}~\textbf{Max Pooling.} } \\
3\,\panel{C} & \ablationReuseCell{\tDS} & \ablationReuseCell{$U$} & \ablationReuseCell{MLP} & \ablationNewCell{$\MAX$} & $V$ & 75 & $\rho$ & 45.7\ablationSD{0.7} & 4.6\ablationSD{0.1} & $-$0.94\ablationSD{0.05} & 48.5\ablationSD{5.4} & 77.4\ablationSD{9.4} & 95.5\ablationSD{2.6} & 76.2\ablationSD{18.0} & 99.5\ablationSD{0.4} \\
\ablationPillarRule
\multicolumn{16}{@{}l}{\stepref{3}~\textbf{Open-Vocabulary Encoder.}} \\
4 & \ablationReuseCell{\tDS} & \ablationNewCell{$e$} & \ablationReuseCell{MLP} & \ablationReuseCell{$\MAX$} & $V$ & 75 & $\rho$ & 42.9\ablationSD{1.0} & 4.7\ablationSD{0.1} & $-$0.85\ablationSD{0.11} & 45.4\ablationSD{8.9} & 63.7\ablationSD{8.7} & 86.4\ablationSD{3.8} & 32.4\ablationSD{9.2} & 97.1\ablationSD{2.3} \\
5 & \ablationReuseCell{\tDS} & \ablationReuseCell{$e$} & \ablationReuseCell{MLP} & \ablationReuseCell{$\MAX$} & \ablationNewCell{$\Sigma^{*}$} & \ablationNewCell{0} & $\rho$ & \ablationBest{48.1}\ablationSD{0.8} & \ablationBest{4.2}\ablationSD{0.2} & $-$0.82\ablationSD{0.05} & 46.9\ablationSD{3.2} & 54.1\ablationSD{4.5} & 71.0\ablationSD{1.9} & 31.7\ablationSD{8.3} & 96.5\ablationSD{2.1} \\
\addlinespace[2.5pt]
\multicolumn{16}{@{}l}{\quad\itshape \stepref{2} on the encoder: $\SUM$ in place of $\MAX$.} \\
6 & \ablationReuseCell{\tDS} & \ablationReuseCell{$e$} & \ablationReuseCell{MLP} & $\SUM$ & \ablationReuseCell{$\Sigma^{*}$} & \ablationReuseCell{0} & $\rho$ & 37.8\ablationSD{1.3} & 6.5\ablationSD{0.2} & $-$1.05\ablationSD{0.16} & 91.9\ablationSD{3.4} & 98.3\ablationSD{1.7} & 98.1\ablationSD{0.8} & 75.1\ablationSD{7.2} & 100.0\ablationSD{0.1} \\
\ablationPillarRule
\multicolumn{16}{@{}l}{\stepref{4}~\textbf{Monotone Head.} } \\
7 & \tDD$^*$ & $\mathbf x$ &  & $\SUM$ & $V$ & 75 & \ablationNewCell{$\rho_{\geq0}$} & 17.5\ablationSD{1.1} & 4.5\ablationSD{0.1} & $-$1.22\ablationSD{0.18} & 51.3 & 70.2 & 98.9 & 37.9 & 100.0 \\
8 & \tDD+ & $\mathbf x$ &  & $\SUM$ & $V$ & 75 & \ablationNewCell{$g_{\text{lin}}$} & 13.1\ablationSD{0.6} & 8.0\ablationSD{0.1} & $-$0.69\ablationSD{0.05} & \ablationBest{0.0}\ablationSD{0.0} & \ablationBest{0.0}\ablationSD{0.0} & \ablationBest{0.0}\ablationSD{0.0} & \ablationBest{0.0}\ablationSD{0.0} & \ablationBest{0.0}\ablationSD{0.0} \\
9 & \ablationReuseCell{\tDS} & \ablationReuseCell{$U$} & \ablationReuseCell{MLP} & \ablationReuseCell{$\MAX$} & $V$ & 75 & \ablationReuseCell{$g_{\text{lin}}$} & 36.9\ablationSD{1.2} & 9.8\ablationSD{1.4} & \ablationBest{\boldmath$-$0.66}\ablationSD{0.13} & \ablationBest{0.0}\ablationSD{0.0} & \ablationBest{0.0}\ablationSD{0.0} & \ablationBest{0.0}\ablationSD{0.0} & \ablationBest{0.0}\ablationSD{0.0} & \ablationBest{0.0}\ablationSD{0.0} \\
10 & \ablationReuseCell{\tDS} & \ablationReuseCell{$e$} & \ablationReuseCell{MLP} & \ablationReuseCell{$\MAX$} & $V$ & 75 & \ablationReuseCell{$g_{\text{lin}}$} & 35.0\ablationSD{0.8} & 10.4\ablationSD{0.8} & $-$0.76\ablationSD{0.10} & \ablationBest{0.0}\ablationSD{0.0} & \ablationBest{0.0}\ablationSD{0.0} & \ablationBest{0.0}\ablationSD{0.0} & \ablationBest{0.0}\ablationSD{0.0} & \ablationBest{0.0}\ablationSD{0.0} \\
11\,\panel{D} & \ablationNewCell{\textbf{\tSYS}} & \ablationReuseCell{$e$} & \ablationReuseCell{MLP} & \ablationReuseCell{$\MAX$} & \ablationReuseCell{$\Sigma^{*}$} & \ablationReuseCell{0} & \ablationReuseCell{$g_{\text{lin}}$} & 38.4\ablationSD{1.1} & 6.7\ablationSD{0.1} & $-$0.80\ablationSD{0.16} & \ablationBest{0.0}\ablationSD{0.0} & \ablationBest{0.0}\ablationSD{0.0} & \ablationBest{0.0}\ablationSD{0.0} & \ablationBest{0.0}\ablationSD{0.0} & \ablationBest{0.0}\ablationSD{0.0} \\
\addlinespace[2.5pt]
\multicolumn{16}{@{}l}{\quad\itshape Control: $\SUM$ in place of $\MAX$.} \\
12 & \ablationReuseCell{\tDS} & \ablationReuseCell{$e$} & \ablationReuseCell{MLP} & $\SUM$ & \ablationReuseCell{$\Sigma^{*}$} & \ablationReuseCell{0} & \ablationReuseCell{$g_{\text{lin}}$} & 36.3\ablationSD{0.9} & 8.6\ablationSD{0.9} & $-$0.76\ablationSD{0.07} & \ablationBest{0.0}\ablationSD{0.0} & \ablationBest{0.0}\ablationSD{0.0} & \ablationBest{0.0}\ablationSD{0.0} & \ablationBest{0.0}\ablationSD{0.0} & \ablationBest{0.0}\ablationSD{0.0} \\
\bottomrule
\end{tabular}
}
\label{tab:ablation}
\end{table*}

\textbf{\texorpdfstring{(\stepref{1})~Features as Sets.}{(1) Features as Sets.}}
\label{sec:ev-set}
For \stepref{1} (\panel{A} and \panel{B} in Figure~\ref{fig:ffn_deepsets_equivalence}), we show that \dd{} is an implicitly $\SUM$-pooled \ds{} model in two ways.
First, we convert trained \dd{} models into a \ds{} model by moving the activation and the bias into the classifier, $\rho$. Without retraining (row~1a of Table~\ref{tab:ablation}), it reproduces \dd{}'s results exactly.
Second, we train a \ds{} model matching the network architecture and hyperparameters of \dd{} (row~1b): F1 $37.3\pm0.5$ against $37.5\pm1.0$, E-AURC $4.33$ against $4.36$, which is within normal seed variance (paired per-seed differences, $p\geq$$0.30$).
Lastly, in row~2, we move the activation function and bias from $\rho$ into $\phi$, as suggested by~\cite{deep_sets} which yields an improvement in F1 to $40.9\pm1.4$.
In the remainder of our experiments, all our \ds{} models share the same hyperparameters with \dd{}, barring the following four exceptions, which were a result of hyperparameter tuning following~\cite{chen2023continuous}.
First, $\phi$ has a hidden layer of $512$ units and ends in $200$ units, the width of \dd{}'s hidden layers, and $\rho$ is a single linear layer. Second, $\phi$ uses a dropout of $0.3$. Third, $\phi$ and $\rho$ are trained with Adam on class-weighted binary cross-entropy since \dd{}'s hyperparameters do not translate well to \ds{} (Appendix~\ref{app:impl-detectors}). Fourth, for AL, $\phi$ and $\rho$ are both fine-tuned on the $400$ new labels instead of all remaining labels (calibration F1 raises by $3.9$ points).

\textbf{\texorpdfstring{(\stepref{2})~Max Pooling.}{(2) Max Pooling.}}
\label{sec:ev-join}
\label{sec:ev-p2}
We then replace $\SUM$- with $\MAX$-pooling (row~3).
$\MAX$-pooling lowers evasion (AUT-ASR) under three of the five attacks: Blind Random ($66.1\to48.5\%$), Informed Random ($97.1\to77.4\%$), and EvD ($95.9\to95.5\%$). However, APG still evades at $99.5\%$ and ADZ increases from $61.1$ to $76.2\%$.
Under clean AL, F1 rises by $4.8$ ($40.9\to45.7$) and E-AURC falls from $5.49$ to $4.62$.
In \dd{}, the attack lowers the malware-polarity channels ($28\to13$\% active).
In \dsv{\MAX, e, \Sigma^{*}}, it cannot lower any channel and succeeds by raising the benign-polarity channels alone ($63\to71$\%), while the malware-polarity channels remain unchanged.
Hence, $\MAX$-pooling narrows the attack surface; however, the non-monotone head still allows raised benign-polarity channels to outweigh malicious evidence.

\textbf{\texorpdfstring{(\stepref{3})~Open-Vocabulary Encoder.}{(3) Open-Vocabulary Encoder.}}
\label{sec:ev-domain}
\label{sec:ev-p1}
Next, we replace the one-hot encoding with a general text encoder $e$ on the same $10{,}000$ strings of $V$ (row~4).
At a fixed vocabulary, the encoder does not raise clean AL F1 ($45.7\%\to42.9\%$, E-AURC $4.62\to4.73$), but it lowers evasion under all five attackers (APG $99.5\to97.1\%$, ADZ $76.2\to32.4\%$).
\ds{}'s advantage over \dd{} under attack, therefore, stems from combining $\MAX$-pooling and $e$.
We then open the vocabulary to $\Sigma^{*}$, eliminating the OOV problem (row~5).
Using $\Sigma^{*}$ improves clean AL performance by $5.2$ points ($42.9\%\to48.1\%$). An open vocabulary also widens the attack surface, as the effective organ pool increases from $3{,}143$ to $7{,}694$ (see Table~\ref{tab:attackers}). However, four out of the five attackers evade row~5 less than row~4, and only Blind Random gains ($45.4\to46.9\%$).
The performance of \ds{} does not depend on the choice of $e$, as we tested three other encoders that pre-date our dataset, which achieved between $48.1\%$ and $49.1\%$ in clean AL F1, compared to the $48.1\%$ of the OpenAI text embedding model used in this work (see Appendix~\ref{app:ablation-encoder} for a comparison). Equally, robustness does not depend on $e$ either, since APG evades all encoders similarly.

\textbf{\texorpdfstring{(\stepref{4})~Monotone Head.}{(4) Monotone Head.}}
\label{sec:ev-cert}
\label{sec:ev-p3}
\label{sec:ev-p4}
Lastly, we replace the classifier $\rho$ with a monotone head $g_{\text{lin}}$ (see rows~8--11), providing complete robustness to additive attacks; no adversarial true positives were evaded.
On \dd{}, a non-negative output layer alone (\dd{}$^{*}$, row~7) is not sufficient.
Since added features can lower the channels prior the output layer (Section~\ref{sec:p-cert}), APG still achieves an average AUT-ASR of $100.0\%$, while costing F1 $20.0$ points ($37.5\to17.5$).
With all weights being non-negative (\ddplus{}, row~8), no attack evades, but the cost of security is higher at F1 $24.4$ points ($37.5\to13.1$).
In contrast, robustness is achieved for \ds{} using $\MAX$-pooling and a monotone head, $\rho=g_{\text{lin}}$ (row~9), while only lowering F1 by $0.6$ points on average ($37.5\to36.9$, relative to \dd{}).
Although $e$ on $V$ (row~10) further increases the cost of security to $2.5$ points on average ($37.5\to$$35.0$), when combined with an open vocabulary $\Sigma^*$ (\sys{}, row~11), it increases F1 to $38.4$ while improving the reliability under a monotone head ($10.38\to$$6.67$ E-AURC). We provide a comparison on a larger fixed vocabulary for $V$ in Section~\ref{sec:why-pays-less}.

\textbf{MAX or SUM Pooling under a Monotone Head.}
Whether a \ds{}-based model with a monotone head, $\rho$, remains robust depends on: (i) the choice of the activation function for $\phi$; or (ii) the choice of the pooling operator $\oplus$. Since we activate $\phi$ with ReLU, monotonicity is guaranteed because $\phi$ encodes features individually; hence, values in $H$ cannot be negative (\emph{cancellation}); however, \emph{accumulation} may still impact performance.
Therefore, in row~12, we evaluate $\oplus=\SUM$, which on average drops F1 by $2.1$ points ($38.4\to36.3$) and reliability by $1.92$ points ($6.67\to8.59$ E-AURC).
Thus, as both pooling operators maintain robustness, under a monotone head, $\oplus=\MAX$ provides performance and reliability.

\subsection{Comparison with State-of-the-Art}
\label{sec:results-main}

\providecommand{\tH}[2]{\shortstack{\rule[-0.6ex]{0pt}{2.6ex}#1\\$#2$}}
\providecommand{\tSD}[1]{\ensuremath{_{\pm#1}}}%
\providecommand{\tRuleGroup}{\specialrule{0.4pt}{1.5pt}{1.5pt}}
\providecolor{tabbaseband}{RGB}{240,240,240}
\providecolor{tabreliband}{RGB}{230,245,220}
\providecolor{tabstabband}{RGB}{235,240,250}
\providecolor{basetext}{RGB}{90,90,90}
\providecolor{relitext}{RGB}{46,125,50}
\providecolor{stabtext}{RGB}{40,53,147}
\providecolor{tabevband}{RGB}{238,229,248}
\providecolor{tabbasebest}{RGB}{216,216,216}
\providecolor{tabrelibest}{RGB}{196,230,178}
\providecolor{tabevbest}{RGB}{216,198,242}
\providecolor{evtext}{RGB}{106,27,154}
\providecommand{\tBand}[3]{\multicolumn{#1}{>{\columncolor{#2}[\tabcolsep][\tabcolsep]}c}{#3}}
\providecommand{\tBandL}[3]{\multicolumn{#1}{>{\columncolor{#2}[0pt][\tabcolsep]}c}{#3}}%
\providecommand{\tBandR}[3]{\multicolumn{#1}{>{\columncolor{#2}[\tabcolsep][0pt]}c}{#3}}%
\providecommand{\tBandV}[3]{\multicolumn{#1}{>{\columncolor{#2}[\tabcolsep][\tabcolsep]}c@{\hspace{\tabcolsep}\vrule width 0.4pt\hspace{\tabcolsep}}}{#3}}%
\providecommand{\tBandF}[3]{\multicolumn{#1}{>{\columncolor{#2}[2\tabcolsep][\tabcolsep]}c}{#3}}%
\begin{table*}[t]
\centering
\color{black}
\setlength{\tabcolsep}{3.8pt}%
\renewcommand{\arraystretch}{1.3}

\caption[Comparison with State-of-the-Art]{\textbf{Comparison with State-of-the-Art} (\setting{1}, \setting{3}). Perf.: performance. Rel.: reliability. Ev.: evasion. F1 and ASR in \%. $\pm$: SD over five seeds. ASR under \setting{1}, \setting{2}: AUT-ASR of APG over time (Table~\ref{tab:main-attack}). Shaded cells: best value per column within each block. Grey: year of publication. \emph{Rob.}: built for robustness to additive attacks, (\checkmark) to norm-bounded attacks. Columns are defined in Section~\ref{sec:ev-setup}.}
\adjustbox{width=\textwidth}{
\begin{tabular}{@{}cl c c c@{\hspace{\tabcolsep}\vrule width 0.4pt\hspace{\tabcolsep}}ccc@{\hspace{\tabcolsep}\vrule width 0.4pt\hspace{\tabcolsep}}ccc@{\hspace{\tabcolsep}\vrule width 0.4pt\hspace{\tabcolsep}}ccc}
\toprule
\multicolumn{5}{@{}c}{\textbf{Method}} & \multicolumn{2}{c}{\setting{1}} & \multicolumn{1}{c}{\setting{2}} & \multicolumn{3}{c}{\setting{3\,\textbar\,A}} & \multicolumn{3}{c@{}}{\setting{3\,\textbar\,B}} \\
\cmidrule(r){1-5}\cmidrule(l{\dimexpr-\tabcolsep+2pt\relax}r{2pt}){6-7}\cmidrule(l{2pt}r{\dimexpr-\tabcolsep+2pt\relax}){8-8}\cmidrule(l{\dimexpr-\tabcolsep+2pt\relax}r{\dimexpr-\tabcolsep+2pt\relax}){9-11}\cmidrule(l{\dimexpr-\tabcolsep+2pt\relax}){12-14}%
 & & & & & \tBandF{1}{tabbaseband}{\color{basetext}\textbf{Perf.}} & \tBand{1}{tabreliband}{\color{relitext}\textbf{Rel.}} & \tBandV{1}{tabevband}{\color{evtext}\textbf{Ev.}} & \tBandF{1}{tabbaseband}{\color{basetext}\textbf{Perf.}} & \tBand{1}{tabreliband}{\color{relitext}\textbf{Rel.}} & \tBandV{1}{tabevband}{\color{evtext}\textbf{Ev.}} & \tBandF{1}{tabbaseband}{\color{basetext}\textbf{Perf.}} & \tBand{1}{tabreliband}{\color{relitext}\textbf{Rel.}} & \tBand{1}{tabevband}{\color{evtext}\textbf{Ev.}} \\
 & Detector & \rotatebox{90}{Domain} & \rotatebox{90}{Head} & \rotatebox{90}{Rob.} & \tBandF{1}{tabbaseband}{\tH{F1}{\uparrow}} & \tBand{1}{tabreliband}{\tH{E-AURC}{\downarrow}} & \tBandV{1}{tabevband}{\tH{ASR}{\downarrow}} & \tBandF{1}{tabbaseband}{\tH{F1}{\uparrow}} & \tBand{1}{tabreliband}{\tH{E-AURC}{\downarrow}} & \tBandV{1}{tabevband}{\tH{ASR}{\downarrow}} & \tBandF{1}{tabbaseband}{\tH{F1}{\uparrow}} & \tBand{1}{tabreliband}{\tH{E-AURC}{\downarrow}} & \tBand{1}{tabevband}{\tH{ASR}{\downarrow}} \\
\midrule
\multirow{3}{*}{\rotatebox[origin=c]{90}{\footnotesize\textbf{Linear}}} & \dr{}\,{\color{black!45}\footnotesize 2014} & $V$ &  &  & 30.5\tSD{0.0} & 5.2\tSD{0.0} & 100.0\tSD{0.0} & 0.0\tSD{0.0} & 5.2\tSD{0.0} & 100.0\tSD{0.0} & 0.0\tSD{0.0} & 6.8\tSD{0.1} & 100.0\tSD{0.0} \\
 & \secsvm{}\,{\color{black!45}\footnotesize 2019} & $V$ &  & \checkmark & 26.9\tSD{0.0} & 5.2\tSD{0.0} & 100.0\tSD{0.0} & 5.5\tSD{0.0} & 5.3\tSD{0.0} & 85.8\tSD{0.0} & 3.4\tSD{0.0} & 5.8\tSD{0.0} & 90.0\tSD{0.0} \\
 & \drplus{}\,{\color{black!45}\footnotesize ours} & $V$ & {\bfseries\boldmath$g_{\text{lin}}$} & \pmb{\checkmark} & 23.6\tSD{0.0} & 9.5\tSD{0.0} & \cellcolor{tabevbest}\textbf{0.0}\tSD{0.0} & 23.6\tSD{0.0} & 9.5\tSD{0.0} & \cellcolor{tabevbest}\textbf{0.0}\tSD{0.0} & 23.6\tSD{0.0} & 9.5\tSD{0.0} & \cellcolor{tabevbest}\textbf{0.0}\tSD{0.0} \\
\tRuleGroup
\multirow{4}{*}{\rotatebox[origin=c]{90}{\footnotesize\textbf{Trees}}} & \rf{}\,{\color{black!45}\footnotesize 2019} & $V$ &  &  & 38.3\tSD{0.3} & 4.9\tSD{0.1} & 100.0\tSD{0.0} & 7.5\tSD{0.1} & 4.8\tSD{0.1} & 79.6\tSD{0.4} & 7.1\tSD{0.6} & 5.0\tSD{0.1} & 72.7\tSD{3.0} \\
 & \lsef{}\,{\color{black!45}\footnotesize 2023} & $V$ &  & (\checkmark) & 24.2\tSD{0.6} & 6.4\tSD{0.1} & 100.0\tSD{0.0} & 0.2\tSD{0.1} & 6.3\tSD{0.1} & 99.3\tSD{0.3} & 0.5\tSD{0.3} & 6.8\tSD{0.1} & 97.6\tSD{1.5} \\
 & \lgbm{}\,{\color{black!45}\footnotesize 2017} & $V$ &  &  & 38.1\tSD{0.0} & 4.3\tSD{0.0} & 100.0\tSD{0.0} & 9.1\tSD{0.0} & \cellcolor{tabrelibest}\textbf{4.3}\tSD{0.0} & 80.7\tSD{0.0} & 5.4\tSD{0.0} & 4.5\tSD{0.0} & 82.9\tSD{0.0} \\
 & \monogbt{}\,{\color{black!45}\footnotesize 2018} & $V$ &  & \checkmark & 23.9\tSD{0.0} & 6.1\tSD{0.0} & \cellcolor{tabevbest}\textbf{0.0}\tSD{0.0} & 23.9\tSD{0.0} & 6.1\tSD{0.0} & \cellcolor{tabevbest}\textbf{0.0}\tSD{0.0} & 23.9\tSD{0.0} & 6.1\tSD{0.0} & \cellcolor{tabevbest}\textbf{0.0}\tSD{0.0} \\
\tRuleGroup
\multirow{7}{*}{\rotatebox[origin=c]{90}{\footnotesize\textbf{MLPs}}} & \malcertain{}\,{\color{black!45}\footnotesize 2024} & $V$ &  &  & 29.9\tSD{0.8} & 6.5\tSD{0.1} & 95.7\tSD{0.6} & 2.7\tSD{0.5} & 8.2\tSD{0.1} & 90.7\tSD{1.9} & 3.8\tSD{0.9} & 9.0\tSD{0.1} & 87.9\tSD{3.1} \\
 & \ramda{}\,{\color{black!45}\footnotesize 2021} & $V$ &  & \checkmark & 36.3\tSD{0.3} & 8.4\tSD{0.2} & 64.4\tSD{0.9} & 25.2\tSD{0.5} & 9.6\tSD{0.2} & 36.2\tSD{2.3} & 25.8\tSD{0.9} & 9.2\tSD{0.1} & 34.4\tSD{1.9} \\
 & \hcc{}\,{\color{black!45}\footnotesize 2023} & $V$ &  &  & 38.1\tSD{0.8} & 6.4\tSD{0.3} & 100.0\tSD{0.0} & 7.2\tSD{1.1} & 7.4\tSD{0.2} & 82.9\tSD{3.3} & 2.8\tSD{1.1} & 10.4\tSD{0.3} & 90.9\tSD{4.0} \\
 & \tif{}\,{\color{black!45}\footnotesize 2026} & $V$ &  &  & 39.6\tSD{1.0} & 6.4\tSD{0.3} & 100.0\tSD{0.0} & 1.9\tSD{0.4} & 7.9\tSD{0.3} & 96.7\tSD{0.7} & 0.4\tSD{0.3} & 9.7\tSD{0.9} & 99.4\tSD{0.4} \\
 & \drmd{}\,{\color{black!45}\footnotesize 2026} & $V$ &  &  & 43.6\tSD{1.0} & 9.4\tSD{1.0} & 100.0\tSD{0.0} & 32.3\tSD{2.1} & 17.7\tSD{3.1} & 26.0\tSD{1.7} & 34.1\tSD{1.2} & 10.0\tSD{1.7} & 20.7\tSD{3.6} \\
 & \dd{}\,{\color{black!45}\footnotesize 2017} & $V$ & $\rho$ &  & 37.5\tSD{1.0} & 4.4\tSD{0.0} & 100.0\tSD{0.0} & 17.4\tSD{2.0} & 4.4\tSD{0.0} & 55.4\tSD{3.2} & 17.7\tSD{0.7} & 4.6\tSD{0.1} & 49.5\tSD{1.2} \\
 & \ddplus{}\,{\color{black!45}\footnotesize ours} & $V$ & {\bfseries\boldmath$g_{\text{lin}}$} & \pmb{\checkmark} & 13.1\tSD{0.6} & 8.0\tSD{0.1} & \cellcolor{tabevbest}\textbf{0.0}\tSD{0.0} & 13.1\tSD{0.6} & 8.0\tSD{0.1} & \cellcolor{tabevbest}\textbf{0.0}\tSD{0.0} & 13.1\tSD{0.6} & 8.0\tSD{0.1} & \cellcolor{tabevbest}\textbf{0.0}\tSD{0.0} \\
\tRuleGroup
\multirow{4}{*}{\rotatebox[origin=c]{90}{\footnotesize\textbf{MIL}}} & \ds{}\,{\color{black!45}\footnotesize ours} & $V$ & $\rho$ &  & 42.9\tSD{1.0} & 4.7\tSD{0.1} & 97.1\tSD{2.3} & 33.7\tSD{1.9} & 5.1\tSD{0.2} & 19.7\tSD{1.7} & 31.7\tSD{2.8} & 4.6\tSD{0.1} & 15.4\tSD{4.4} \\
 & \ds{}\,{\color{black!45}\footnotesize ours} & $V$ & {\bfseries\boldmath$g_{\text{lin}}$} & \pmb{\checkmark} & 35.0\tSD{0.8} & 10.4\tSD{0.8} & \cellcolor{tabevbest}\textbf{0.0}\tSD{0.0} & 35.0\tSD{0.8} & 10.4\tSD{0.8} & \cellcolor{tabevbest}\textbf{0.0}\tSD{0.0} & 35.0\tSD{0.8} & 10.4\tSD{0.8} & \cellcolor{tabevbest}\textbf{0.0}\tSD{0.0} \\
 & \ds{}\,{\color{black!45}\footnotesize ours} & {\bfseries\boldmath$\Sigma^{*}$} & $\rho$ &  & \cellcolor{tabbasebest}\textbf{48.1}\tSD{0.8} & \cellcolor{tabrelibest}\textbf{4.2}\tSD{0.2} & 96.5\tSD{2.1} & 36.7\tSD{1.9} & 4.4\tSD{0.2} & 24.5\tSD{2.7} & 30.3\tSD{0.6} & \cellcolor{tabrelibest}\textbf{4.3}\tSD{0.1} & 24.5\tSD{2.4} \\
 & \textbf{\sys{}}\,{\color{black!45}\footnotesize ours} & {\bfseries\boldmath$\Sigma^{*}$} & {\bfseries\boldmath$g_{\text{lin}}$} & \pmb{\checkmark} & 38.4\tSD{1.1} & 6.7\tSD{0.1} & \cellcolor{tabevbest}\textbf{0.0}\tSD{0.0} & \cellcolor{tabbasebest}\textbf{38.4}\tSD{1.1} & 6.7\tSD{0.1} & \cellcolor{tabevbest}\textbf{0.0}\tSD{0.0} & \cellcolor{tabbasebest}\textbf{38.4}\tSD{1.1} & 6.7\tSD{0.1} & \cellcolor{tabevbest}\textbf{0.0}\tSD{0.0} \\
\bottomrule
\end{tabular}
}
\label{tab:main-postattack}
\end{table*}

\noindent We compare our models with twelve baselines from three architecture families: linear models, tree ensembles, and MLPs. Our \ds{} models form a fourth family, MIL.
Table~\ref{tab:main-postattack} reports clean AL (\setting{1}) and attacked AL (\setting{3}).
While Table~\ref{tab:main-attack} in Appendix~\ref{app:main-attack} reports attack performance over time without AL (\setting{2}).

\textbf{\setting{1}.}
Under clean active learning, MLPs are the strongest baselines, with
\drmd{} ($43.6$) and \tif{} ($39.6$) having the highest performance.
Following the top performing MLPs, are \rf{} ($38.3$), \hcc{} ($38.1$), and \lgbm{} ($38.1$).
With the remaining baselines from those families, alongside linear models, trailing behind ($13.1$--$37.5$).
Our non-monotone MIL model, \dsv{\MAX, e, \Sigma^{*}}, reaches the highest F1 of all detectors ($48.1$), surpassing \drmd{} while having the best reliability (E-AURC, $4.22$).

\textit{Cost of Security under Clean AL.} The monotone models of \drplus{}, \monogbt{}, \ddplus{}, \dsv{\MAX, e, V, g_{\text{lin}}}, and \sys{} remain unevaded at the \textit{cost for security}: \drplus{} loses $22.8\%$ of \dr{}'s F1 performance; \monogbt{} loses $37.2\%$ of \lgbm{}'s F1; and, \ddplus{} loses $64.9\%$ of \dd{}'s F1. On the same $V$, \dsv{\MAX, e, V, g_{\text{lin}}} loses $18.4\%$ of \dsv{\MAX, e, V, \rho}'s F1.
While \sys{} does still pay the cost of security ($20.2\%$ of \dsv{\MAX, e, \rho}'s F1), it still outperforms ten of the baselines, including \dd{} by $0.9$ points in F1.

\textit{Vocabulary.} For our non-monotone \ds{} models, comparing \dsv{\MAX, e, V, \rho} with \dsv{\MAX, e, \Sigma^{*}, \rho}, the open vocabulary $\Sigma^{*}$ improves both F1 by $5.2$ points and reliability by $0.5$. Similarly, \sys{}, which is \dsv{\MAX, e, \Sigma^{*}, g_{\text{lin}}}, improves F1 by $3.4$ points and reliability by $3.7$ points compared to \dsv{\MAX, e, V, g_{\text{lin}}}.

\textbf{\setting{2}.} On all non-monotone baselines, APG achieves an AUT-ASR of at least $95.7\%$, with the exception of \ramda{} at $64.4\%$. For our \ds{}-based approaches, APG also evades ($96.5$-$97.1\%$) only against the non-monotone models.
As shown in Appendix Table~\ref{tab:main-attack}, while the robustness of monotone detectors remains across attackers, the average AUT-ASR across all other detectors for each attacker is as follows:
$62.3\%$ for \textsc{B-Rnd};
$82.8\%$ for \textsc{I-Rnd};
$61.3\%$ for \textsc{ADZ},
$89.5\%$ for \textsc{EvD},
and $96.4\%$ for \textsc{APG}.
Crucially, the comparison across all attacks highlights that while \dsv{\MAX, e, \Sigma^{*}, \rho} exposes a larger attack surface, the average AUT-ASR across attackers is $5.0$ points lower than \dsv{\MAX, e, V, \rho}.

\textbf{\setting{3}.}
While all monotone models are un-evaded, AL on attacked malware recovers evasion for some detectors across both \setting{3\,\textbar\,A} and \setting{3\,\textbar\,B}. While \dr{} still remains fully evaded, \secsvm{}'s evasion decreases by an average of $12.1$ points. For tree-based detectors, only \rf{} and \lgbm{} meaningfully decrease evasion by $23.9$ and $18.2$ points on average.
AL has the largest improvement on \drmd{}, \dsv{\MAX, e, V, \rho}, and \dsv{\MAX, e, \Sigma^{*}, \rho} as they similarly recover $74.5$ points on average under \setting{3\,\textbar\,A}.
For \setting{3\,\textbar\,B}, \drmd{} and \dsv{\MAX, e, V, \rho} gain an additional $4.8$ points on average, whereas \dsv{\MAX, e, \Sigma^{*}, \rho}, which exposes a larger attack surface, does not.
Although \ramda{} was the least evaded detector under \setting{2}, it only improves by an average of $29.1$ points, ranking 4th under both. Notably, our \dsv{\MAX, e, V, \rho} model is the least evaded under both \setting{3\,\textbar\,A} and \setting{3\,\textbar\,B}.

\textit{Reliability.} Both Settings impact reliability for all non-monotone detectors, where under \setting{3\,\textbar\,A}, \lgbm{} is the least affected. While \lgbm{} is closely followed by \dsv{\MAX, e, \Sigma^{*}, \rho}, which shows an advantage of $27.6$ F1 points and recovers to rank 1 under \setting{3\,\textbar\,B}. Under \setting{3\,\textbar\,A}, \drmd{} has the worst affected reliability, as its E-AURC rises from $9.42$ to $17.75$, while only declining by $0.55$ points under \setting{3\,\textbar\,B}.

\textit{Monotone Detectors.} All monotone models remain un-evaded, and \sys{} retains the highest performance under both settings. \dsv{\MAX, e, V, g_{\text{lin}}} trails by $3.4$ F1 points, and the monotone baselines follow by a wide margin. Here, \monogbt{} trails by $14.5$, and \drplus{} by $14.8$ F1 points, respectively. \ddplus{} ranks last with $13.1$ F1 points.

\subsection{Why \sys{} Pays Less}
\label{sec:why-pays-less}

\noindent With \ds{} models, we match the architecture to the set-shaped structure of the \drfe{} feature space, achieving the highest performance of any detector in our benchmark (row~5). While \sys{} loses $20.2\%$ of its F1 (row~11 versus row~5 of Table~\ref{tab:ablation}), it pays the smallest cost of security compared to the other monotone detectors, which lose $22.8\%$ for \drplus{}, $37.2\%$ for \monogbt{}, and $64.9\%$ for \ddplus{}. Therefore, \sys{} remains $14.5$ points F1 above the best monotone baseline.

\textbf{Monotonicity.} Linear models and MLPs are only robust if all weights are constrained. For example, APG still fully evades \dd{}$^{*}$ with a monotone output layer (row~7 in Table~\ref{tab:ablation}), and \ddplus{} costs a $64.9\%$ reduction in F1 to completely eliminate APG evasions (row~8 in Table~\ref{tab:ablation}).
In contrast, \sys{} only requires the classifier $\rho$ to be monotone, since adding features can only raise channels by construction.

\textbf{Pooling Operator.} Compared to $\MAX$ pooling, $\SUM$ pooling leads to feature accumulation (Figure~\ref{fig:polarity}), reducing F1 by $2.1$ points while decreasing reliability by $1.92$ (row~12, Table~\ref{tab:ablation}). A monotone head for \sys{} is not a performance bottleneck. Three alternative heads score within $2.1$ F1 points, and we select linear+ for simplicity (Table~\ref{tab:heads}).

\textbf{Open Vocabulary.} Feature selection is an explicit design choice for \sys{}. If all malware-features of a malignant sample were transplanted to a benign application, monotone detectors would score this sample at least as high. This holds for $32\%$ of malware under $V$, for $14\%$ under $V_{100\text{k}}$ (Appendix~\ref{app:ablation-vocab}), and reduces to $3\%$ under $\Sigma^{*}$. A larger vocabulary raises F1 by $3.5$ to $4.3$ points for all but \drplus{}, yet \ds{} models stay at least $10.3$ points above \monogbt{} (Table~\ref{tab:vocab100k}). $\Sigma^{*}$ improves the reliability of \sys{}, improving E-AURC to $6.7$ from the $10.4$ of $V$ and $7.9$ of $V_{100\text{k}}$, while reaching a similar F1 to that of $V_{100\text{k}}$ ($38.6$ versus $38.4$).

In summary, \ds{} models match the architecture to the set-shaped problem, \sys{} constrains only the head, $\MAX$ pooling prevents features from accumulating, while feature selection becomes an optional design-choice.

\section{Limitations}
\label{sec:limitations}

\noindent In our design and evaluation of \sys{}, we made a concerted effort to ablate every step of the process; however, there remain the following natural limitations to this work.
First, our experiments were conducted only on the Hypercube dataset; however, this choice was deliberate, as it constitutes the current standard for unbiased dataset construction~\cite{hypercube} with no comparable standards for other marketplaces.
Second, all our baselines share the same top-$10{,}000$ vocabulary, $V$, for consistency. Although this selection is in line with~\cite{kan2024tesseract,mcfadden2026drmd,mcfadden2026replicant,aurora}, there have been other proposed approaches for the selection of \drfe{} features; however, as \sys{} uses $\Sigma^{*}$, it is independent of feature selection.
Third, we focus on additive benign attacks, as they are the predominant threat in the literature. Due to monotonicity, \sys{} is robust to the removal of benign features; whereas the removal of malware features in the problem-space, without breaking malicious functionality, remains an open problem.
Fourth, we focus on the threat model of an attacker aiming to evade a target system; however, monotonicity does not prevent data poisoning~\cite{suciu2018fail}. Therefore, the addition of malicious features to benign samples could facilitate the poisoning of \sys{}, which we believe would be a natural direction for future work aiming to improve the resilience of \sys{}-based detectors.

\section{Related Work}
\label{sec:related}

\textbf{MIL in Security.} MIL is not new to the security domain: Jorgensen et al.\ encode emails via BoW to counter good-word attacks on spam filters~\cite{jorgensen2008mil}, and Pevn\'y et al.\ model network traffic and JSON documents, such as static reports of executables, as nested bags pooled by mean and maximum~\cite{pevny2016discriminative,pevny2020nested,mandlik2022jsongrinder}.
Our work differs in the choice of instance and pooling operator.
Their instances are text segments, network flows, or JSON values.
Words appended to a segment can still lower that segment's score, and added values shift the mean.
Whereas, our instances are single statically extracted strings, to which we only apply $\MAX$ pooling, so an added string cannot dilute the evidence of the others.
In combination with a monotone head, this hardens \sys{} to be robust against additive attacks (Section~\ref{sec:p-cert})

\textbf{Monotone and Certified Detectors.} Monotone classification has long been proposed as a defense against additive attacks.
Incer et al.\ train monotone gradient-boosted trees on Boolean features of Windows executables, and report a drop of approximately 13\% in detection rate, which they interpret as the cost of security~\cite{incer2018monotonic}.
Fleshman et al.\ constrain linear models and neural networks to non-negative weights, which eliminates evasions for the linear models and nearly eliminates them for the neural network~\cite{fleshman2019nonnegative}, while Chen et al.\ train PDF classifiers with formally verified robustness properties, for instance against inserting pages of benign documents~\cite{chen2020robustpdf}.
Gimenez et al.\ show that a monotone feature mapping combined with a monotone classifier is robust by design and, conversely, that every robust detector decomposes into a monotone classifier on a latent feature space~\cite{gimenez2025erdalt}.
A second line of work certifies robustness by randomized or derandomized smoothing~\cite{huang2023rsdel,saha2024drsm,gibert2024smoothing}, which covers edits of the raw bytes, but only up to a bounded size.
All of these detectors operate on a fixed feature space, whereas \sys{} applies monotonicity to an open-vocabulary set representation and remains robust to any number of added strings.

\textbf{Concept Drift \& Active Learning.}
Malware detectors degrade over time as malware and benign applications evolve, and evaluations that ignore the temporal order of samples overestimate their performance~\cite{pendlebury2019tesseract,kan2024tesseract}. Detectors cope with this concept drift by rejecting samples that are likely to be misclassified based on conformal evaluation~\cite{barbero2022transcendent}; detecting and explaining drifting samples~\cite{yang2021cade}; and using active learning, where analysts label the most uncertain samples each month to retrain the model~\cite{chen2023continuous}. Recent detectors learn representations that remain stable over time~\cite{zheng2026tif}, correct unreliable predictions~\cite{li2024malcertain}, or train a DRL agent that jointly optimizes classification and rejection decisions into one policy~\cite{mcfadden2026drmd}. Alternatively, Aurora evaluates whether their confidence estimates remain reliable and stable under drift~\cite{aurora}. Replicant learns evasion policies under a label-only black-box threat model and uses them to harden detectors~\cite{mcfadden2026replicant}. We evaluate \sys{} in this monthly active-learning setting, both clean and under attack, so that robustness is measured on every model the operator deploys.

\section{Conclusion}
\label{sec:conclusion}

\noindent Under set-shaped feature representations encoded as multi-hot vectors, linear models and MLPs are sum-pooled set models in disguise. Multiple-instance learning matches the model architecture to this modality and turns the encoder, pooling operator, and head into explicit design choices. A general text encoder maps each feature string to an embedding, including strings that first appear after training.
$\MAX$ pooled \ds{} models with a monotone head retain malicious evidence regardless of the set encoders robustness or how many benign organs an attacker transplants.
Altogether, these produce \sys{}, which pays the smallest relative \textit{cost of security} for remaining completely unevaded by state-of-the-art black-box and white-box additive attacks.

\section*{Acknowledgments}
This research was partially funded and supported by: 
the Defence Science and Technology Laboratory (DSTL), an executive agency of the UK Ministry of Defence, supporting the Autonomous Resilient Cyber Defence (ARCD) project within the DSTL Cyber Defence Enhancement programme; Google ASPIRE and Google Academics Research Award unrestricted gifts.

\bibliographystyle{plain}
\bibliography{refs}

@inproceedings{deep_sets,
  author    = {Manzil Zaheer and Satwik Kottur and Siamak Ravanbakhsh and
               Barnabas Poczos and Ruslan Salakhutdinov and Alexander J. Smola},
  title     = {Deep Sets},
  booktitle = {Advances in Neural Information Processing Systems 30},
  year      = {2017},
  pages     = {3391--3401}
}

@article{onwuzurike2019mamadroid,
  title={Mamadroid: Detecting android malware by building markov chains of behavioral models (extended version)},
  author={Onwuzurike, Lucky and Mariconti, Enrico and Andriotis, Panagiotis and Cristofaro, Emiliano De and Ross, Gordon and Stringhini, Gianluca},
  journal={ACM Transactions on Privacy and Security (TOPS)},
  volume={22},
  number={2},
  pages={1--34},
  year={2019},
  publisher={ACM, New York, NY, USA}
}

@inproceedings{wu2019malscan,
  title     = {{MalScan}: Fast Market-Wide Mobile Malware Scanning by Social-Network Centrality Analysis},
  author    = {Wu, Yueming and Li, Xiaodi and Zou, Deqing and Yang, Wei and Zhang, Xin and Jin, Hai},
  booktitle = {Proc. IEEE/ACM Int. Conf. Automated Software Engineering (ASE)},
  pages     = {139--150},
  year      = {2019},
  url       = {https://ieeexplore.ieee.org/document/8952382}
}

@inproceedings{zhang2020apigraph,
  title     = {Enhancing State-of-the-art Classifiers with {API} Semantics to Detect Evolved {Android} Malware},
  author    = {Zhang, Xiaohan and Zhang, Yuan and Zhong, Ming and Ding, Daizong and Cao, Yinzhi and Zhang, Yukun and Zhang, Mi and Yang, Min},
  booktitle = {Proc. ACM CCS},
  pages     = {757--770},
  year      = {2020},
  url       = {https://dl.acm.org/doi/10.1145/3372297.3417291}
}

@article{gao2021gdroid,
  title     = {{GDroid}: {Android} malware detection and classification with graph convolutional network},
  author    = {Gao, Han and Cheng, Shaoyin and Zhang, Weiming},
  journal   = {Computers \& Security},
  volume    = {106},
  pages     = {102264},
  year      = {2021},
  url       = {https://www.sciencedirect.com/science/article/abs/pii/S0167404821000882}
}

@inproceedings{daoudi2021dexray,
  title={Dexray: a simple, yet effective deep learning approach to android malware detection based on image representation of bytecode},
  author={Daoudi, Nadia and Samhi, Jordan and Kabore, Abdoul Kader and Allix, Kevin and Bissyand{\'e}, Tegawend{\'e} F and Klein, Jacques},
  booktitle={International workshop on deployable machine learning for security defense},
  pages={81--106},
  year={2021},
  organization={Springer}
}

@inproceedings{arp2014drebin,
  title={{DREBIN}: Effective and Explainable Detection of Android Malware in Your Pocket},
  author={Arp, Daniel and Spreitzenbarth, Michael and H{\"u}bner, Malte and Gascon, Hugo and Rieck, Konrad},
  booktitle={Proc. NDSS},
  year={2014},
}

@inproceedings{hypercube,
  title = {{Beyond the TESSERACT: Trustworthy Dataset Curation for Sound Evaluations of Android Malware Classifiers}},
  author = {Chow, Theo and D'Onghia, Mario and Linhardt, Lorenz and Kan, Zeliang and Arp, Daniel and Cavallaro, Lorenzo and Pierazzi, Fabio},
  year = {2026},
  booktitle = {{IEEE} Conf. Secure and Trustworthy Machine Learning ({SaTML})},
}

@inproceedings{grosse2017adversarial,
  title={Adversarial Examples for Malware Detection},
  author={Grosse, Kathrin and Papernot, Nicolas and Manoharan, Praveen and Backes, Michael and McDaniel, Patrick},
  booktitle={Computer Security -- ESORICS 2017},
  pages={62--79},
  year={2017},
}

@inproceedings{li2021ramda,
  title={Robust Android Malware Detection Against Adversarial Example Attacks},
  author={Li, Heng and Zhou, Shiyao and Yuan, Wei and Luo, Xiapu and Gao, Cuiying and Chen, Shuiyan},
  booktitle={Proc. The Web Conference 2021},
  pages={3603--3612},
  year={2021},
}

@article{demontis2019secureml,
  title={Yes, Machine Learning Can Be More Secure! A Case Study on Android Malware Detection},
  author={Demontis, Ambra and Melis, Marco and Biggio, Battista and Maiorca, Davide and Arp, Daniel and Rieck, Konrad and Corona, Igino and Giacinto, Giorgio and Roli, Fabio},
  journal={IEEE Trans. Dependable and Secure Computing},
  volume={16},
  number={4},
  pages={711--724},
  year={2019},
}

@article{pintor2022secml,
  title={{secml}: Secure and Explainable Machine Learning in {Python}},
  author={Pintor, Maura and Demetrio, Luca and Sotgiu, Angelo and Melis, Marco and Demontis, Ambra and Biggio, Battista},
  journal={SoftwareX},
  volume={18},
  pages={101095},
  year={2022}
}

@inproceedings{chen2023continuous,
  title={Continuous Learning for Android Malware Detection},
  author={Chen, Yizheng and Ding, Zhoujie and Wagner, David},
  booktitle={Proc. USENIX Security},
  pages={1127--1144},
  year={2023}
}

@inproceedings{yang2021cade,
  title={{CADE}: Detecting and Explaining Concept Drift Samples for Security Applications},
  author={Yang, Limin and Guo, Wenbo and Hao, Qingying and Ciptadi, Arridhana and Ahmadzadeh, Ali and Xing, Xinyu and Wang, Gang},
  booktitle={Proc. USENIX Security},
  pages={2327--2344},
  year={2021}
}

@article{breiman2001random,
  title={Random Forests},
  author={Breiman, Leo},
  journal={Machine Learning},
  volume={45},
  number={1},
  pages={5--32},
  year={2001},
}

@article{mcfadden2026drmd, title={DRMD: Deep Reinforcement Learning for Malware Detection Under Concept Drift}, volume={40}, url={https://ojs.aaai.org/index.php/AAAI/article/view/37053}, DOI={10.1609/aaai.v40i2.37053}, number={2}, journal={Proceedings of the AAAI Conference on Artificial Intelligence}, author={McFadden, Shae and Foley, Myles and D’Onghia, Mario and Hicks, Chris and Mavroudis, Vasilios and Paoletti, Nicola and Pierazzi, Fabio}, year={2026}, month={Mar.}, pages={854–862}}

@inproceedings{pierazzi2020intriguing,
  title={Intriguing Properties of Adversarial {ML} Attacks in the Problem Space},
  author={Pierazzi, Fabio and Pendlebury, Feargus and Cortellazzi, Jacopo and Cavallaro, Lorenzo},
  booktitle={Proc. IEEE S\&P},
  pages={1332--1349},
  year={2020},
}

@inproceedings{incer2018monotonic,
  title     = {Adversarially Robust Malware Detection Using Monotonic Classification},
  author    = {Incer, Inigo and Theodorides, Michael and Afroz, Sadia and Wagner, David},
  booktitle = {Proc. ACM Int. Workshop Security and Privacy Analytics (IWSPA)},
  year      = {2018},
  organization = {ACM},
}

@inproceedings{ke2017lightgbm,
  title     = {{LightGBM}: A Highly Efficient Gradient Boosting Decision Tree},
  author    = {Ke, Guolin and Meng, Qi and Finley, Thomas and Wang, Taifeng and Chen, Wei and Ma, Weidong and Ye, Qiwei and Liu, Tie-Yan},
  booktitle = {Advances in Neural Information Processing Systems 30},
  year      = {2017},
}

@article{gibert2024smoothing,
  title   = {Certified Adversarial Robustness of Machine Learning-based Malware Detectors via (De)Randomized Smoothing},
  author  = {Gibert, Daniel and Demetrio, Luca and Zizzo, Giulio and Le, Quan and Planes, Jordi and Biggio, Battista},
  journal = {arXiv preprint arXiv:2405.00392},
  year    = {2024}
}

@article{bostani2024evadedroid,
  title={{EvadeDroid}: A Practical Evasion Attack on Machine Learning for Black-box {Android} Malware Detection},
  author={Bostani, Hamid and Moonsamy, Veelasha},
  journal={Computers \& Security},
  volume={139},
  pages={103676},
  year={2024},
}

@inproceedings{he2023advdroidzero,
  title={Efficient Query-Based Attack against {ML}-Based {Android} Malware Detection under Zero Knowledge Setting},
  author={He, Ping and Xia, Yifan and Zhang, Xuhong and Ji, Shouling},
  booktitle={Proc. ACM CCS},
  pages={90--104},
  year={2023},
}

@inproceedings{pendlebury2019tesseract,
  title={{TESSERACT}: Eliminating Experimental Bias in Malware Classification across Space and Time},
  author={Pendlebury, Feargus and Pierazzi, Fabio and Jordaney, Roberto and Kinder, Johannes and Cavallaro, Lorenzo},
  booktitle={Proc. USENIX Security},
  pages={729--746},
  year={2019}
}

@inproceedings{jordaney2017transcend,
  title={Transcend: Detecting Concept Drift in Malware Classification Models},
  author={Jordaney, Roberto and Sharad, Kumar and Dash, Santanu K. and Wang, Zhi and Papini, Davide and Nouretdinov, Ilia and Cavallaro, Lorenzo},
  booktitle={Proc. USENIX Security},
  pages={625--642},
  year={2017}
}

@misc{aurora,
      title={On the Reliability and Stability of Selective Methods in Malware Classification Tasks},
      author={Alexander Herzog and Aliai Eusebi and Lorenzo Cavallaro},
      year={2025},
  note = {arXiv:2505.22843},
      eprint={2505.22843},
      archivePrefix={arXiv},
      primaryClass={cs.CR},
      url={https://arxiv.org/abs/2505.22843},
}

@article{settles2009active,
  title={Active learning literature survey},
  author={Settles, Burr},
  year={2009},
  publisher={University of Wisconsin-Madison Department of Computer Sciences}
}

@inproceedings{barr2015transplantation,
  title={Automated Software Transplantation},
  author={Barr, Earl T. and Harman, Mark and Jia, Yue and Marginean, Alexandru and Petke, Justyna},
  booktitle={Proc. Int. Symposium on Software Testing and Analysis (ISSTA)},
  pages={257--269},
  year={2015},
}

@inproceedings{barbero2022transcendent,
  author    = {Federico Barbero and Feargus Pendlebury and Fabio Pierazzi and Lorenzo Cavallaro},
  title     = {Transcending Transcend: Revisiting Malware Classification in the Presence of Concept Drift},
  booktitle = {{IEEE} Symposium on Security and Privacy},
  year      = {2022}
}

@article{scano2026droidbreaker,
  title={DroidBreaker: Practical and Functional Problem-Space Attacks on Machine-Learning Android Malware Detectors},
  author={Scano, Christian and Soi, Diego and Sotgiu, Angelo and Demetrio, Luca and Maiorca, Davide and Giacinto, Giorgio and Roli, Fabio and Biggio, Battista},
  journal={arXiv preprint arXiv:2606.26707},
  year={2026}
}

@article{maiorca2015stealth,
  title     = {Stealth attacks: An extended insight into the obfuscation effects on {Android} malware},
  author    = {Maiorca, Davide and Ariu, Davide and Corona, Igino and Aresu, Marco and Giacinto, Giorgio},
  journal   = {Computers \& Security},
  volume    = {51},
  pages     = {16--31},
  year      = {2015},
}

@article{hendrycks2017baseline,
  title={A baseline for detecting misclassified and out-of-distribution examples in neural networks},
  author={Hendrycks, Dan and Gimpel, Kevin},
  journal={arXiv preprint arXiv:1610.02136},
  year={2016}
}

@article{cortellazzi2025intriguing,
  title={Intriguing properties of adversarial ml attacks in the problem space [extended version]},
  author={Cortellazzi, Jacopo and Quiring, Erwin and Arp, Daniel and Pendlebury, Feargus and Pierazzi, Fabio and Cavallaro, Lorenzo},
  journal={ACM Transactions on Privacy and Security},
  volume={28},
  number={4},
  pages={1--37},
  year={2025},
  publisher={ACM, New York, NY}
}

@article{zheng2026tif,
  title={Learning Temporal Invariance in {Android} Malware Detectors},
  author={Zheng, Xinran and Yang, Shuo and Ngai, Edith C.H. and Jana, Suman and Cavallaro, Lorenzo},
  journal={IEEE Trans. Software Engineering},
  year={2026},
  note={Early access. arXiv:2502.05098}
}

@misc{kan2024tesseract,
  title         = {{TESSERACT}: Eliminating Experimental Bias in Malware Classification across Space and Time (Extended Version)},
  author        = {Kan, Zeliang and McFadden, Shae and Arp, Daniel and Pendlebury, Feargus and Jordaney, Roberto and Kinder, Johannes and Pierazzi, Fabio and Cavallaro, Lorenzo},
  year          = {2024},
  note = {arXiv:2402.01359},
  eprint        = {2402.01359},
  archivePrefix = {arXiv},
  primaryClass  = {cs.LG},
}

@inproceedings{arp2022dos,
  title     = {Dos and Don'ts of Machine Learning in Computer Security},
  author    = {Arp, Daniel and Quiring, Erwin and Pendlebury, Feargus and Warnecke, Alexander and Pierazzi, Fabio and Wressnegger, Christian and Cavallaro, Lorenzo and Rieck, Konrad},
  booktitle = {Proc. USENIX Security},
  pages     = {3971--3988},
  year      = {2022}
}

@article{mcfadden2026replicant,
  title={REPLICANT: Learning Policies for Evading and Hardening Malware Detectors},
  author={McFadden, Shae and Tsingenopoulos, Ilias and D'Onghia, Mario and Herzog, Alexander and Foley, Myles and Hicks, Chris and Cavallaro, Lorenzo and Pierazzi, Fabio},
  journal={arXiv preprint arXiv:2608.28499},
  year={2026}
}

@inproceedings{qi2017pointnet,
  title     = {{PointNet}: Deep Learning on Point Sets for {3D} Classification and Segmentation},
  author    = {Qi, Charles R. and Su, Hao and Mo, Kaichun and Guibas, Leonidas J.},
  booktitle = {Proc. IEEE Conf. Computer Vision and Pattern Recognition (CVPR)},
  pages     = {652--660},
  year      = {2017}
}

@book{davey2002lattices,
  title     = {Introduction to Lattices and Order},
  author    = {Davey, Brian A. and Priestley, Hilary A.},
  edition   = {2nd},
  publisher = {Cambridge University Press},
  year      = {2002}
}

@inproceedings{biggio2013evasion,
  title     = {Evasion Attacks against Machine Learning at Test Time},
  author    = {Biggio, Battista and Corona, Igino and Maiorca, Davide and Nelson, Blaine and {\v{S}}rndi{\'c}, Nedim and Laskov, Pavel and Giacinto, Giorgio and Roli, Fabio},
  booktitle = {Machine Learning and Knowledge Discovery in Databases (ECML PKDD 2013), Part III},
  series    = {LNCS},
  volume    = {8190},
  pages     = {387--402},
  year      = {2013},
}

@inproceedings{hu2017malgan,
  title     = {Generating Adversarial Malware Examples for Black-Box Attacks Based on {GAN}},
  author    = {Hu, Weiwei and Tan, Ying},
  booktitle = {Data Mining and Big Data (DMBD 2022), Part II},
  series    = {CCIS},
  volume    = {1745},
  pages     = {409--423},
  year      = {2022},
  note      = {Preprint arXiv:1702.05983 (2017)}
}

@inproceedings{aldujaili2018adversarial,
  title     = {Adversarial Deep Learning for Robust Detection of Binary Encoded Malware},
  author    = {Al-Dujaili, Abdullah and Huang, Alex and Hemberg, Erik and O'Reilly, Una-May},
  booktitle = {IEEE Security and Privacy Workshops (SPW)},
  pages     = {76--82},
  year      = {2018},
}

@inproceedings{liang2016cracking,
  title     = {Cracking Classifiers for Evasion: A Case Study on the {Google}'s Phishing Pages Filter},
  author    = {Liang, Bin and Su, Miaoqiang and You, Wei and Shi, Wenchang and Yang, Gang},
  booktitle = {Proc. Int. Conf. World Wide Web (WWW)},
  pages     = {345--356},
  year      = {2016},
}

@inproceedings{srndic2014practical,
  title     = {Practical Evasion of a Learning-Based Classifier: A Case Study},
  author    = {{\v{S}}rndi{\'c}, Nedim and Laskov, Pavel},
  booktitle = {IEEE S\&P},
  pages     = {197--211},
  year      = {2014},
}

@inproceedings{ma2009identifying,
  title     = {Identifying Suspicious {URLs}: An Application of Large-Scale Online Learning},
  author    = {Ma, Justin and Saul, Lawrence K. and Savage, Stefan and Voelker, Geoffrey M.},
  booktitle = {Proc. Int. Conf. Machine Learning (ICML)},
  pages     = {681--688},
  year      = {2009},
}

@inproceedings{wressnegger2013ngrams,
  title     = {A Close Look on n-Grams in Intrusion Detection: Anomaly Detection vs. Classification},
  author    = {Wressnegger, Christian and Schwenk, Guido and Arp, Daniel and Rieck, Konrad},
  booktitle = {Proc. ACM Workshop Artificial Intelligence and Security (AISec)},
  pages     = {67--76},
  year      = {2013},
}

@article{rosenzweig2025noteasy,
  title     = {It's not Easy: Applying Supervised Machine Learning to Detect Malicious Extensions in the {Chrome Web Store}},
  author    = {Rosenzweig, Ben and Dalla Valle, Valentino and Apruzzese, Giovanni and Fass, Aurore},
  journal   = {ACM Trans. Web},
  volume    = {20},
  number    = {1},
  pages     = {1--37},
  year      = {2026},
  note      = {Preprint arXiv:2509.21590 (2025)}
}

@inproceedings{schultz2001datamining,
  title     = {Data Mining Methods for Detection of New Malicious Executables},
  author    = {Schultz, Matthew G. and Eskin, Eleazar and Zadok, Erez and Stolfo, Salvatore J.},
  booktitle = {Proc. IEEE S\&P},
  pages     = {38--49},
  year      = {2001},
}

@misc{anderson2018ember,
  title         = {{EMBER}: An Open Dataset for Training Static {PE} Malware Machine Learning Models},
  author        = {Anderson, Hyrum S. and Roth, Phil},
  year          = {2018},
  note = {arXiv:1804.04637},
  eprint        = {1804.04637},
  archivePrefix = {arXiv}
}

@inproceedings{srndic2013hidost,
  title     = {Detection of Malicious {PDF} Files Based on Hierarchical Document Structure},
  author    = {{\v{S}}rndi{\'c}, Nedim and Laskov, Pavel},
  booktitle = {Proc. NDSS},
  year      = {2013},
  url       = {https://www.ndss-symposium.org/ndss2013/ndss-2013-programme/detection-malicious-pdf-files-based-hierarchical-document-structure/}
}

@inproceedings{chen2020robustpdf,
  title     = {On Training Robust {PDF} Malware Classifiers},
  author    = {Chen, Yizheng and Wang, Shiqi and She, Dongdong and Jana, Suman},
  booktitle = {Proc. USENIX Security},
  pages     = {2343--2360},
  year      = {2020},
  url       = {https://www.usenix.org/conference/usenixsecurity20/presentation/chen-yizheng}
}

@inproceedings{curtsinger2011zozzle,
  title     = {{ZOZZLE}: Fast and Precise In-Browser {JavaScript} Malware Detection},
  author    = {Curtsinger, Charlie and Livshits, Benjamin and Zorn, Benjamin and Seifert, Christian},
  booktitle = {Proc. USENIX Security},
  year      = {2011},
  url       = {https://www.usenix.org/legacy/event/sec11/tech/full_papers/Curtsinger.pdf}
}

@inproceedings{lowd2005good,
  title     = {Good Word Attacks on Statistical Spam Filters},
  author    = {Lowd, Daniel and Meek, Christopher},
  booktitle = {Proc. Conf. Email and Anti-Spam (CEAS)},
  year      = {2005},
  url       = {https://ix.cs.uoregon.edu/~lowd/ceas05lowd.pdf}
}

@inproceedings{sejfia2022amalfi,
  title     = {Practical Automated Detection of Malicious npm Packages},
  author    = {Sejfia, Adriana and Sch{\"a}fer, Max},
  booktitle = {Proc. Int. Conf. Software Engineering (ICSE)},
  pages     = {1681--1692},
  year      = {2022},
}

@inproceedings{jagpal2015webeval,
  title     = {Trends and Lessons from Three Years Fighting Malicious Extensions},
  author    = {Jagpal, Nav and Dingle, Eric and Gravel, Jean-Philippe and Mavrommatis, Panayiotis and Provos, Niels and Abu Rajab, Moheeb and Thomas, Kurt},
  booktitle = {Proc. USENIX Security},
  pages     = {579--593},
  year      = {2015},
  url       = {https://www.usenix.org/conference/usenixsecurity15/technical-sessions/presentation/jagpal}
}

@inproceedings{weinberger2009hashing,
  title     = {Feature Hashing for Large Scale Multitask Learning},
  author    = {Weinberger, Kilian and Dasgupta, Anirban and Langford, John and Smola, Alex and Attenberg, Josh},
  booktitle = {Proc. Int. Conf. Machine Learning (ICML)},
  pages     = {1113--1120},
  year      = {2009},
}

@article{kolter2006learning,
  title={Learning to detect and classify malicious executables in the wild.},
  author={Kolter, J Zico and Maloof, Marcus A},
  journal={Journal of machine learning research},
  volume={7},
  number={12},
  year={2006}
}

@inproceedings{xu2016evademl,
  title     = {Automatically Evading Classifiers: A Case Study on {PDF} Malware Classifiers},
  author    = {Xu, Weilin and Qi, Yanjun and Evans, David},
  booktitle = {Proc. NDSS},
  year      = {2016},
}

@inproceedings{ladisa2023crosslang,
  title     = {On the Feasibility of Cross-Language Detection of Malicious Packages in npm and {PyPI}},
  author    = {Ladisa, Piergiorgio and Ponta, Serena Elisa and Ronzoni, Nicola and Martinez, Matias and Barais, Olivier},
  booktitle = {Proc. Computer Security Applications Conference (ACSAC)},
  pages     = {71--82},
  year      = {2023},
  note      = {Preprint arXiv:2310.09571}
}

@inproceedings{canali2011prophiler,
  title     = {Prophiler: A Fast Filter for the Large-Scale Detection of Malicious Web Pages},
  author    = {Canali, Davide and Cova, Marco and Vigna, Giovanni and Kruegel, Christopher},
  booktitle = {Proc. Int. Conf. World Wide Web (WWW)},
  pages     = {197--206},
  year      = {2011},
}

@article{dietterich1997mil,
  title     = {Solving the Multiple Instance Problem with Axis-Parallel Rectangles},
  author    = {Dietterich, Thomas G. and Lathrop, Richard H. and Lozano-P{\'e}rez, Tom{\'a}s},
  journal   = {Artificial Intelligence},
  volume    = {89},
  number    = {1--2},
  pages     = {31--71},
  year      = {1997},
}

@inproceedings{pevny2016discriminative,
  title     = {Discriminative Models for Multi-Instance Problems with Tree Structure},
  author    = {Pevn{\'y}, Tom{\'a}{\v{s}} and Somol, Petr},
  booktitle = {Proc. ACM Workshop Artificial Intelligence and Security (AISec)},
  year      = {2016},
}

@article{demetrio2021gamma,
  author  = {Luca Demetrio and Battista Biggio and Giovanni Lagorio and Fabio Roli and Alessandro Armando},
  title   = {Functionality-Preserving Black-Box Optimization of Adversarial Windows Malware},
  journal = {IEEE Trans. Inf. Forensics Security},
  volume  = {16},
  pages   = {3469--3478},
  year    = {2021},
}

@inproceedings{li2024malcertain,
  title     = {{MalCertain}: Enhancing Deep Neural Network Based {Android} Malware Detection by Tackling Prediction Uncertainty},
  author    = {Li, Haodong and Xu, Guosheng and Wang, Liu and Xiao, Xusheng and Luo, Xiapu and Xu, Guoai and Wang, Haoyu},
  booktitle = {Proc. IEEE/ACM Int. Conf. Software Engineering (ICSE)},
  pages     = {1--13},
  year      = {2024},
}

@article{bojanowski2017enriching,
  title={Enriching word vectors with subword information},
  author={Bojanowski, Piotr and Grave, Edouard and Joulin, Armand and Mikolov, Tomas},
  journal={Transactions of the association for computational linguistics},
  volume={5},
  pages={135--146},
  year={2017},
  publisher={MIT Press, One Rogers Street, Cambridge, MA 02142-1209, USA journals-info~…}
}

@inproceedings{grave2018learning,
  title={Learning word vectors for 157 languages},
  author={Grave, Edouard and Bojanowski, Piotr and Gupta, Prakhar and Joulin, Armand and Mikolov, Tom{\'a}{\v{s}}},
  booktitle={Proceedings of the eleventh international conference on language resources and evaluation (LREC 2018)},
  year={2018}
}

@inproceedings{reimers2019sentence,
  title     = {{Sentence-BERT}: Sentence Embeddings using {Siamese} {BERT}-Networks},
  author    = {Reimers, Nils and Gurevych, Iryna},
  booktitle = {Proc. Conf. Empirical Methods in Natural Language Processing (EMNLP-IJCNLP)},
  pages     = {3982--3992},
  year      = {2019}
}

@misc{openai2024embeddings,
  title        = {New embedding models and {API} updates},
  author       = {{OpenAI}},
  year         = {2024},
  month        = jan,
  howpublished = {\url{https://openai.com/index/new-embedding-models-and-api-updates/}}
}

@inproceedings{calzavara2023verifiable,
  title     = {Verifiable Learning for Robust Tree Ensembles},
  author    = {Calzavara, Stefano and Cazzaro, Lorenzo and Pibiri, Giulio Ermanno and Prezza, Nicola},
  booktitle = {Proc. ACM CCS},
  pages     = {1850--1864},
  year      = {2023},
}

@article{jorgensen2008mil,
  author  = {Jorgensen, Zach and Zhou, Yan and Inge, Meador},
  title   = {A Multiple Instance Learning Strategy for Combating Good Word Attacks on Spam Filters},
  journal = {Journal of Machine Learning Research},
  volume  = {9},
  number  = {38},
  pages   = {1115--1146},
  year    = {2008}
}

@misc{pevny2020nested,
  author        = {Pevn{\'y}, Tom{\'a}{\v{s}} and D{\v{e}}di{\v{c}}, Marek},
  title         = {Nested Multiple Instance Learning in Modelling of {HTTP} Network Traffic},
  year          = {2020},
  note = {arXiv:2002.04059},
  eprint        = {2002.04059},
  archivePrefix = {arXiv}
}

@article{mandlik2022jsongrinder,
  title={JsonGrinder{.}jl: automated differentiable neural architecture for embedding arbitrary JSON data},
  author={Mandl{\'\i}k, {\v{S}}imon and Ra{\v{c}}insk{\`y}, Mat{\v{e}}j and Lis{\`y}, Viliam and Pevn{\`y}, Tom{\'a}{\v{s}}},
  journal={Journal of Machine Learning Research},
  volume={23},
  number={298},
  pages={1--5},
  year={2022}
}

@misc{anderson2018rl,
  author        = {Hyrum S. Anderson and Anant Kharkar and Bobby Filar and David Evans and Phil Roth},
  title         = {Learning to Evade Static {PE} Machine Learning Malware Models via Reinforcement Learning},
  year          = {2018},
  note = {arXiv:1801.08917},
  eprint        = {1801.08917},
  archivePrefix = {arXiv}
}

@misc{gimenez2025erdalt,
  author        = {Pierre-Fran{\c{c}}ois Gimenez and Sarath Sivaprasad and Mario Fritz},
  title         = {Certifiably Robust Malware Detectors by Design},
  year          = {2025},
  note = {arXiv:2508.10038},
  eprint        = {2508.10038},
  archivePrefix = {arXiv}
}

@article{fleshman2019nonnegative,
  title={Non-negative networks against adversarial attacks},
  author={Fleshman, William and Raff, Edward and Sylvester, Jared and Forsyth, Steven and McLean, Mark},
  journal={arXiv preprint arXiv:1806.06108},
  year={2018}
}

@article{huang2023rsdel,
  title={Rs-del: Edit distance robustness certificates for sequence classifiers via randomized deletion},
  author={Huang, Zhuoqun and Marchant, Neil G and Lucas, Keane and Bauer, Lujo and Ohrimenko, Olga and Rubinstein, Benjamin},
  journal={Advances in Neural Information Processing Systems},
  volume={36},
  pages={18676--18711},
  year={2023}
}

@inproceedings{saha2024drsm,
  author    = {Saha, Shoumik and Wang, Wenxiao and Kaya, Yigitcan and Feizi, Soheil and Dumitras, Tudor},
  title     = {{DRSM}: De-Randomized Smoothing on Malware Classifier Providing Certified Robustness},
  booktitle = {Int. Conf. Learning Representations (ICLR)},
  year      = {2024}
}

@article{olah2020zoom,
  title={Zoom in: An introduction to circuits},
  author={Olah, Chris and Cammarata, Nick and Schubert, Ludwig and Goh, Gabriel and Petrov, Michael and Carter, Shan},
  journal={Distill},
  volume={5},
  number={3},
  pages={e00024--001},
  year={2020}
}

@inproceedings{sill1997monotonic,
  title={Monotonic Networks},
  author={Sill, Joseph},
  booktitle={Advances in Neural Information Processing Systems (NIPS)},
  volume={10},
  year={1997}
}

@article{gupta2016monotonic,
  title={Monotonic Calibrated Interpolated Look-Up Tables},
  author={Gupta, Maya and Cotter, Andrew and Pfeifer, Jan and Voevodski, Konstantin and Canini, Kevin and Mangylov, Alexander and Moczydlowski, Wojciech and van Esbroeck, Alexander},
  journal={Journal of Machine Learning Research},
  volume={17},
  number={109},
  pages={1--47},
  year={2016}
}

@article{daniels2010monotone,
  title={Monotone and Partially Monotone Neural Networks},
  author={Daniels, Hennie and Velikova, Marina},
  journal={IEEE Transactions on Neural Networks},
  volume={21},
  number={6},
  pages={906--917},
  year={2010},
  doi={10.1109/TNN.2010.2044803}
}

@inproceedings{mcfadden2024impact,
  title={The impact of active learning on availability data poisoning for android malware classifiers},
  author={McFadden, Shae and Kan, Zeliang and Cavallaro, Lorenzo and Pierazzi, Fabio},
  booktitle={2024 Annual Computer Security Applications Conference Workshops (ACSAC Workshops)},
  pages={73--84},
  year={2024},
  organization={IEEE}
}

@inproceedings{mcfadden2023poster,
  title={Poster: Rpal-recovering malware classifiers from data poisoning using active learning},
  author={McFadden, Shae and Kan, Zeliang and Cavallaro, Lorenzo and Pierazzi, Fabio},
  booktitle={Proceedings of the 2023 ACM SIGSAC Conference on Computer and Communications Security},
  pages={3561--3563},
  year={2023}
}

@article{bostani2024effectiveness,
  title={On the effectiveness of adversarial training on malware classifiers},
  author={Bostani, Hamid and Cortellazzi, Jacopo and Arp, Daniel and Pierazzi, Fabio and Moonsamy, Veelasha and Cavallaro, Lorenzo},
  journal={arXiv preprint arXiv:2412.18218},
  year={2024}
}

@article{jigsaw,
  title={{Jigsaw Puzzle: Selective Backdoor Attack to Subvert Malware Classifiers}},
  author={Yang, Limin and Chen, Zhi and Cortellazzi, Jacopo and Pendlebury, Feargus and Tu, Kevin and Pierazzi, Fabio and Cavallaro, Lorenzo and Wang, Gang},
  journal={arXiv preprint arXiv:2202.05470},
  year={2022}
}

@inproceedings{suciu2018fail,
  title     = {When Does Machine Learning {FAIL}? Generalized Transferability for Evasion and Poisoning Attacks},
  author    = {Suciu, Octavian and Marginean, Radu and Kaya, Yigitcan and Daum{\'e} III, Hal and Dumitras, Tudor},
  booktitle = {27th USENIX Security Symposium (USENIX Security 18)},
  pages     = {1299--1316},
  year      = {2018}
}

\appendix


\providecommand{\tH}[2]{\shortstack{\rule[-0.6ex]{0pt}{2.6ex}#1\\$#2$}}
\providecommand{\tSD}[1]{\ensuremath{_{\pm#1}}}%
\providecommand{\tRuleGroup}{\specialrule{0.4pt}{1.5pt}{1.5pt}}
\providecolor{tabbaseband}{RGB}{240,240,240}
\providecolor{tabreliband}{RGB}{230,245,220}
\providecolor{tabstabband}{RGB}{235,240,250}
\providecolor{basetext}{RGB}{90,90,90}
\providecolor{relitext}{RGB}{46,125,50}
\providecolor{stabtext}{RGB}{40,53,147}
\providecolor{tabevband}{RGB}{238,229,248}
\providecolor{tabbasebest}{RGB}{216,216,216}
\providecolor{tabrelibest}{RGB}{196,230,178}
\providecolor{tabevbest}{RGB}{216,198,242}
\providecolor{evtext}{RGB}{106,27,154}
\providecommand{\tBand}[3]{\multicolumn{#1}{>{\columncolor{#2}[\tabcolsep][\tabcolsep]}c}{#3}}
\providecommand{\tBandL}[3]{\multicolumn{#1}{>{\columncolor{#2}[0pt][\tabcolsep]}c}{#3}}%
\providecommand{\tBandR}[3]{\multicolumn{#1}{>{\columncolor{#2}[\tabcolsep][0pt]}c}{#3}}%
\providecommand{\tBandV}[3]{\multicolumn{#1}{>{\columncolor{#2}[\tabcolsep][\tabcolsep]}c@{\hspace{\tabcolsep}\vrule width 0.4pt\hspace{\tabcolsep}}}{#3}}%
\providecommand{\tBandF}[3]{\multicolumn{#1}{>{\columncolor{#2}[2\tabcolsep][\tabcolsep]}c}{#3}}%
\begin{table*}[t]
\centering
\setlength{\tabcolsep}{2.6pt}%
\color{black}
\renewcommand{\arraystretch}{1.3}

\caption[Attacks on the active-learning checkpoints]{\textbf{Attacks on the active-learning checkpoints} (\setting{2}). F1 (\%, mean over months) counting evaded apps as misses. \emph{ASR}: AUT-ASR (\%). $\pm$: SD over five seeds. Shaded cells are the best value per column. Grey: year of publication. \emph{Rob.}: authors mention robustness to additive attacks, (\checkmark) to norm-bounded attacks. Columns are defined in Section~\ref{sec:ev-setup}}
\adjustbox{width=\textwidth}{
\begin{tabular}{@{}cl c c c@{\hspace{\tabcolsep}\vrule width 0.4pt\hspace{\tabcolsep}}cc@{\hspace{\tabcolsep}\vrule width 0.4pt\hspace{\tabcolsep}}cc@{\hspace{\tabcolsep}\vrule width 0.4pt\hspace{\tabcolsep}}cc@{\hspace{\tabcolsep}\vrule width 0.4pt\hspace{\tabcolsep}}cc@{\hspace{\tabcolsep}\vrule width 0.4pt\hspace{\tabcolsep}}cc}
\toprule
\multicolumn{5}{@{}c}{\textbf{Method}} & \multicolumn{10}{c@{}}{\setting{2}} \\
\cmidrule(r){1-5}\cmidrule(l){6-15}
\multicolumn{5}{@{}c}{} & \multicolumn{2}{c}{B-Rnd} & \multicolumn{2}{c}{I-Rnd} & \multicolumn{2}{c}{EvD} & \multicolumn{2}{c}{ADZ} & \multicolumn{2}{c}{APG} \\
\cmidrule(l{\dimexpr-\tabcolsep+2pt\relax}r{\dimexpr-\tabcolsep+2pt\relax}){6-7}\cmidrule(l{\dimexpr-\tabcolsep+2pt\relax}r{\dimexpr-\tabcolsep+2pt\relax}){8-9}\cmidrule(l{\dimexpr-\tabcolsep+2pt\relax}r{\dimexpr-\tabcolsep+2pt\relax}){10-11}\cmidrule(l{\dimexpr-\tabcolsep+2pt\relax}r{\dimexpr-\tabcolsep+2pt\relax}){12-13}\cmidrule(l{\dimexpr-\tabcolsep+2pt\relax}){14-15}
 & & & & & \tBandF{1}{tabbaseband}{\color{basetext}\textbf{Perf.}} & \tBandV{1}{tabevband}{\color{evtext}\textbf{Ev.}} & \tBandF{1}{tabbaseband}{\color{basetext}\textbf{Perf.}} & \tBandV{1}{tabevband}{\color{evtext}\textbf{Ev.}} & \tBandF{1}{tabbaseband}{\color{basetext}\textbf{Perf.}} & \tBandV{1}{tabevband}{\color{evtext}\textbf{Ev.}} & \tBandF{1}{tabbaseband}{\color{basetext}\textbf{Perf.}} & \tBandV{1}{tabevband}{\color{evtext}\textbf{Ev.}} & \tBandF{1}{tabbaseband}{\color{basetext}\textbf{Perf.}} & \tBand{1}{tabevband}{\color{evtext}\textbf{Ev.}} \\
 & Detector & \rotatebox{90}{Domain} & \rotatebox{90}{Head} & \rotatebox{90}{Rob.} & \tBandF{1}{tabbaseband}{\tH{F1}{\uparrow}} & \tBandV{1}{tabevband}{\tH{ASR}{\downarrow}} & \tBandF{1}{tabbaseband}{\tH{F1}{\uparrow}} & \tBandV{1}{tabevband}{\tH{ASR}{\downarrow}} & \tBandF{1}{tabbaseband}{\tH{F1}{\uparrow}} & \tBandV{1}{tabevband}{\tH{ASR}{\downarrow}} & \tBandF{1}{tabbaseband}{\tH{F1}{\uparrow}} & \tBandV{1}{tabevband}{\tH{ASR}{\downarrow}} & \tBandF{1}{tabbaseband}{\tH{F1}{\uparrow}} & \tBand{1}{tabevband}{\tH{ASR}{\downarrow}} \\
\midrule
\multirow{3}{*}{\rotatebox[origin=c]{90}{\footnotesize\textbf{Linear}}} & \dr{}\,{\color{black!45}\footnotesize 2014} & $V$ &  &  & 12.2\tSD{0.4} & 66.8\tSD{1.4} & 1.4\tSD{0.1} & 97.0\tSD{0.1} & 0.0\tSD{0.0} & 100.0\tSD{0.0} & 7.9\tSD{0.3} & 78.4\tSD{1.0} & 0.0\tSD{0.0} & 100.0\tSD{0.0} \\
 & \secsvm{}\,{\color{black!45}\footnotesize 2019} & $V$ &  & \checkmark & 10.3\tSD{0.6} & 68.5\tSD{2.9} & 0.2\tSD{0.0} & 99.6\tSD{0.1} & 0.0\tSD{0.0} & 100.0\tSD{0.0} & 10.0\tSD{0.3} & 72.9\tSD{1.1} & 0.0\tSD{0.0} & 100.0\tSD{0.0} \\
 & \drplus{}\,{\color{black!45}\footnotesize ours} & $V$ & {\bfseries\boldmath$g_{\text{lin}}$} & \pmb{\checkmark} & 23.6\tSD{0.0} & \cellcolor{tabevbest}\textbf{0.0}\tSD{0.0} & 23.6\tSD{0.0} & \cellcolor{tabevbest}\textbf{0.0}\tSD{0.0} & 23.6\tSD{0.0} & \cellcolor{tabevbest}\textbf{0.0}\tSD{0.0} & 23.6\tSD{0.0} & \cellcolor{tabevbest}\textbf{0.0}\tSD{0.0} & 23.6\tSD{0.0} & \cellcolor{tabevbest}\textbf{0.0}\tSD{0.0} \\
\tRuleGroup
\multirow{4}{*}{\rotatebox[origin=c]{90}{\footnotesize\textbf{Trees}}} & \rf{}\,{\color{black!45}\footnotesize 2019} & $V$ &  &  & 11.0\tSD{0.4} & 78.7\tSD{0.7} & 4.1\tSD{0.6} & 92.6\tSD{1.0} & 4.6\tSD{0.6} & 91.8\tSD{1.0} & 10.5\tSD{0.9} & 79.8\tSD{1.4} & 0.0\tSD{0.0} & 100.0\tSD{0.0} \\
 & \lsef{}\,{\color{black!45}\footnotesize 2023} & $V$ &  & (\checkmark) & 3.8\tSD{0.8} & 87.8\tSD{2.6} & 0.7\tSD{0.3} & 97.8\tSD{0.8} & 0.5\tSD{0.2} & 97.9\tSD{1.0} & 3.6\tSD{1.1} & 87.9\tSD{3.2} & 0.0\tSD{0.0} & 100.0\tSD{0.0} \\
 & \lgbm{}\,{\color{black!45}\footnotesize 2017} & $V$ &  &  & 15.2\tSD{0.4} & 67.6\tSD{1.2} & 2.1\tSD{0.1} & 96.0\tSD{0.3} & 0.0\tSD{0.0} & 100.0\tSD{0.0} & 5.7\tSD{0.2} & 90.0\tSD{0.3} & 0.0\tSD{0.0} & 100.0\tSD{0.0} \\
 & \monogbt{}\,{\color{black!45}\footnotesize 2018} & $V$ &  & \checkmark & 23.9\tSD{0.0} & \cellcolor{tabevbest}\textbf{0.0}\tSD{0.0} & 23.9\tSD{0.0} & \cellcolor{tabevbest}\textbf{0.0}\tSD{0.0} & 23.9\tSD{0.0} & \cellcolor{tabevbest}\textbf{0.0}\tSD{0.0} & 23.9\tSD{0.0} & \cellcolor{tabevbest}\textbf{0.0}\tSD{0.0} & 23.9\tSD{0.0} & \cellcolor{tabevbest}\textbf{0.0}\tSD{0.0} \\
\tRuleGroup
\multirow{7}{*}{\rotatebox[origin=c]{90}{\footnotesize\textbf{MLPs}}} & \malcertain{}\,{\color{black!45}\footnotesize 2024} & $V$ &  &  & 10.4\tSD{1.4} & 82.6\tSD{2.8} & 4.9\tSD{0.7} & 96.3\tSD{1.1} & 10.3\tSD{1.3} & 81.7\tSD{2.6} & 8.6\tSD{2.3} & 85.7\tSD{4.2} & 5.0\tSD{0.6} & 95.7\tSD{0.6} \\
 & \ramda{}\,{\color{black!45}\footnotesize 2021} & $V$ &  & \checkmark & 32.7\tSD{0.4} & 12.2\tSD{0.6} & 31.8\tSD{0.3} & 15.3\tSD{0.8} & 25.2\tSD{0.5} & 36.5\tSD{0.9} & 31.9\tSD{0.6} & 15.3\tSD{1.4} & 15.2\tSD{0.4} & 64.4\tSD{0.9} \\
 & \hcc{}\,{\color{black!45}\footnotesize 2023} & $V$ &  &  & 1.6\tSD{0.8} & 97.6\tSD{2.3} & 0.5\tSD{0.3} & 99.9\tSD{0.1} & 0.6\tSD{0.3} & 99.7\tSD{0.1} & 0.5\tSD{0.3} & 99.8\tSD{0.2} & 0.5\tSD{0.3} & 100.0\tSD{0.0} \\
 & \tif{}\,{\color{black!45}\footnotesize 2026} & $V$ &  &  & 19.7\tSD{1.7} & 57.4\tSD{3.3} & 5.5\tSD{1.2} & 90.2\tSD{2.3} & 0.3\tSD{0.2} & 99.5\tSD{0.3} & 24.7\tSD{1.5} & 46.5\tSD{3.4} & 0.0\tSD{0.0} & 100.0\tSD{0.0} \\
 & \drmd{}\,{\color{black!45}\footnotesize 2026} & $V$ &  &  & 25.4\tSD{2.9} & 49.2\tSD{7.6} & 1.9\tSD{1.2} & 97.0\tSD{2.1} & 0.0\tSD{0.0} & 99.9\tSD{0.1} & 26.7\tSD{6.6} & 46.3\tSD{17.0} & 0.0\tSD{0.0} & 100.0\tSD{0.0} \\
 & \dd{}\,{\color{black!45}\footnotesize 2017} & $V$ & $\rho$ &  & 22.1\tSD{1.0} & 49.5\tSD{2.6} & 11.8\tSD{0.6} & 76.8\tSD{1.0} & 0.4\tSD{0.2} & 99.4\tSD{0.1} & 28.9\tSD{1.4} & 30.2\tSD{3.4} & 0.0\tSD{0.0} & 100.0\tSD{0.0} \\
 & \ddplus{}\,{\color{black!45}\footnotesize ours} & $V$ & {\bfseries\boldmath$g_{\text{lin}}$} & \pmb{\checkmark} & 13.1\tSD{0.6} & \cellcolor{tabevbest}\textbf{0.0}\tSD{0.0} & 13.1\tSD{0.6} & \cellcolor{tabevbest}\textbf{0.0}\tSD{0.0} & 13.1\tSD{0.6} & \cellcolor{tabevbest}\textbf{0.0}\tSD{0.0} & 13.1\tSD{0.6} & \cellcolor{tabevbest}\textbf{0.0}\tSD{0.0} & 13.1\tSD{0.6} & \cellcolor{tabevbest}\textbf{0.0}\tSD{0.0} \\
\tRuleGroup
\multirow{4}{*}{\rotatebox[origin=c]{90}{\footnotesize\textbf{MIL}}} & \ds{}\,{\color{black!45}\footnotesize ours} & $V$ & $\rho$ &  & 26.5\tSD{4.0} & 45.4\tSD{8.9} & 18.7\tSD{4.4} & 63.7\tSD{8.7} & 7.9\tSD{2.3} & 86.4\tSD{3.8} & 31.8\tSD{4.2} & 32.4\tSD{9.2} & 1.9\tSD{1.4} & 97.1\tSD{2.3} \\
 & \ds{}\,{\color{black!45}\footnotesize ours} & $V$ & {\bfseries\boldmath$g_{\text{lin}}$} & \pmb{\checkmark} & 35.0\tSD{0.8} & \cellcolor{tabevbest}\textbf{0.0}\tSD{0.0} & 35.0\tSD{0.8} & \cellcolor{tabevbest}\textbf{0.0}\tSD{0.0} & 35.0\tSD{0.8} & \cellcolor{tabevbest}\textbf{0.0}\tSD{0.0} & 35.0\tSD{0.8} & \cellcolor{tabevbest}\textbf{0.0}\tSD{0.0} & 35.0\tSD{0.8} & \cellcolor{tabevbest}\textbf{0.0}\tSD{0.0} \\
 & \ds{}\,{\color{black!45}\footnotesize ours} & {\bfseries\boldmath$\Sigma^{*}$} & $\rho$ &  & 29.2\tSD{1.2} & 46.9\tSD{3.2} & 25.7\tSD{2.2} & 54.1\tSD{4.5} & 17.6\tSD{0.8} & 71.0\tSD{1.9} & 36.0\tSD{3.5} & 31.7\tSD{8.3} & 2.3\tSD{1.0} & 96.5\tSD{2.1} \\
 & \textbf{\sys{}}\,{\color{black!45}\footnotesize ours} & {\bfseries\boldmath$\Sigma^{*}$} & {\bfseries\boldmath$g_{\text{lin}}$} & \pmb{\checkmark} & \cellcolor{tabbasebest}\textbf{38.4}\tSD{1.1} & \cellcolor{tabevbest}\textbf{0.0}\tSD{0.0} & \cellcolor{tabbasebest}\textbf{38.4}\tSD{1.1} & \cellcolor{tabevbest}\textbf{0.0}\tSD{0.0} & \cellcolor{tabbasebest}\textbf{38.4}\tSD{1.1} & \cellcolor{tabevbest}\textbf{0.0}\tSD{0.0} & \cellcolor{tabbasebest}\textbf{38.4}\tSD{1.1} & \cellcolor{tabevbest}\textbf{0.0}\tSD{0.0} & \cellcolor{tabbasebest}\textbf{38.4}\tSD{1.1} & \cellcolor{tabevbest}\textbf{0.0}\tSD{0.0} \\
\bottomrule
\end{tabular}
}
\label{tab:main-attack}
\end{table*}

\begin{figure*}[t]
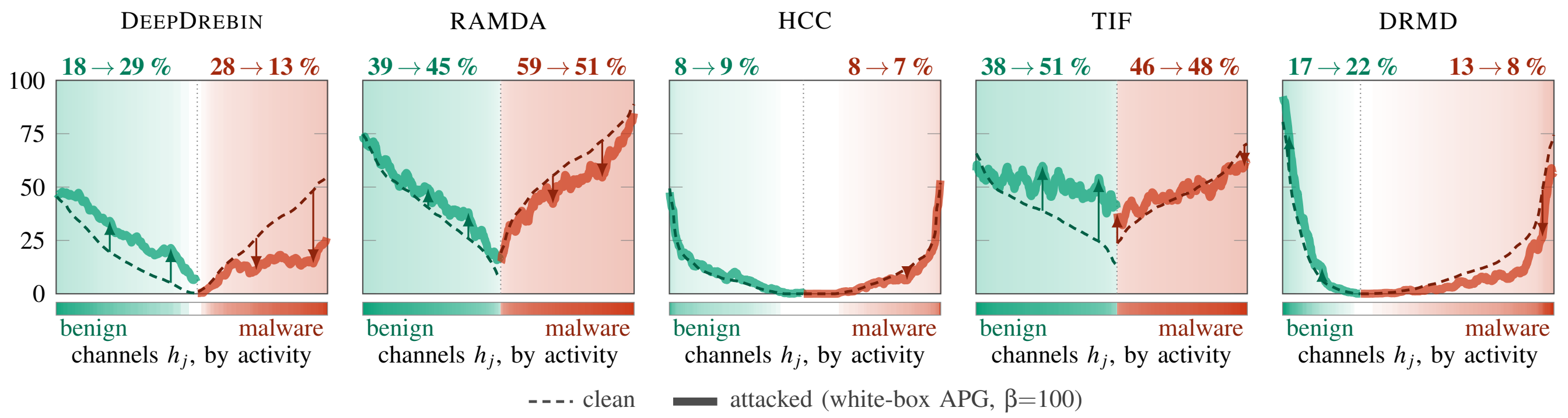

    \centering
    \def\polData{figures/polarity_baselines_data}%
\input{\polData/summary.tex}
\input{figures/polarity_style}
\begin{tikzpicture}
\begin{groupplot}[group style={group size=5 by 1, horizontal sep=0.40cm, y descriptions at=edge left},
                  polax, width=0.262\textwidth, height=4.0cm,
                  xlabel={channels $h_j$, by activity}]
\nextgroupplot[title={\dd{}}]    \polPanel{dd}{DD}
\nextgroupplot[title={\ramda{}}] \polPanel{ramda}{RAM}
\nextgroupplot[title={\hcc{}}]   \polPanel{hcc}{HCC}
\nextgroupplot[title={\tif{}}]   \polPanel{tif}{TIF}
\nextgroupplot[title={\drmd{}}]  \polPanel{drmd}{DRMD}
\end{groupplot}
\polLegend{(group c3r1.south)}
\end{tikzpicture}
    \caption[Channel polarity of the neural baselines]{\textbf{More can be less for MLPs.} Same construction as Figure~\ref{fig:polarity}. Share of applications (\%) in which a channel of the first hidden layer $\mathbf h$ is active. $\mathbf h$ sums one weight vector per present feature, an implicit $\SUM$ pooling. Hence, added benign features can switch off malware-polarity channels as seen in four of the five baselines. }
    \label{fig:polarity-baselines}
\end{figure*}

\section{Open-Vocabulary Encoders}
\label{app:ablation-domain}
\label{app:ablation-encoder}
\noindent Table~\ref{tab:encoders} replaces the general encoder $e$ of \dsv{\MAX, e, \Sigma^{*}} (row~5 of Table~\ref{tab:ablation}) and keeps everything else fixed. Besides OpenAI's text-embedding-3-large from 2024~\cite{openai2024embeddings}, we use three encoders that pre-date the dataset, namely hashed character n-grams without a learned model~\cite{bojanowski2017enriching,weinberger2009hashing}, fastText~\cite{grave2018learning} and Sentence-BERT~\cite{reimers2019sentence}. Lower dimensions are prefixes, hash widths or a PCA fitted on 2021 strings.

\begin{table}[t]
\centering
\providecommand{\tH}[2]{\shortstack{\rule[-0.6ex]{0pt}{2.6ex}#1\\$#2$}}
\providecommand{\tBand}[3]{\multicolumn{#1}{>{\columncolor{#2}[\tabcolsep][\tabcolsep]}c}{#3}}
\providecolor{tabbaseband}{RGB}{240,240,240}\providecolor{tabreliband}{RGB}{230,245,220}
\providecolor{tabstabband}{RGB}{235,240,250}\providecolor{tabevband}{RGB}{238,229,248}
\providecolor{basetext}{RGB}{90,90,90}\providecolor{relitext}{RGB}{46,125,50}
\providecolor{stabtext}{RGB}{40,53,147}\providecolor{evtext}{RGB}{106,27,154}
\providecolor{tabbasebest}{RGB}{216,216,216}\providecolor{tabrelibest}{RGB}{196,230,178}\providecolor{tabstabbest}{RGB}{208,218,243}\providecolor{tabevbest}{RGB}{216,198,242}

\caption{\textbf{Element Encoder Ablation} (\setting{1}, \setting{2}) on \dsv{\MAX, e, \Sigma^{*}}. $d$: encoder dimension. Perf.: performance. Rel.: reliability. Stab.: stability. Ev.: evasion, the AUT-ASR of white-box APG over the 30 test months. F1 reported with $\pm$SD and mean over five seeds. Shaded cells: best value per column. $^{\dagger}$Row~5 of Table~\ref{tab:ablation}.}

\adjustbox{max width=\columnwidth}{
\begin{tabular}{@{}lccc>{\columncolor{white}[\tabcolsep][0pt]}c@{}}
\toprule
 & \multicolumn{3}{c}{\setting{1}} & \multicolumn{1}{c@{}}{\setting{2}} \\
\cmidrule(lr){2-4}\cmidrule(l){5-5}
 & \tBand{1}{tabbaseband}{\color{basetext}\textbf{Perf.}} & \tBand{1}{tabreliband}{\color{relitext}\textbf{Rel.}} & \tBand{1}{tabstabband}{\color{stabtext}\textbf{Stab.}} & \multicolumn{1}{>{\columncolor{tabevband}[\tabcolsep][0pt]}c@{}}{\color{evtext}\textbf{Ev.} (\%)} \\
$d$ & \tBand{1}{tabbaseband}{\tH{F1}{\uparrow}} & \tBand{1}{tabreliband}{\tH{E-AURC}{\downarrow}} & \tBand{1}{tabstabband}{\tH{Sen[F1]}{\uparrow}} & \multicolumn{1}{>{\columncolor{tabevband}[\tabcolsep][0pt]}c@{}}{\tH{APG}{\downarrow}} \\
\midrule
\multicolumn{5}{@{}l}{\itshape OpenAI} \\
128 & 42.5$_{\pm1.9}$ & 5.4$_{\pm0.6}$ & $-$0.73$_{\pm0.07}$ & \cellcolor{tabevbest}\textbf{88.5}$_{\pm4.6}$ \\
256 & 44.2$_{\pm0.7}$ & 4.9$_{\pm0.3}$ & \cellcolor{tabstabbest}\textbf{\boldmath$-$0.55}$_{\pm0.11}$ & 89.4$_{\pm2.0}$ \\
512 & 47.2$_{\pm0.9}$ & 4.2$_{\pm0.2}$ & $-$0.76$_{\pm0.04}$ & 95.3$_{\pm3.4}$ \\
3072$^{\dagger}$ & 48.1$_{\pm0.8}$ & 4.2$_{\pm0.2}$ & $-$0.82$_{\pm0.05}$ & 96.5$_{\pm2.1}$ \\
\addlinespace[2pt]
\multicolumn{5}{@{}l}{\itshape hashed n-grams} \\
128 & \cellcolor{tabbasebest}\textbf{49.1}$_{\pm0.5}$ & 4.0$_{\pm0.1}$ & $-$0.88$_{\pm0.08}$ & 99.8$_{\pm0.1}$ \\
256 & 48.3$_{\pm0.8}$ & 4.2$_{\pm0.2}$ & $-$0.90$_{\pm0.07}$ & 99.2$_{\pm0.3}$ \\
512 & 48.0$_{\pm0.7}$ & 4.2$_{\pm0.1}$ & $-$0.88$_{\pm0.08}$ & 97.9$_{\pm1.1}$ \\
3072 & 47.0$_{\pm1.5}$ & 4.4$_{\pm0.2}$ & $-$0.92$_{\pm0.09}$ & 95.4$_{\pm2.5}$ \\
\addlinespace[2pt]
\multicolumn{5}{@{}l}{\itshape fastText} \\
128 & 49.0$_{\pm1.0}$ & \cellcolor{tabrelibest}\textbf{3.9}$_{\pm0.1}$ & $-$0.86$_{\pm0.04}$ & 99.9$_{\pm0.1}$ \\
256 & 48.2$_{\pm0.6}$ & 4.0$_{\pm0.2}$ & $-$0.81$_{\pm0.06}$ & 97.3$_{\pm2.6}$ \\
300 & 46.4$_{\pm1.2}$ & 4.5$_{\pm0.4}$ & $-$0.74$_{\pm0.17}$ & 97.7$_{\pm1.3}$ \\
\addlinespace[2pt]
\multicolumn{5}{@{}l}{\itshape SBERT (NLI)} \\
128 & 48.1$_{\pm0.4}$ & 4.1$_{\pm0.1}$ & $-$0.85$_{\pm0.07}$ & 99.7$_{\pm0.3}$ \\
256 & 47.4$_{\pm0.9}$ & 4.2$_{\pm0.1}$ & $-$0.80$_{\pm0.07}$ & 96.5$_{\pm1.7}$ \\
512 & 47.2$_{\pm0.4}$ & 4.2$_{\pm0.1}$ & $-$0.80$_{\pm0.04}$ & 95.1$_{\pm2.1}$ \\
768 & 37.9$_{\pm2.4}$ & 8.5$_{\pm2.7}$ & $-$0.68$_{\pm0.07}$ & 92.3$_{\pm4.6}$ \\
\bottomrule
\end{tabular}
}
\label{tab:encoders}
\end{table}

\section{Monotone Heads}
\label{app:ablation-cert}
\label{app:ablation-heads}
\noindent Table~\ref{tab:heads} compares a linear head with non-negative weights (linear$^{+}$) on $V$ alone and in addition to three monotone head variants for \dd{} and \ds{}. These are (i) calmm (per-coordinate piecewise-linear calibrators~\cite{gupta2016monotonic} followed by a min-max network~\cite{sill1997monotonic}), (ii) min-max~\cite{sill1997monotonic}, and (iii) an MLP with non-negative weights (MLP$^{+}$)~\cite{daniels2010monotone}. Our main results report \sys{} and \ddplus{} using the linear$^{+}$, the simplest of the four. On \ds{}, they reach similar F1 (linear$^{+}$ $38.4$, calmm $38.5$, min-max $37.3$, MLP$^{+}$ $36.4$; seed standard deviation $0.8$ to $1.6$). On \dd{}, linear$^{+}$ reaches $13.1$ F1, min-max $10.7$, calmm $11.5$, and MLP$^{+}$ $12.6$. On \ds{}, every monotone head has AUT-ASR $0.0\%$ under all five attacks. On \dd{}, it follows from monotonicity.

\begin{table}[t]
\centering
\providecommand{\tH}[2]{\shortstack{\rule[-0.6ex]{0pt}{2.6ex}#1\\$#2$}}
\providecommand{\tBand}[3]{\multicolumn{#1}{>{\columncolor{#2}[\tabcolsep][\tabcolsep]}c}{#3}}
\providecolor{tabbaseband}{RGB}{240,240,240}\providecolor{tabreliband}{RGB}{230,245,220}
\providecolor{tabstabband}{RGB}{235,240,250}\providecolor{tabevband}{RGB}{238,229,248}
\providecolor{basetext}{RGB}{90,90,90}\providecolor{relitext}{RGB}{46,125,50}
\providecolor{stabtext}{RGB}{40,53,147}\providecolor{evtext}{RGB}{106,27,154}
\providecolor{tabbasebest}{RGB}{216,216,216}\providecolor{tabrelibest}{RGB}{196,230,178}\providecolor{tabstabbest}{RGB}{208,218,243}\providecolor{tabevbest}{RGB}{216,198,242}

\caption{\textbf{Monotone Heads Ablation} (\setting{1}), grouped by family as in Table~\ref{tab:main-postattack}. Body: the layers below the head (---: none, the head reads $\mathbf x$ directly). Element and Domain as in Table~\ref{tab:ablation}. Perf.: performance. Rel.: reliability. Stab.: stability. Under \setting{2}, the AUT-ASR of white-box APG is $0.0\%$ for every row (\dd{} body: by monotonicity). Subscripts: $\pm$SD over five seeds. Shaded cells: best value per column within each block. $^{\dagger}$The head $g_{\text{lin}}$ of \drplus{}, \ddplus{} and \sys{}.}

\setlength{\tabcolsep}{4.2pt}%
\adjustbox{max width=\columnwidth}{
\begin{tabular}{@{}lccl cc>{\columncolor{white}[\tabcolsep][0pt]}c@{}}
\toprule
 & & & & \multicolumn{3}{c@{}}{\setting{1}} \\
\cmidrule(l){5-7}
 & & & & \tBand{1}{tabbaseband}{\color{basetext}\textbf{Perf.}} & \tBand{1}{tabreliband}{\color{relitext}\textbf{Rel.}} & \multicolumn{1}{>{\columncolor{tabstabband}[\tabcolsep][0pt]}c@{}}{\color{stabtext}\textbf{Stab.}} \\
Body & \rotatebox{90}{Element} & \rotatebox{90}{Domain} & Head & \tBand{1}{tabbaseband}{\tH{F1}{\uparrow}} & \tBand{1}{tabreliband}{\tH{E-AURC}{\downarrow}} & \multicolumn{1}{>{\columncolor{tabstabband}[\tabcolsep][0pt]}c@{}}{\tH{Sen[F1]}{\uparrow}} \\
\midrule
\multicolumn{7}{@{}l}{\textbf{Linear}} \\
--- & $\mathbf x$ & $V$ & linear$^{+}$$^{\dagger}$ & 23.6$_{\pm0.0}$ & 9.5$_{\pm0.0}$ & $-$0.83$_{\pm0.00}$ \\
\addlinespace[2pt]
\multicolumn{7}{@{}l}{\textbf{MLPs}} \\
\dd{} & $\mathbf x$ & $V$ & linear$^{+}$$^{\dagger}$ & \cellcolor{tabbasebest}\textbf{13.1}$_{\pm0.6}$ & 8.0$_{\pm0.1}$ & $-$0.69$_{\pm0.05}$ \\
 & & & calmm & 11.5$_{\pm0.6}$ & \cellcolor{tabrelibest}\textbf{7.5}$_{\pm0.1}$ & $-$0.53$_{\pm0.03}$ \\
 & & & min-max & 10.7$_{\pm0.3}$ & \cellcolor{tabrelibest}\textbf{7.5}$_{\pm0.1}$ & \cellcolor{tabstabbest}\textbf{\boldmath$-$0.42}$_{\pm0.07}$ \\
 & & & MLP$^{+}$ & 12.6$_{\pm0.3}$ & 7.7$_{\pm0.0}$ & $-$0.62$_{\pm0.06}$ \\
\addlinespace[2pt]
\multicolumn{7}{@{}l}{\textbf{MIL}} \\
\ds{} & $e$ & $\Sigma^{*}$ & linear$^{+}$$^{\dagger}$ & 38.4$_{\pm1.1}$ & \cellcolor{tabrelibest}\textbf{6.7}$_{\pm0.1}$ & \cellcolor{tabstabbest}\textbf{\boldmath$-$0.80}$_{\pm0.16}$ \\
 & & & calmm & \cellcolor{tabbasebest}\textbf{38.5}$_{\pm0.8}$ & 6.9$_{\pm0.3}$ & $-$0.88$_{\pm0.12}$ \\
 & & & min-max & 37.3$_{\pm0.8}$ & \cellcolor{tabrelibest}\textbf{6.7}$_{\pm0.2}$ & $-$0.89$_{\pm0.09}$ \\
 & & & MLP$^{+}$ & 36.4$_{\pm1.6}$ & \cellcolor{tabrelibest}\textbf{6.7}$_{\pm0.4}$ & $-$0.94$_{\pm0.09}$ \\
\bottomrule
\end{tabular}
}
\label{tab:heads}
\end{table}

\section{\texorpdfstring{Training on a Larger Vocabulary}{Larger Fixed Vocabulary}}
\label{app:ablation-vocab}
\noindent Table~\ref{tab:vocab100k} trains monotone detectors on $V_{100\text{k}}$. The $100{,}000$ most important features identified over the training period follow the same approach for $V$ for $10{,}000$ features.

\begin{table}[t]
\centering
\caption{\textbf{Larger Fixed Vocabulary} (\setting{1}, \setting{2}). Monotone detectors trained on $V_{100\text{k}}$. Perf.: performance. Rel.: reliability. Ev.: evasion, the AUT-ASR of white-box APG over the $30$ test months, here $0.0$ by monotonicity. F1 reported with mean$\pm$SD over five seeds.}

\begin{tabular}{@{}lcc>{\columncolor{white}[\tabcolsep][0pt]}c@{}}
\toprule
 & \multicolumn{2}{c}{\setting{1}} & \multicolumn{1}{c@{}}{\setting{2}} \\
\cmidrule(lr){2-3}\cmidrule(l){4-4}
 & \tBand{1}{tabbaseband}{\color{basetext}\textbf{Perf.}} & \tBand{1}{tabreliband}{\color{relitext}\textbf{Rel.}} & \multicolumn{1}{>{\columncolor{tabevband}[\tabcolsep][0pt]}c@{}}{\color{evtext}\textbf{Ev.} (\%)} \\
Detector & \tBand{1}{tabbaseband}{\tH{F1}{\uparrow}} & \tBand{1}{tabreliband}{\tH{E-AURC}{\downarrow}} & \multicolumn{1}{>{\columncolor{tabevband}[\tabcolsep][0pt]}c@{}}{\tH{APG}{\downarrow}} \\
\midrule
\drplus{} & 20.4$_{\pm0.0}$ & 10.0$_{\pm0.0}$ & 0.0$_{\pm0.0}$ \\
\monogbt{} & 28.2$_{\pm0.0}$ & 6.6$_{\pm0.0}$ & 0.0$_{\pm0.0}$ \\
\ddplus{} & 16.7$_{\pm0.7}$ & 8.0$_{\pm0.1}$ & 0.0$_{\pm0.0}$ \\
\dsv{\MAX, U, V_{100\text{k}}}, $g_{\text{lin}}$ & 40.5$_{\pm0.5}$ & 8.6$_{\pm0.6}$ & 0.0$_{\pm0.0}$ \\
\dsv{\MAX, e, V_{100\text{k}}}, $g_{\text{lin}}$ & 38.6$_{\pm0.9}$ & 7.9$_{\pm0.5}$ & 0.0$_{\pm0.0}$ \\
\bottomrule
\end{tabular}
\label{tab:vocab100k}
\end{table}

\section{Neural Baselines Channel Polarities}
\label{app:polarity-baselines}
\noindent Figure~\ref{fig:polarity-baselines} repeats Figure~\ref{fig:polarity} for our MLP-based baselines. Under APG:
\dd{} switches off malware-polarity channels (28\%$\to$13\%) and switches on benign-polarity channels (18\%$\to$29\%), evading 109 of 109;
\ramda{} switches off malware-polarity channels (59\%$\to$51\%), while its benign-polarity channels rise (39\%$\to$45\%; 307 of 466);
\hcc{} switches off malware-polarity channels (8\%$\to$7\%) and slightly raises benign-polarity channels (8\%$\to$9\%; 221 of 221);
\tif{} raises both malware (46\%$\to$48\%) and benign-polarity channels (38\%$\to$51\%; 128 of 128);
\drmd{} switches off malware-polarity channels (13\%$\to$8\%) and switches on benign-polarity channels (17\%$\to$22\%; 549 of 549).

\section{Attacks Under Active Learning}
\label{app:main-attack}
\noindent Table~\ref{tab:main-attack} attacks every detector monthly following \setting{2} (Section~\ref{sec:results-main}). APG achieves an ASR of $95.7\%$ or higher on every non-monotone baseline except \ramda{ ($64.4\%$)}. The non-monotone \ds{} models are evaded less than every other baseline except \ramda{ and \malcertain{}} under Informed Random, EvadeDroid and APG, however, APG still evades at $96.5\%$--$97.1\%$. Only the monotone detectors, \drplus{,} \monogbt{}, \ddplus{}, \dsv{\MAX, e, V, g_{\text{lin}}} and \sys{}, are never evaded, with \sys{} having the highest clean F1 of $38.4$.

\newcommand{\impl}[1]{\par\textbf{#1}.~}

\section{Hyperparameters}
\label{app:impl-detectors}
\noindent This appendix provides the hyperparameters for all detectors in our benchmarks, which were tuned on the calibration months (\ulcal{Jan--Jun 2022}), following~\cite{chen2023continuous} .

\impl{\dr{}~\cite{arp2014drebin}} scikit-learn LinearSVC with max iterations $50{,}000$ and $C{=}1$ as in~\cite{pendlebury2019tesseract}; on the grid $C\in\{10^{-3},\dots,10^{3}\}$ it is within $0.002$ calibration F1 of the best value.

\impl{\secsvm{}~\cite{demontis2019secureml}} Full-batch projected sub-gradient descent as implemented by the authors~\cite{pintor2022secml}, with $C{=}1$ as in~\cite{demontis2019secureml} and weight bounds $[-0.5,0.5]$ from its candidate set; on the grid $K\in\{0.125,0.25,\dots,8\}\times C\in\{0.1,1,10\}$ this setting is within $0.001$ calibration F1 of the best value. Line search with base $\eta{=}0.5$, convergence threshold $\epsilon{=}10^{-4}$ and at most $10{,}000$ iterations.

\impl{\drplus{}~\cite{fleshman2019nonnegative}} The optimizer of \secsvm{} with weight bounds $[0,2]$ and $C{=}1$ (grid $K\in\{0.5,1,2,4,8\}\times C\in\{0.1,1,10\}$).

\impl{\rf{}~\cite{onwuzurike2019mamadroid}} scikit-learn implementation with $101$ trees of maximum depth $64$ as in~\cite{onwuzurike2019mamadroid,pendlebury2019tesseract}.

\impl{\dd{}~\cite{grosse2017adversarial}} Two-hidden-layer MLP ($200{\times}200$, ReLU, dropout $0.5$) trained with SGD at batch size $1{,}000$ as in~\cite{grosse2017adversarial}. A grid of $19$ settings over lr $0.01$ to $0.8$ and $10$ to $80$ epochs selects lr${=}0.4$ and $40$ epochs. The published $10$ epochs with lr${=}0.1$ from~\cite{pendlebury2019tesseract} score $0.06$ lower calibration F1. \ddplus{}: the same network with all weights kept non-negative by projection after every step~\cite{fleshman2019nonnegative}. The same grid selects lr${=}0.8$ and $40$ epochs.

\impl{\ramda{}~\cite{li2021ramda}} Re-implementation of the authors' public code. VAE ($n_{\text{features}}{\to}600{\to}600{\to}80$ latent, dropout $0.1$), $50$ epochs, and detection MLP ($160{\to}40{\to}40{\to}2$), $30$ epochs, both with Adam at batch size $128$ on class-balanced batches. The grid lr $\{5{\times}10^{-5},10^{-4},10^{-3}\}\times$ VAE epochs $\{25,50,100\}$ selects lr${=}10^{-4}$ and $50$ epochs. The authors' lr${=}10^{-3}$ scores $0.07$ lower calibration F1. Their threshold $v{=}30$ is tied to their $379$ features, so we set $v$ at every fit such that $1\%$ of the benign training apps are flagged. For label selection, we combine both rules into one score, $\max\big(\tfrac{r-v}{r+v},\,P(\text{mal})-P(\text{ben})\big)$ with reconstruction error $r$, which is zero at the decision boundary.

\impl{\hcc{}~\cite{chen2023continuous}} Encoder ($512{\to}384{\to}256{\to}128$, ReLU) and MLP head ($128{\to}100{\to}100{\to}2$, dropout $0.2$), trained as published (APIGraph configuration) with the learning rate selected on the calibration months: SGD at lr${=}$$0.005$, decayed by $0.95$ every $10$ epochs, for $250$ epochs (batch size $1{,}024$), contrastive loss weight $\lambda{=}100$, margin $10$, AVClass2 families with generic tags as one \emph{unknown} family; monthly warm start with Adam at $5\%$ of the lr for $100$ epochs and pseudo-loss selection (grid: $8$ variants of the two published configurations). Without family labels the calibration F1 is unchanged ($78.4$ vs.\ $78.1$).

\impl{\drmd{}~\cite{mcfadden2026drmd}} PPO bandit with $4{\times}512$-unit actor and critic MLPs (LeakyReLU, dropout $0.5$), lr${=}2.5{\times}10^{-4}$ (annealed), clip $0.2$, entropy coefficient $0.01$, minibatch size $100$, and $5$ epochs on the newest $5{,}000$ labels, with temporal rewards, as published. The reject action is off since every detector classifies every app. The setting that a grid over lr $\{10^{-4},2.5{\times}10^{-4},5{\times}10^{-4}\}\times$ epochs $\{5,10\}$ selects scores no higher calibration F1.

\impl{\tif{}~\cite{zheng2026tif}} Authors' public code. Encoder ($512{\to}256{\to}200$) trained in two stages with Adam, first with a multi-proxy contrastive loss and then for up to $20$ epochs at lr${=}10^{-4}$ with an IRM penalty. The grid stage-1 lr $\{5{\times}10^{-5},10^{-4},10^{-3}\}\times$ epochs $\{20,30\}$ selects lr${=}5{\times}10^{-5}$ and $20$ epochs. The authors' lr${=}10^{-4}$ with $30$ epochs scores $0.009$ lower calibration F1.

\impl{\malcertain{}~\cite{li2024malcertain}} Base MLP ($200{\times}200$, dropout $0.4$, $30$ epochs) and a linear SVM ($C{=}1{,}000$) on $45$ uncertainty features that correct predicted benign apps, trained on the latest $20\%$ of the labeled pool.

\impl{\lsef{}~\cite{calzavara2023verifiable}} Authors' code with $k{=}0.5$. On binary features, the large-spread condition holds only for trees on disjoint features, so the model has $11$ trees of max depth $64$, one per disjoint feature subset. No setting is published for Android malware detection, so we select from the calibration months among the $10$ best settings from a search in 2021. The published size of $101$ trees with depth $6$ scores no F1 above $0$ in that search.

\impl{\lgbm{}~\cite{ke2017lightgbm}} Version $4.6.0$ with $31$ leaves, $500$ rounds, and lr${=}0.1$, without depth limit, subsampling, or class weighting (grid leaves $\{15,31,63,127,255\}\times$ rounds $\{50,100,200,500,1000\}\times$ lr $\{0.05,0.1\}$). Among the settings within $0.005$, we take the one with the best calibration F1 that has the fewest leaves and then the fewest rounds.

\impl{\monogbt{}~\cite{incer2018monotonic}} The \lgbm{} model above, with all features constrained to be non-decreasing, is an adaptation of~\cite{incer2018monotonic}, which uses XGBoost on a subset of the features. The same grid and selection rule give $31$ leaves, $1{,}000$ rounds, and lr${=}0.1$.

\impl{\ds{}, \sys{}} Frozen text-embedding-3-large ($3{,}072$ dims) over one set of all strings, $\phi$: $3{,}072{\to}512{\to}200$ (ReLU, dropout $0.3$), head $\rho$: linear, $200{\to}1$, monotone head $g_{\text{lin}}(\mathbf r){=}\mathrm{softplus}(\boldsymbol\theta)^{\top}\mathbf r{+}b$. Weighted BCE, Adam at lr${=}0.002$, batch size $64$, $10$ epochs, gradient clipping at $1$, then $5$ warm-up epochs in 2021; monthly fine-tuning for $10$ epochs at lr${=}10^{-3}$; the encoder is selected in the same way among $14$ candidates. With the optimizer and loss of \dd{} (SGD at lr${=}0.4$, batch size $1{,}000$, $40$ epochs, unweighted cross-entropy), calibration F1 falls from $79.3$ to $53.1$ for \dsv{\MAX, e, \Sigma^{*}, \rho} and from $76.9$ to $65.8$ for \sys{}.

\clearpage
\balance
\end{document}